\documentclass[aps,prd,preprint,draft,showpacs,eqsecnum,nofootinbib]{revtex4}
\usepackage{graphicx,epsf,amssymb}
\newcommand{\be}{\begin{equation}}
\newcommand{\ee}{\end{equation}}
\newcommand{\ba}{\begin{eqnarray}}
\newcommand{\ea}{\end{eqnarray}}

\newcommand{\bea}{\begin{eqnarray}}
\newcommand{\eea}{\end{eqnarray}}

\newcommand{\nn}{\nonumber}

\newcommand{\NeqFour}{{\cal N} =4}

\def\comma{\,,}
\def\fstop{\,.}

\def\outdent{\hskip -22mm}

\newif\ifdraft
\drafttrue
\newif\ifpreprint
\preprinttrue

\def\sect#1{section~{\ref{#1}}}
\def\app#1{appendix~{\ref{#1}}}

\def\fig#1{fig.~{\ref{#1}}}

\def\NeqFour{{\cal N}=4}
\def\NeqOne{{\cal N}=1}
\def\NeqZero{{\cal N}=0}
\def\ns{n_{\mskip-2mu s}}
\def\nf{n_{\mskip-2mu f}}
\def\Nc{N_{c}}

\def\Shift#1#2{{[#1,#2\rangle}}

\def\sandp#1.#2.#3{%
\left\langle\smash{#1}{\vphantom1}^{-}\right|{#2}%
\left|\smash{#3}{\vphantom1}^{+}\right\rangle}
\def\sandpp#1.#2.#3{%
\left\langle\smash{#1}{\vphantom1}^{+}\right|{#2}%
\left|\smash{#3}{\vphantom1}^{+}\right\rangle}
\def\sandmm#1.#2.#3{%
\left\langle\smash{#1}{\vphantom1}^{-}\right|{#2}%
\left|\smash{#3}{\vphantom1}^{-}\right\rangle}
\def\spab#1.#2.#3{\sandmm#1.#2.#3}
\def\spba#1.#2.#3{\sandpp#1.#2.#3}
\def\spaa#1.#2.#3.#4{\sandmp#1.{#2#3}.#4}
\def\spbb#1.#2.#3.#4{\sandpm#1.{#2#3}.#4}
\def\spa#1.#2{\langle#1\,#2\rangle}
\def\spb#1.#2{[#1\,#2]}
\def\spash#1.#2{\vphantom{\hat K}\spa{\smash{#1}}.{\smash{#2}}}
\def\spbsh#1.#2{\vphantom{\hat K}\spb{\smash{#1}}.{\smash{#2}}}
\def\lor#1.#2{\left(#1\,#2\right)}
\def\sand#1.#2.#3{%
\left\langle\smash{#1}{\vphantom1}^{-}\right|{#2}%
\left|\smash{#3}{\vphantom1}^{-}\right\rangle}
\def\sandpp#1.#2.#3{%
\left\langle\smash{#1}{\vphantom1}^{+}\right|{#2}%
\left|\smash{#3}{\vphantom1}^{+}\right\rangle}
\def\sandpm#1.#2.#3{%
\left\langle\smash{#1}{\vphantom1}^{+}\right|{#2}%
\left|\smash{#3}{\vphantom1}^{-}\right\rangle}
\def\sandmp#1.#2.#3{%
\left\langle\smash{#1}{\vphantom1}^{-}\right|{#2}%
\left|\smash{#3}{\vphantom1}^{+}\right\rangle}

\def\MSbar{$\overline{\rm MS}$}

\newbox\SlashedBox
\def\slashed#1{\setbox\SlashedBox=\hbox{#1}
\hbox to 0pt{\hbox to 1\wd\SlashedBox{\hfil/\hfil}\hss}#1}
\def\hboxtosizeof#1#2{\setbox\SlashedBox=\hbox{#1}
\hbox to 1\wd\SlashedBox{#2}}

\newbox\charbox
\newbox\slabox
\def\s#1{{      
        \setbox\charbox=\hbox{$#1$}
        \setbox\slabox=\hbox{$/$}
        \dimen\charbox=\ht\slabox
        \advance\dimen\charbox by -\dp\slabox
        \advance\dimen\charbox by -\ht\charbox
        \advance\dimen\charbox by \dp\charbox
        \divide\dimen\charbox by 2
        \raise-\dimen\charbox\hbox to \wd\charbox{\hss/\hss}
        \llap{$#1$}
}}

\def\eqn#1{eq.~(\ref{#1})}
\def\Eqn#1{Equation~(\ref{#1})}
\def\eqns#1#2{eqs.~(\ref{#1}) and~(\ref{#2})}

\def\e{\epsilon}
\def\eps{\epsilon}

\def\Li{\mathop{\rm Li}\nolimits}
\def\Ls{\mathop{\rm Ls}\nolimits}

\def\tree{{\rm tree}}
\def\oneloop{{1 \mbox{-} \rm loop}}

\def\cg{c_\Gamma}

\def\Kh{{\hat K}}

\def\sandp#1.#2.#3{%
\left\langle\smash{#1}{\vphantom1}^{+}\right|{#2}%
\left|\smash{#3}{\vphantom1}^{+}\right\rangle}
\def\ksl{\s{k}}
\def\Ksl{\s{K}}

\def\Den#1#2 {\prod\limits_{k=#1}^{#2} \spa{k}.{(k+1)}}

\def\Fact{{\cal F}}

\def\Res{\mathop{\rm Res}}
\def\tlambda{{\tilde\lambda}}

\def\S{{\chi }}

\def\Ll{\mathop{\rm L{}}\nolimits}
\def\Lnl{\mathop{\rm \widehat{L}{}}\nolimits}

\def\Kz{\mathop{\hbox{\rm K}}\nolimits_0}

\def\Mz{\mathop{\hbox{\rm M}}\nolimits_0}

\def\Cuth{{\widehat {C}}}
\def\CuthRat{{\widehat {CR}}}
\def\Remaining{{\widehat {R}}}

\def\Ph{{\hat K}}
\def\Pb{{\overline P}}
\def\Vertex{R}
\def\Rational{R}
\def\DiagrammaticRational{D}
\def\PureCut{C}
\def\Res{\mathop{\rm Res}}

\def\Overlap{O}
\def\Inf{\mathop{\rm Inf}}
\def\InfPart#1#2{\mathop{\rm Inf}_{#1}{#2}}

\newcommand{\Bmp}[1]{\langle #1\rangle}

\newcommand{\Asdef}{A_s}
\newcommand{\At}{A^{\tree}}
\newcommand{\Al}{A^{\NeqZero}}
\newcommand{\hR}{\hat{R}}
\newcommand{\hK}{\hat{K}}

\newcommand{\Gdef}[1]{\mathcal{G}_{#1}}
\newcommand{\Gdefd}[1]{\mathcal{G}^{\dag}_{#1}}
\newcommand{\Gbdef}[1]{\overline{\mathcal{G}}_{#1}}
\newcommand{\Gbdefd}[1]{\overline{\mathcal{G}}^{\dag}_{#1}}
\newcommand{\Uc}[2]{U^{#1}_{#2}}
\newcommand{\Kdef}[1]{\mathcal{K}(#1)}
\newcommand{\Hdef}{\mathcal{H}}
\newcommand{\Fc}{\mathcal{F}}
\newcommand{\Fb}{\overline{\mathcal{F}}}
\newcommand{\Ft}[2]{\Fc^{#1}_{#2}}
\newcommand{\Fbt}[2]{\Fb^{#1}_{#2}}
\newcommand{\Cc}{\mathcal{C}}
\newcommand{\Sc}{\mathcal{S}}

\def\fb{\overline{f}}
\def\Hb{\overline{H}}

\newcommand{\Asdefp}{A_s'}

\def\Nccoeff{{\cal{N}}}
\def\SelectSeta{{\S_j}}
\def\SelectSetb{{\hat\S_j}}

\newbox\ourfigbox
\def\SizedFigureWithCaption#1#2#3{%
\setbox\ourfigbox
  \hbox{\hss\epsfxsize #1 \epsfbox{#2}\hss}
\hbox to \wd\ourfigbox{\vbox{\noindent\copy\ourfigbox\break
\vskip -6mm      \hbox to \wd\ourfigbox{\hss#3\hss}}}
}

\def\llongrightarrow{%
\relbar\mskip-0.5mu\joinrel\mskip-0.5mu\relbar
     \mskip-0.5mu\joinrel\longrightarrow}
\def\inlimit^#1{\buildrel#1\over\llongrightarrow}
\def\dash{\hbox{-\kern-.02em}}

\begin{document}
\hfuzz 10 pt


\ifpreprint
\noindent
UCLA/06/TEP/27
\hfill SLAC--PUB--11936
\hfill SPhT-T06/070
\hfill hep-ph/0607014
\fi

\title{All One-loop Maximally Helicity Violating Gluonic Amplitudes in QCD%
\footnote{Research supported in part by the US Department of
 Energy under contracts DE--FG03--91ER40662 and DE--AC02--76SF00515}}

\author{Carola F. Berger}
\affiliation{Stanford Linear Accelerator Center \\
             Stanford University\\
             Stanford, CA 94309, USA
}

\author{Zvi Bern}
\affiliation{ Department of Physics and Astronomy, UCLA\\
\hbox{Los Angeles, CA 90095--1547, USA}
}

\author{Lance J. Dixon}
\affiliation{ Stanford Linear Accelerator Center \\
              Stanford University\\
             Stanford, CA 94309, USA
}

\author{Darren Forde and David A. Kosower}
\affiliation{Service de Physique Th\'eorique\footnote{Laboratory
   of the {\it Direction des Sciences de la Mati\`ere\/}
   of the {\it Commissariat \`a l'Energie Atomique\/} of France},
   CEA--Saclay\\
          F--91191 Gif-sur-Yvette cedex, France
}

\date{July 2006}

\begin{abstract}
We use on-shell recursion relations to compute analytically the one-loop
corrections to maximally-helicity-violating $n$-gluon amplitudes in QCD.
The cut-containing parts have been computed previously;
our work supplies the remaining rational parts for these amplitudes,
which contain two gluons of negative helicity and the rest positive,
in an arbitrary color ordering.  
We also present formulae specific to the six-gluon cases with helicities
$({-}{+}{-}{+}{+}{+})$ and $({-}{+}{+}{-}{+}{+})$,
as well as numerical results for six, seven, and eight gluons.
Our construction of the $n$-gluon amplitudes illustrates the relatively
modest growth in complexity of the on-shell-recursive calculation
as the number of external legs increases.  These amplitudes add to the growing
body of one-loop amplitudes known for all $n$, which are useful for
studies of general properties of amplitudes, including their 
twistor-space structure.
\end{abstract}

\pacs{11.15.Bt, 11.25.Db, 11.25.Tq, 11.55.Bq, 12.38.Bx \hspace{1cm}}

\maketitle


\renewcommand{\thefootnote}{\arabic{footnote}}
\setcounter{footnote}{0}

\section{Introduction}
\label{IntroSection}

The forthcoming experimental program at CERN's Large Hadron Collider
(LHC) will place new demands on theoretical calculations.  Finding and
understanding new physics in this environment will require the study
of processes with higher multiplicity than at the Tevatron. For
example, it is important to improve our understanding of missing
transverse energy in association with leptons and multiple jets,
arising from Standard-Model production of $W,Z\, +$ multi-jets.  Such
event classes form backgrounds to searches for supersymmetry
and other models of new electroweak physics.
In order to reach the precision required by searches for and measurements
of new physics, these processes need to be
computed to next-to-leading order (NLO), which entails the computation
of one-loop amplitudes.  The crucial case of $W,Z+4$ jet production
--- a background to supersymmetry searches when the $Z$ decays 
to a pair of neutrinos ---
involves the computation of amplitudes with seven external
particles, including the vector boson.  These are challenging
calculations.  State-of-the-art Feynman-diagrammatic
computations have only recently reached six-point
amplitudes~\cite{Denner,GieleGloverNumerical,EGZ,EGZ06,OtherNumerical}.

In this paper, we instead use on-shell methods to compute loop
amplitudes. On-shell methods were first developed at loop level ---
in the unitarity method --- providing the first practical method for
obtaining complete amplitudes using previously computed on-shell
amplitudes~\cite{Neq4Oneloop, Neq1Oneloop,BernMorgan, OneLoopReview,
UnitarityMachinery}%
\footnote{Unitarity has of course been a fundamental concept in quantum
field theory since its inception (see {\it e.g.} ref.~\cite{Eden}).
In more recent years, it has become a practical and efficient
computational method for reconstructing dimensionally
regularized~\cite{HV} amplitudes containing massless particles and
multiple kinematic invariants.}.
With a few
exceptions~\cite{BernMorgan,BMSTUnitarity}, applications of the unitarity
method were generally restricted to supersymmetric theories or to the
(poly)logarithmic part of QCD amplitudes. This limitation arose from the
greater complexity of calculations using $D$-dimensional
unitarity~\cite{vanNeerven}, required for reconstructing complete
QCD amplitudes including rational terms.  In the
past two years a number of related techniques have emerged, including
the application of maximally-helicity-violating (MHV)
vertices~\cite{CSW,Risager} to loop calculations~\cite{BST,BBSTQCD},
and the use~\cite{CachazoAnomalyBCF7} of the holomorphic anomaly.
However, these techniques also suffer from the same limitations.
Recent improvements to the unitarity
method~\cite{BCFUnitarity,BMSTUnitarity} use {\it complex\/} momenta
within generalized
unitarity~\cite{ZFourPartons,TwoLoopSplit,NeqFourSevenPoint}, 
allowing, for example, a simple and
purely algebraic determination of all box integral coefficients.  In
ref.~\cite{BBCFSQCD}, Britto, Buchbinder, Cachazo and Feng developed
efficient techniques for evaluating generic one-loop unitarity cuts,
by using spinor variables and performing the cut integration via
residue extraction.  Applying these ideas, Britto,
Feng and Mastrolia~\cite{BFM} computed the
cut-containing terms for the most complex six-gluon helicity amplitudes,
the terms with a scalar circulating in the loop and gluon helicity
assignments $({-}{+}{-}{+}{-}{+})$ and $({-}{-}{+}{-}{+}{+})$.
The cut-containing terms for other helicity configurations, and for other
components of the amplitudes (within a supersymmetric decomposition~\cite{GGGGG}), 
were obtained in refs.~\cite{Neq4Oneloop,Neq1Oneloop,NeqOneNMHVSixPt,DunbarBoxN1,%
RecurCoeff,BBCFSQCD}.

The first use of on-shell methods to obtain state-of-the art
QCD amplitudes dates to the construction of the
$Z\rightarrow 4$~parton one-loop matrix elements~\cite{ZFourPartons}
(or equivalently, by crossing, the virtual corrections for $pp
\rightarrow W, Z + 2 \,{\rm jets}$).
These matrix elements have been incorporated into several
numerical programs for NLO corrections to a variety of collider
processes~\cite{ZFourPartonsNLODS,ZFourPartonsNLO,MCFM}.
The technique used in ref.~\cite{ZFourPartons} was to first obtain
the (poly)logarithmic terms in the amplitudes via the unitarity method.
The rational terms were found by constructing functions with the proper
on-shell factorization properties, by matching the known behavior as
two partons become parallel (collinear singularities),
or as invariants formed out of three or
more partons vanish (multi-particle poles).
This rational-term reconstruction method did not
achieve widespread application because it was unclear how to make it
systematic.  In particular, as the number of external legs $n$ increases,
it becomes very difficult to build properly factorizing functions;
the number of candidate terms in an ansatz for the
rational terms grows rapidly with $n$.

One-loop on-shell recursion relations~\cite{Bootstrap,Genhel} provide
a means for overcoming this difficulty.  They allow for a practical and
systematic construction of the rational terms of loop amplitudes.
Such recursion relations were written down at tree level by
Britto, Cachazo and Feng~\cite{BCFRecurrence}.  Their work was
stimulated by the compact forms of seven- and higher-point tree
amplitudes~\cite{NeqFourSevenPoint,NeqFourNMHV,RSVNewTree} that
emerged from studying infrared consistency
equations~\cite{UniversalIR} for one-loop amplitudes (computed using
the unitarity-based method), and by the connection between
twistor space and complexified momenta noted in
ref.~\cite{WittenTopologicalString}.
Britto, Cachazo, Feng and Witten~\cite{BCFW} then proved the
tree-level on-shell recursion relations using a special
continuation, or `shift', of the external momenta by a complex variable
$z$. The proof essentially involves only Cauchy's residue theorem,
where the residues in the $z$ plane are determined by the
amplitudes' factorization properties (for complex momenta).
The remarkable generality and simplicity of the
proof enabled widespread application at tree
level~\cite{LuoWen,TreeRecurResults,MoreTreeRecurResults},
including to theories with massive
particles~\cite{GloverMassive,Massive,FMassive}.

The extension of on-shell recursive methods to one-loop
amplitudes was developed in
refs.~\cite{OnShellRecurrenceI,Qpap,Bootstrap,FordeKosower,Genhel}.
The essential inputs to this approach are the cut-constructible parts 
of a desired amplitude --- those terms containing polylogarithms,
logarithms, and associated $\pi^2$ terms ---
and appropriate tree and loop amplitudes with fewer legs,
which appear in factorization limits of the amplitude.
Closed-form expressions were found for several sequences of
one-loop amplitudes with an arbitrary number of gluons.
In particular, the amplitudes for `MHV' helicity configurations,
in which two of the gluons have negative helicity and the
rest have positive helicity, were computed for the special case
that the two negative-helicity gluons are adjacent in the
color ordering~\cite{Bootstrap,FordeKosower}.

In this paper, we use the same methods to compute the
rational parts, and thereby the complete expressions,
for all remaining $n$-gluon MHV amplitudes, those for
which the two negative-helicity gluons are {\it not} color-adjacent.
Under a supersymmetric decomposition~\cite{GGGGG}, $n$-gluon
amplitudes may be thought of as composed of $\NeqFour$ and $\NeqOne$
supersymmetric pieces together with a non-supersymmetric ($\NeqZero$)
scalar loop contribution.  Using the unitarity method, the $\NeqFour$
MHV contributions were computed in ref.~\cite{Neq4Oneloop}, and the
$\NeqOne$ terms in ref.~\cite{Neq1Oneloop}.  The polylogarithmic
contributions to the scalar-loop components of the MHV amplitudes
were obtained in ref.~\cite{BBSTQCD} using MHV vertices~\cite{CSW,BST};
these contributions serve as direct input to our recursive construction
of the rational terms.

A number of new aspects must be considered in order to
construct loop-level on-shell recursion relations.
The most obvious feature is that the shifted amplitude $A(z)$
contains branch cuts in the $z$ plane.  We will handle this feature
as in refs.~\cite{Bootstrap,Genhel}, by subtracting the cut-containing
terms in the amplitude, along with certain additional `cut completions'
that cancel spurious singularities, before applying Cauchy's theorem
to the remaining rational part.

Secondly, we must ensure that the residues
of all poles at finite $z$ are known.
The factorization properties of amplitudes with real external
momenta are well understood~\cite{TreeReview, Neq4Oneloop,
BernChalmers, OneLoopSplitUnitarity}.  They completely determine the
complex factorization properties of tree amplitudes~\cite{BCFW}.
However, the same statement is not true at loop level.
In general, non-supersymmetric loop
amplitudes contain `unreal poles' which appear for complex momenta but
are absent for real momenta.  Since these unreal poles are not yet
fully understood, the best strategy is to consider complex-momentum
shifts that avoid channels with unknown factorization properties.
As discussed in ref.~\cite{Genhel}, for $n$-gluon amplitudes
the problematic channels always have precisely two identical-helicity
gluons on one side of the factorization.  In this paper, we consider
MHV amplitudes with two negative-helicity gluons, and we elect to
shift the momenta of these two gluons.  Because these gluons are
color-adjacent to positive-helicity gluons, the two-particle channels
will always have opposite gluon helicities.  Hence there are no
problematic channels in our construction.

The straightforward application of Cauchy's theorem underlying our
computation requires that the amplitude fall off as the shift parameter
$z$ is taken to infinity.  A more elaborate construction is
needed~\cite{Genhel} when this requirement is not satisfied.  We
can see from known four- and five-point one-loop amplitudes that the
requirement is indeed satisfied for our choices of shift momenta.  We
assume that the same holds for $n>5$. We can confirm this assumption 
at the end of the calculation (and we have done so for $n$ up to 8), 
by checking both factorization properties 
and reflection or `flip' symmetries of the computed amplitudes.

In addition to determining rational terms in amplitudes efficiently,
on-shell recursion relations can in some cases provide an
effective way to compute the (rational) coefficients of
integral functions appearing in amplitudes~\cite{RecurCoeff}.
Indeed, this method was used to determine the coefficients of all
integral functions appearing in the split-helicity $n$-gluon amplitude,
for which identical-helicity gluons are nearest neighbors in the
color ordering.

Our interest in constructing all-multiplicity amplitudes
stems partly from the desire to study the growth in complexity of the
amplitudes as the number of external partons increases.  A difficulty
with previous computational methods has been the rather rapid growth
in complexity with the number of external legs. In
contrast, with our bootstrap construction we find only mild growth,
at least for the MHV and split-helicity configurations,
allowing us to obtain closed-form analytic expressions.
Furthermore, experience has shown that analytic all-$n$
expressions provide a wealth of intuition into the general structure
of the scattering amplitudes.

In a very interesting series of recent papers, Xiao, Yang and Zhu have
obtained the rational terms for all independent one-loop six-gluon amplitudes
using Feynman diagrams, by applying spinor simplifications together
with integrations that target only the rational
terms~\cite{XYZgen5,XYZ6}.  When combined with the previously-obtained
cut parts~\cite{NeqFourNMHV,Neq1Oneloop,NeqOneNMHVSixPt,BBSTQCD,%
BBCFSQCD,BFM}, these results provide analytic expressions for all
remaining six-gluon helicity configurations in QCD.  We have compared
our results for the MHV case and find complete agreement.  We have
also compared our results for six gluons with the numerical results of
Ellis, Giele and Zanderighi~\cite{EGZ06} and find agreement.

This paper is organized as follows. In the next section we briefly remind
the reader of our notation. We give an overview of the
on-shell bootstrap approach in section~\ref{RecursionReviewSection}.
We apply this method to the recursive construction of MHV
amplitudes in section~\ref{SixSevenSection}, giving details of
the method for five and six external gluons.
We also give numerical
values at selected momentum configurations
for the six-, seven- and eight-point amplitudes.  These values
can serve as a reference for future phenomenological
implementations. Section~\ref{AllnSection} contains the all-multiplicity
expression for the complete rational parts of one-loop MHV
amplitudes.  We summarize our results and give an outlook in
\sect{ConclusionSection}. In the appendices we quote previously
obtained results for the (poly)logarithmic parts of MHV amplitudes.
We also give analytical expressions for the
newly-computed rational parts of the six-point MHV amplitudes.

\section{Notation}
\label{NotationSection}

In this paper, we shall follow the same notation as defined in
section~II of our companion paper~\cite{Genhel}.  We therefore refer
the reader to that paper for notation.  Our amplitudes will be
expressed in terms of spinor inner products~\cite{SpinorHelicity}.  As
is standard for one-loop QCD amplitudes, we present color-ordered
amplitudes~\cite{TreeColor,TreeReview,BKColor,OneLoopReview}.  We need to present only the
leading-color contributions, because the subleading-color partial
amplitudes for a gluon in the loop are given by a sum over
permutations of the leading-color ones~\cite{Neq4Oneloop}.  (For
fundamental representation particles such as quarks in the loop the
leading contributions give the entire expression directly.)

We shall use the supersymmetric decomposition of $n$-gluon QCD
amplitudes~\cite{GGGGG, Neq1Oneloop} to rewrite the contributions of
different spin states circulating in the loop in terms of
supersymmetric and non-supersymmetric components,
\ba
A_{n;1}^{[1/2]} &=& A^{\NeqOne}_{n;1} - A^{\NeqZero}_{n;1} \,, \label{An1/2}\\
A_{n;1}^{[1]} &=& A^{\NeqFour}_{n;1} -4 A^{\NeqOne}_{n;1}+ A^{\NeqZero}_{n;1}
\,,
\label{An1}
\ea
where $A_{n;1}^{[J]}$ denotes an $n$-gluon amplitude with a particle
of spin $J$ circulating in the loop.
The non-supersymmetric amplitudes, denoted by $\NeqZero$, are
just the contributions of a complex scalar circulating in the loop,
$A^{\NeqZero}_{n;1} \equiv A^{[0]}_{n;1}$. The amplitudes
$A^{\NeqOne}_{n;1}$ and $A^{\NeqFour}_{n;1}$ represent
the contribution of an $\NeqOne$ chiral multiplet consisting of a scalar
and fermion, and an $\NeqFour$ multiplet
consisting of one gluon, four gluinos and six real scalars, respectively.

The supersymmetric decomposition is convenient because it separates the
amplitudes into pieces with differing analytic properties.
The supersymmetric parts
can be constructed completely from unitarity cuts in
four dimensions~\cite{Neq4Oneloop,Neq1Oneloop}, without additional
rational contributions.  The polylogarithms and logarithms of the
$\NeqZero$ non-supersymmetric contributions may also be computed from
the four-dimensional unitarity cuts, but in this case there are in
general additional non-trivial rational contributions.

The leading-color QCD amplitudes are expressible in terms of
the different supersymmetric components~(\ref{An1/2})--(\ref{An1})
via,
\begin{eqnarray}
A_{n;1}^{\rm QCD} &=&
A^{\NeqFour}_{n;1} -4 A^{\NeqOne}_{n;1}+ (1-\eps\delta_R ) A^{\NeqZero}_{n;1}
+ {\nf\over\Nc} \Bigl( A^{\NeqOne}_{n;1}- A^{\NeqZero}_{n;1}\Bigr)
+ {n_{s}\over\Nc} A^{\NeqZero}_{n;1} \,,
\label{AnQCD}
\end{eqnarray}
where $\nf$ is the number of active quark flavors in QCD.  We also
allow for a term proportional to the number of active fundamental
representation scalars $\ns$, which vanishes in QCD.  We regulate
the infrared and ultraviolet divergences of one-loop
amplitudes dimensionally.  The regularization-scheme-dependent
parameter $\delta_R$ specifies the number of helicity states of
internal gluons to be $(2-\epsilon\delta_R)$. For the 't~Hooft-Veltman
scheme~\cite{HV} $\delta_R = 1$, while in the four-dimensional
helicity scheme~\cite{BKStringBased,Neq1Oneloop,OtherFDH} $\delta_R =
0$.

The amplitudes that we shall present in this paper are unrenormalized.
To carry out \MSbar\ renormalization, one should subtract from the
leading-color partial amplitudes $A_{n;1}$ the quantity,
\begin{equation}
  c_\Gamma \left[{n-2\over2}{1\over\e}\left({11\over3}
  - {2\over3} {\nf\over N_c} - {1\over3}{\ns\over N_c} \right) \right]
  A_n^\tree\,,
\label{MSbarsubtraction}
\end{equation}
for $D = 4 - 2 \e$ dimensions,
where,
\begin{equation}
\cg = {1\over(4\pi)^{2-\eps}}
  {\Gamma(1+\eps)\Gamma^2(1-\eps)\over\Gamma(1-2\eps)}\, .
\label{cgdefn}
\end{equation}
%


\section{Review of Recursive Bootstrap Approach}
\label{RecursionReviewSection}

On-shell recursion relations are an efficient way to obtain
the rational parts of amplitudes directly from their analytic properties.
In this section, we first recapitulate the proof~\cite{BCFW} of
the tree-level recursion relations~\cite{BCFRecurrence}.
The proof relies only on elementary complex analysis and on
general factorization properties satisfied by any scattering amplitude.
The generality of the proof permits its extension to loop level,
along the lines of refs.~\cite{Bootstrap,Genhel},
as we summarize in the rest of the section.

The proof of the tree-level relations employs a
parameter-dependent `$[j,l\rangle$' shift of two of the external
massless spinors, $j$ and $l$, in an $n$-point process,
\begin{eqnarray}
[j,l\rangle: \qquad
&\tlambda_j &\rightarrow \tlambda_j - z\tlambda_l \,, \nonumber\\
&\lambda_l &\rightarrow \lambda_l + z\lambda_j \,,
\label{SpinorShift}
\end{eqnarray}
where $z$ is a complex parameter.  This shift in spinor variables
corresponds to shifting two of the momenta to complex values,
\begin{eqnarray}
&k_j^\mu &\rightarrow k_j^\mu(z) = k_j^\mu -
       {z\over2}{\sand{j}.{\gamma^\mu}.{l}},\nonumber\\
&k_l^\mu &\rightarrow k_l^\mu(z) = k_l^\mu +
       {z\over2}{\sand{j}.{\gamma^\mu}.{l}} \,,
\label{MomentumShift}
\end{eqnarray}
so that they remain massless, $k_j^2(z) = k_l^2(z) = 0$, and overall
momentum conservation is maintained.  (Similar considerations
apply to cases with massive particles~\cite{GloverMassive,Massive,FMassive}.)
An on-shell amplitude containing the momenta $k_j$ and $k_l$
then becomes parameter-dependent as well,
\begin{equation}
A(z) = A(k_1,\ldots,k_j(z),k_{j+1},\ldots,k_l(z),k_{l+1},\ldots,k_n)\,,
\end{equation}
where the physical amplitude is recovered by taking $z = 0$.

At tree level, $A(z)$ is an analytic function containing only
poles, so we may exploit Cauchy's theorem to construct it from its
residues. Assuming $A(z)\rightarrow 0$ as $z\rightarrow\infty$,
then there is no `surface term' in the contour integral around
the circle at infinity, so it vanishes,
\begin{equation}
{1\over 2\pi i} \oint_C {dz\over z}\,A(z)  = 0\,.
\label{ContourInt}
\end{equation}
Evaluating the integral as a sum of residues allows us to solve for the
physical amplitude $A(0)$,
\begin{equation}
A(0) = -\sum_{{\rm poles}\ \alpha} \Res_{z=z_\alpha}  {A(z)\over z}\,.
\label{NoSurface}
\end{equation}
At tree level, the requirement that $A(z)$ vanishes as
$z\rightarrow\infty$ is rather mild; wide classes of shifts
exist that satisfy this 
requirement~\cite{BCFRecurrence, BCFW,GloverMassive}.

\begin{figure}[t]
\centerline{\epsfxsize 2 truein\epsfbox{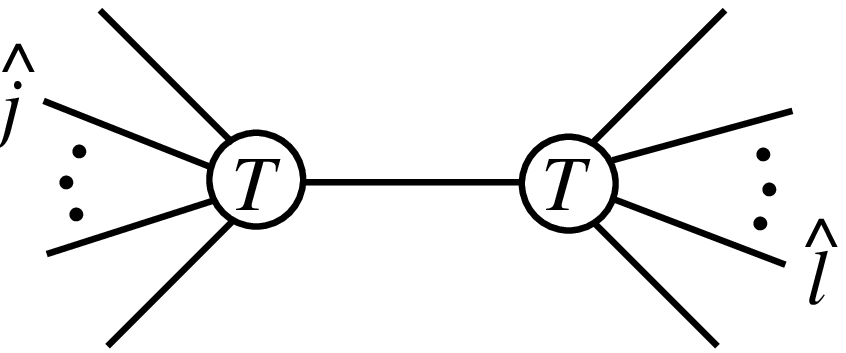}}
\caption{Schematic representation of tree-level recursive
contributions.  The labels `$T$' refer to tree vertices which are
on-shell amplitudes. The momenta $\hat{\jmath}$ and $\hat{l}$
are shifted, on-shell momenta, evaluated according to
\eqns{eq:shifted_mom_over_1}{eq:shifted_mom_over_2}. }
\label{TreeGenericFigure}
\end{figure}

The residues in \eqn{NoSurface} may be obtained using the generic
factorization properties that any amplitude must satisfy.
Consider the factorization channel $s_{\alpha \ldots \beta}
\rightarrow 0$. In this limit, the amplitude behaves
as~\cite{TreeReview},
\def\indentA{\hskip 7mm}
\begin{eqnarray}
A^\tree_n(k_1,\ldots,k_n) &\longrightarrow &
\hskip -.1cm
 \sum_{h = \pm}
A^\tree(k_{\alpha},\ldots, k_{\beta},-K_{\alpha \ldots \beta}^h)
\label{TreeFactorization} \\
&& \null \hskip 2.5 cm \indentA\times
{i\over s_{\alpha \ldots \beta}} \times
A^\tree(k_{\beta+1},\ldots, k_{\alpha -1},K_{\alpha \ldots \beta}^{-h})
 \, , \hskip 1.2 cm \nn
\end{eqnarray}
where $h = \pm 1$ is the helicity of the intermediate gluon.

Now consider the effect of the shift (\ref{MomentumShift}).
If the shifted legs $j$ and $l$ are on opposite sides of the
pole, as depicted in \fig{TreeGenericFigure}, and leg $j$ is between
legs $\alpha$ and
$\beta$ in the color ordering, then under the shift,
\be
s_{\alpha \ldots \beta} \rightarrow s_{\alpha \ldots \beta}
- z \sand{j}.{\Ksl_{\alpha\ldots \beta}}.{l} \,,
\ee
so we have a simple pole in $z$ in this channel.
Using elementary complex variable theory we may then evaluate the
residue, using the rule,
\ba
\Res_{z = z_{\alpha\beta}} {f(z) \over z- z_{\alpha\beta}} =
f(z_{\alpha\beta}) \,.
\ea
This rule holds for any analytic $f(z)$ with no additional
singularities at $z = z_{\alpha\beta}$; in particular, it holds for
the product of shifted tree amplitudes appearing as coefficients of
each pole in $z$.  (Channels with only two isolated legs correspond to
a collinear limit, which has a somewhat different factorization for real
momenta than eq. (3.6).  It turns out that for complex momenta as
needed here, this case can be treated on the same footing as
multiparticle factorization.)

Using this rule and summing over all residues then gives us the tree-level
on-shell recursion relation,
\def\indentA{\hskip 7mm}
\begin{eqnarray}
A^\tree_n(k_1,\ldots,k_n)  &=&
\hskip -.1cm
 \sum_{{\rm partitions}\, P}\, \sum_{h = \pm}
A^\tree(k_{P_1},\ldots,\hat k_j,\ldots,k_{P_{-1}},-\hat K_P^h)
\label{TreeRecursion} \\
&& \null \hskip 2.5 cm \indentA\times
{i\over K_P^2} \times
A^\tree(k_{\Pb_1},\ldots,\hat k_l,\ldots,k_{\Pb_{-1}},\hat K_P^{-h})
 \, , \hskip 1.2 cm \nn
\end{eqnarray}
where
\bea
K_P &\equiv& k_{P_1} + k_{P_2} + \cdots +      k_j  + \cdots + k_{P_{-1}}\,,\\
\hat K_P &\equiv& k_{P_1} + k_{P_2} + \cdots + \hat k_j  + \cdots + k_{P_{-1}}\,,
\eea
and $h = \pm 1$ labels the helicity of the intermediate state. The hatted
momenta are defined below.
We have defined a partition $P$ to be a set of two or more
cyclically-consecutive momentum labels containing $j$, such that the
complementary set $\Pb$ consists of two or more cyclically-consecutive
labels containing $l$,
\begin{eqnarray}
 P &\equiv& \{ P_1, P_2, \ldots, j, \ldots, P_{-1} \} \,,
\label{PartitionDef} \\
 \Pb &\equiv& \{ \Pb_1, \Pb_2,
  \ldots, l, \ldots, \Pb_{-1} \} \,, \nn \\
 P &\cup& \Pb = \{ 1,2,\ldots,n \} \,, \nn
\end{eqnarray}
which ensures that the sum of momenta in each partition is
$z$-dependent.  The shifted complex on-shell momenta
$\hat k_j$, $\hat k_l$ and $\hat K_P$ in \eqn{TreeRecursion}
are determined by solving
the on-shell condition, $\hat K_P^2 = 0$, for $z$, giving the location
of the pole,
\be
z_P =  {K_P^2 \over \sand{j}.{\Ksl_P}.{l} } \,.
\ee
Substituting $z \rightarrow z_P$ in the shifted momenta
(\ref{MomentumShift}) gives,
\ba
\hat k_j^\mu &\equiv& k_j(z_P) =
         k_j^\mu  - {1\over 2} {K_P^2 \over \sand{j}.{\Ksl_P}.{l} }
                                \sand{j}.{\gamma^\mu}.{l} \,,
\label{eq:shifted_mom_over_1}\\
\hat k_l^\mu &\equiv& k_l(z_P) =
         k_l^\mu  + {1\over 2} {K_P^2 \over \sand{j}.{\Ksl_P}.{l} }
                                \sand{j}.{\gamma^\mu}.{l} \,,
\label{eq:shifted_mom_over_2}
\ea
which define $\hat k_j$ and $\hat k_l$ appearing in the recursion
relation (\ref{TreeRecursion}).

At loop level, a number of new features arise.  In particular,
obtaining an on-shell recursion relation requires dealing with branch
cuts, spurious singularities, and in some cases, the treatment of
factorization using complex momenta, which can differ from `ordinary'
factorization using real momenta.  We refer to channels whose
complex factorization properties are non-universal, or at least
not yet fully understood, as `non-standard'.  In non-standard
factorization channels, double poles and unreal poles may appear in
two-particle channels with like-helicity
gluons~\cite{OnShellRecurrenceI,Qpap}.
Fortunately, nonstandard channels for $n$-gluon amplitudes
only occur when two adjacent gluons, one of which is shifted and
one of which is not, have the same helicity.
As mentioned in the introduction, for the MHV amplitudes we
consider here, our electing to shift the momenta
of the two negative-helicity legs avoids all nonstandard channels.

To set up a loop-level on-shell recursion we decompose the amplitude
into `pure-cut' and `rational' pieces,
\begin{equation}
A_n(z) =  \cg {} \Bigl[\PureCut_n(z) + \Vertex_n(z) \Bigr] \,.
\label{ACREq}
\end{equation}
The rational parts $R_n$ are defined
by setting all logarithms, polylogarithms, and associated $\pi^2$
terms to zero,
\begin{equation}
\Vertex_n \equiv {1\over \cg} A_n\Bigr|_{\rm rat} \equiv
{1\over \cg} A_n\biggr|_{\ln, \Li, \pi^2 \rightarrow 0} \,.
\label{RationalDefinition}
\end{equation}
The pure-cut part $C_n$ consists of the remaining terms,
all of which must contain logarithms, polylogarithms, or $\pi^2$ terms.
The cut-containing terms have already been computed for the MHV
case~\cite{BBSTQCD}.
Our task is to obtain the rational terms (\ref{RationalDefinition})
via on-shell recursion.

With the decomposition~(\ref{ACREq}), the evaluation of
a contour integral like~(\ref{ContourInt}), but for $R_n(z)$,
is complicated, in general, by the appearance of spurious singularities.
Such singularities are at kinematical points that do not correspond
to any physical singularity. They arise in the course of integral
reductions and often cannot be removed by algebraic means.  Although the
final amplitude cannot contain unphysical spurious singularities,
in many cases they cancel only between (poly)logarithmic and
rational terms.  If we were to set up a contour integral
(\ref{ContourInt}) over the rational terms $R_n(z)$ instead of $A_n(z)$,
it would pick up contributions from unphysical poles.

As a concrete example of a spurious singularity, consider
the function,
\begin{equation}
{\ln(r)\over (1-r)^2} \,,
\label{sampleL1pure}
\end{equation}
where $r$ is a ratio of momentum invariants. This function
has a spurious singularity as $r\rightarrow 1$.
In the full amplitude this function appears in the combination,
\begin{equation}
{\ln(r)+1-r\over (1-r)^2} \,,
\label{sampleL1}
\end{equation}
which is finite as $r\rightarrow 1$.

To avoid the additional complication of spurious singularities, a
good approach is to instead `complete' the cut contribution by
replacing functions like~\eqn{sampleL1pure} with non-singular
combinations like~\eqn{sampleL1}. Although this procedure is not
unique, any function that is free of unphysical spurious
singularities in $z$ is satisfactory.  We denote this cut completion
by $\Cuth_n$,
\begin{equation}
\Cuth_n(z) = C_n(z) + \widehat{CR}_n(z) \,,
\label{CuthCCR}
\end{equation}
where $\widehat{CR}_n(z)$ are the rational functions added in order to
cancel the unphysical spurious singularities in $z$.  For a given
shift we do not need to remove all spurious singularities, but only
those that depend on $z$.  Since we have added in rational terms to
the completed cuts, we should subtract them out from the remaining
rational terms.  Defining
\be
\Remaining_n(z) =   \Vertex_n(z) - \widehat{CR}_n(z)\, ,
\ee
we have,
\be
A_n(z) = \cg {}\Bigl[ \Cuth_n(z) + \Remaining_n(z) \Bigr] \,.
\label{CompletedCutDecomposition}
\ee

With the decomposition~(\ref{CompletedCutDecomposition}), as shown
in ref.~\cite{Genhel}, the physical amplitude is given by
\be
A_n(0)
= \Inf A_n +  \cg {} \Biggl[ \Cuth_n(0) - \Inf\Cuth_n + R_n^D + O_n
 \Biggr]
\,,
\label{BasicFormula}
\ee
where $\Inf A_n$ is the potential contribution to the amplitude from
large $z$, $\Cuth_n(0)$ is the completed-cut contribution,
$\Inf\Cuth_n$ is the potential large-$z$ spurious behavior of the
completed cut, which must be subtracted off, $R_n^D$ are the recursive
diagram contributions, and the `overlap' terms $O_n$ remove double
counting between the recursive diagrams and the rational terms that
were added to complete the cuts.

Let us discuss each of the contributions to the amplitude in
\eqn{BasicFormula} in turn.  First consider the $\Inf A_n$ terms.
As discussed in the introduction, using the known four- and five-gluon
MHV amplitudes~\cite{BKStringBased,GGGGG}, one can shift the two
negative-helicity gluons and observe that the resulting $A_n(z)$
falls off as $z$ approaches infinity, that is, $\Inf A_n = 0$.
We assume that this property continues to hold for more than
five gluons.  We check this assumption after the fact,
by examining various symmetries and limits of the computed
amplitudes.

Now consider the $\Inf\Cuth_n$ subtraction. In general, since
$\Cuth_n(z)$ is required only to remove unphysical singularities
at finite values of $z$, it is possible that the chosen $\Cuth_n(z)$
has non-vanishing behavior as $z\rightarrow \infty$.  This would introduce
spurious large $z$ behavior in $\hat{R}_n(z)$, so we simply subtract
off this large $z$ behavior with $\Inf\Cuth_n$.  If we assume that
no logarithms of $z$ appear in any surviving terms as $z
\rightarrow \infty$, the procedure for computing $\Inf\Cuth_n$ is
to start with $\Cuth_n(z)$ and series expand around $w = 1/z = 0$,
keeping only the $w^0$ term.  This term is $\Inf\Cuth_n$.  (The
assumption can easily be checked in any given case.)  In many
cases, including all cases in this paper, $\Inf \Cuth_n$ goes to a
constant at large $z$, and it can be obtained directly by taking
the limit,
\be
 \Inf \Cuth_n \equiv \lim\limits_{z \rightarrow \infty}
 \Cuth_n(z) \,.
\ee
%

\begin{figure}[t]
\centerline{\epsfxsize 5 truein\epsfbox{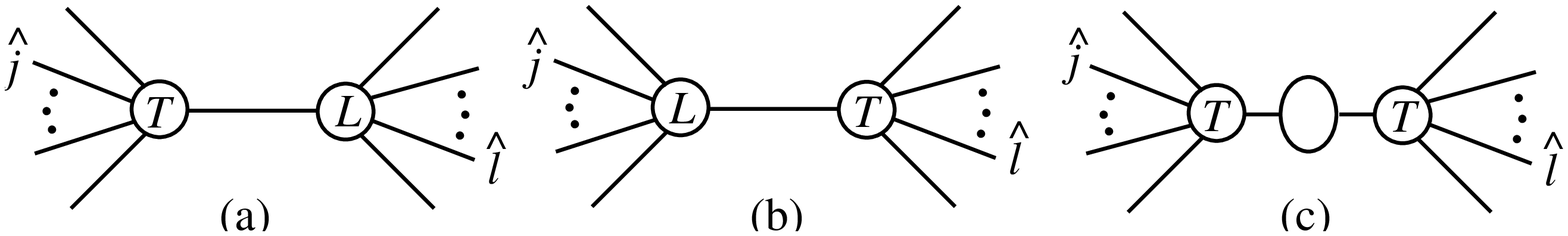}}
\caption{ Schematic representation of one-loop recursive
contributions.  The labels `$T$' and `$L$' refer to tree and loop
vertices.  The factorization-function contribution (c) does not appear
for MHV amplitudes.}
\label{LoopGenericFigure}
\end{figure}

The $R^D_n$ recursive diagram contributions on the right-hand side of
\eqn{BasicFormula} are obtained via a recursion relation, following
similar argumentation as at tree level. The sum over residues
corresponds to a sum over factorization channels. At one loop, as
depicted in \fig{LoopGenericFigure}, there are in general three
contributions to factorization in any given channel
$s_{\alpha \ldots\beta} \rightarrow 0$ ,
\begin{equation}
A_L^\tree \times {i\over s_{\alpha \ldots \beta }} \times A_R^\oneloop
+A_L^\oneloop \times {i\over s_{\alpha\ldots \beta}} \times A_R^\tree
+A_L^\tree \times {i \, \Fact^\oneloop\over s_{\alpha\ldots \beta}}
\times A_R^\tree\,.
\end{equation}
In the first two terms, one of the factorized amplitudes is a one-loop
amplitude and the other is a tree amplitude.
In the $\NeqZero$
scalar loop case, the last term
corresponds to a one-loop correction to the
propagator~\cite{BernChalmers}.  Accordingly, in addition to the sum over
channels, we will have a sum over these additional factorization function
contributions.  Taking the rational parts, we thus obtain, with the
shift~(\ref{SpinorShift}), the sum of recursive diagrams,
\def\indentA{\hskip 7mm}
\begin{eqnarray}
 - \sum_{{\rm poles}\ \alpha} \Res_{z=z_\alpha} {\Vertex_n(z)\over z}
&\equiv&
R^D_n(k_1,\ldots,k_n) \nonumber\\
 &=& \hskip -.1cm
 \sum_{{\rm partitions}\, P}\, \sum_{h = \pm} \Biggl\{
\Vertex(k_{P_1},\ldots,\hat k_j,\ldots,k_{P_{-1}},-\Ph_P^h)  \nonumber\\
&& \null \hskip 2.5 cm \indentA\times
{i\over K_P^2} \times
A^\tree(k_{\Pb_1},\ldots,\hat k_l,\ldots,k_{\Pb_{-1}},\Ph_P^{-h}) \nn\\
&& \null \hskip 2.5 cm
+ A^\tree(k_{P_1},\ldots,\hat k_j,\ldots,k_{P_{-1}},-\Ph_P^h)
   \nonumber \\
&& \null \hskip 2.5 cm \indentA \times
{i\over K_P^2} \times
\Vertex(k_{\Pb_1},\ldots,\hat k_l,\ldots,k_{\Pb_{-1}},\Ph_P^{-h}) \nn\\
&& \null \hskip 2.5 cm
+ A^\tree(k_{P_1},\ldots,\hat k_j,\ldots,k_{P_{-1}},-\Ph_P^h)  \nn\\
&& \null \hskip 2.5 cm \indentA\times
{i \Fact(K_P)\over K_P^2} \times
A^\tree(k_{\Pb_1},\ldots,\hat k_l,\ldots,k_{\Pb_{-1}},\Ph_P^{-h})
\Biggr\}
 \,,  \nn \\
&& \null \label{RationalRecursion}
\end{eqnarray}
where the partition $P$ is given in \eqn{PartitionDef}.
The `vertices' $R$ are one-loop amplitudes, but with all polylogarithms
and $\pi^2$ set to vanish, as in \eqn{RationalDefinition}.
The term containing the factorization function ${\cal F}$ may be found in
ref.~\cite{Genhel}. It only contributes in multi-particle
channels, and only if the tree amplitude contains a pole in that channel.
The MHV tree amplitudes have no multi-particle poles;
hence we encounter no factorization-function contributions in this paper.

The result~(\ref{RationalRecursion}) follows directly from the general
factorization behavior of one-loop amplitudes, plus the separate
factorization of pure-cut and rational terms that was established in
ref.~\cite{Bootstrap}.  Just as in the case of the tree-level
recursion~(\ref{TreeRecursion})~\cite{BCFW}, it exhibits the required
factorization properties in each channel $P$ (dropping the terms with
logarithms, polylogarithms, and $\pi^2$).  Although the $R$ functions
are not complete amplitudes, they can be thought of as vertices from a
diagrammatic perspective, and this equation therefore lends itself to the same
kind of diagrammatic interpretation available for \eqn{TreeRecursion}.

\begin{figure}[t]
\centerline{\epsfxsize 2 truein\epsfbox{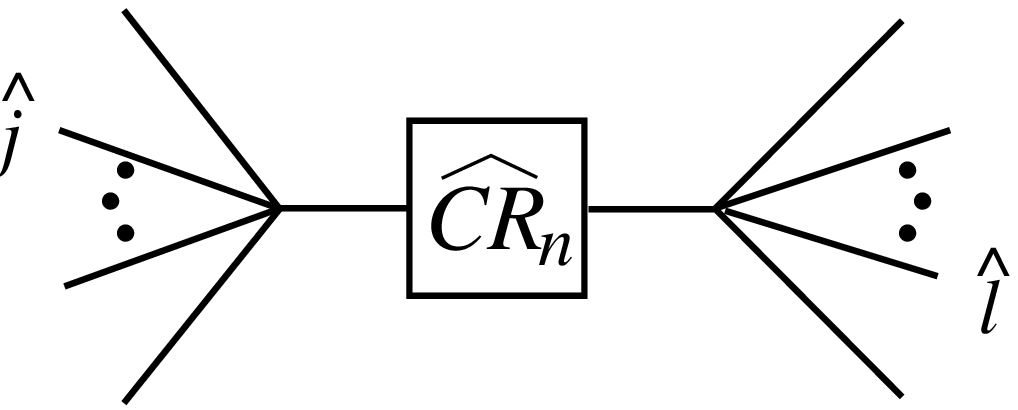}}
\caption{Diagrammatic representation of overlap contributions.
Each overlap diagram corresponds to a physical channel. }
\label{OverlapGenericFigure}
\end{figure}

Finally, consider the overlap terms $O_n$ in \eqn{BasicFormula}.
The recursive diagrams already encode the complete residues on the
physical poles. Therefore, if the $\CuthRat_n$ rational terms in the
completed-cuts $\Cuth_n$ have any poles in the physical channels, we
must subtract these out in order to remove this double-count.
Since we know $\CuthRat_n$ explicitly, it
is straightforward to compute the overlap by performing the
shift~(\ref{SpinorShift}) and extracting the residues on the physical
poles, {\it i.e.},
\begin{equation}
\Overlap_n \equiv \sum_{{\rm poles}\,\alpha} \Res_{z=z_\alpha}
                 {\CuthRat_n(z)\over z} \, .
\end{equation}
These overlap contributions may be assigned a diagrammatic
interpretation, as depicted in \fig{OverlapGenericFigure}, with each
diagram corresponding to a different physical factorization.  Although
the definition of the completed-cut terms $\Cuth_n$ is not unique, the
ambiguity cancels between $\Cuth_n(0)$ and the overlap terms.


\section{Five-, Six-, Seven-, and Eight-Point Amplitudes}
\label{SixSevenSection}

We will now apply this formalism to the computation of $\NeqZero$
(scalar loop) amplitudes with two negative-helicity gluons.
The logarithmic and
polylogarithmic terms $\PureCut_n$ in the $\NeqZero$ case are
known for all $n$-gluon amplitudes with two negative-helicity
gluons.  They were computed first for the case where the two
negative-helicity gluons are color-adjacent~\cite{Neq1Oneloop},
and more recently for the general case~\cite{BBSTQCD}. We quote
the results for the cut parts in appendix~\ref{CutAppendix}, including
the specific cut completion we employ in this publication,
\eqn{BBST}.  In appendix~\ref{CutAppendix} we also explain why
this cut completion is suitable:  after performing the complex shift
it contains no $z$-dependent spurious singularities.

The rational remainders for all-multiplicity MHV amplitudes were
computed previously for the special case where the two
negative-helicity gluons are adjacent~\cite{FordeKosower}.
Below we will first illustrate
our bootstrap approach by recomputing the rational part of the
five-point amplitude with two non-adjacent negative
helicities~\cite{GGGGG}.  This result also serves as one of the inputs
for the recursive construction of amplitudes with more than five gluons.
Then we will compute the remaining, previously unknown, rational terms
of all $n$-gluon $\NeqZero$ amplitudes with two negative-helicity
gluons in arbitrary positions.  We start in this section with the
computation of the six-point amplitudes.  In the next section we outline
the all-$n$ calculation.

Our bootstrap approach relies on previously-computed
amplitudes. A list of amplitudes that enter our calculation
can be found in the appendix of our companion paper~\cite{Genhel}.

\subsection{Recomputation of the five-point amplitude
$A_{5;1}^{\NeqZero}(1^-,2^+,3^-,4^+,5^+)$}
\label{5ptSection}

The five-point amplitude $A_{5;1}^{\NeqZero}(1^-,2^+,3^-,4^+,5^+)$
is given by~\cite{GGGGG}
\be
A^{\NeqZero}_{5;1}(1^-,2^+,3^-,4^+,5^+) =
      \cg \textrm{} \left( \Cuth_5 +  \Remaining_5 \right)
\, , \label{A5Neq0mfull}
\ee
where
\ba
\Cuth_5 & = &
 {1\over3 \cg} A^{\NeqOne}_{5;1}(1^-,2^+,3^-,4^+,5^+)
 \nn \\
&& \null
+
i \Biggl[
      - {{\spa1.2} {\spa2.3} {\spa3.4}
          {\spa4.1}^2 {\spb2.4}^2
         \over {\spa4.5} {\spa5.1} {\spa2.4}^2}\,
          {2 \, \Ls_1\bigl( {-s_{23}\over -s_{51}},
            \,{-s_{34}\over -s_{51}}\bigr)
           + \Ll_1\bigl( {-s_{23}\over -s_{51}}\bigr)
           + \Ll_1\bigl( {-s_{34}\over -s_{51}}\bigr)  \over s_{51}^2}
\nn \\
&& \quad
      + {{\spa3.2} {\spa2.1} {\spa1.5}
          {\spa5.3}^2 {\spb2.5}^2
         \over {\spa5.4} {\spa4.3} {\spa2.5}^2}\,
          {2 \, \Ls_1\bigl( {-s_{12}\over -s_{34}},
            \,{-s_{51}\over -s_{34}}\bigr)
           + \Ll_1\bigl( {-s_{12}\over -s_{34}}\bigr)
           + \Ll_1\bigl( {-s_{51}\over -s_{34}}\bigr)  \over s_{34}^2} \nn \\
&& \quad
      +{2\over 3} {{\spa2.3}^2 {\spa4.1}^3 {\spb2.4}^3
          \over {\spa4.5} {\spa5.1} {\spa2.4}}
          {\Ll_2\bigl( {-s_{23}\over -s_{51}}\bigr)  \over s_{51}^3}
      -{2\over 3} {{\spa2.1}^2 {\spa5.3}^3 {\spb2.5}^3
          \over {\spa5.4} {\spa4.3} {\spa2.5}}
          {\Ll_2\bigl( {-s_{12}\over -s_{34}}\bigr)  \over s_{34}^3} \nn \\
&& \quad
 + {\Ll_2\bigl( {-s_{34}\over -s_{51}}\bigr)\over s_{51}^3}\,
     \Biggl( {1\over3} { \spa1.3\spb2.4\spb2.5
     (\spa1.5 \spb5.2 \spa2.3-\spa3.4 \spb4.2 \spa2.1) \over \spa4.5}
  \nn \\
&& \quad
    +{2\over 3} {{\spa1.2}^2{\spa3.4}^2\spa4.1{\spb2.4}^3
            \over \spa4.5\spa5.1\spa2.4}
    -{2\over 3} {{\spa3.2}^2{\spa1.5}^2\spa5.3{\spb2.5}^3
            \over \spa5.4\spa4.3\spa2.5}  \Biggr) \nn \\
&& \quad
   +{1\over 6} {{\spa1.3}^3
        \bigl( \spa1.5\spb5.2\spa2.3 - \spa3.4\spb4.2\spa2.1\bigr)
          \over \spa1.2\spa2.3\spa3.4\spa4.5\spa5.1}\,
            {\Ll_0\bigl( {-s_{34}\over -s_{51}}\bigr)\over s_{51}}
\Biggr]\, ,
\nn \\
&&\outdent\null \qquad \label{A5Neq0m}
\\
\Remaining_5 & = &
  {2 i\over9} {\spa1.3^4 \over \spa1.2\spa2.3\spa3.4\spa4.5\spa5.1}
 +{i\over 3} {{\spb2.4}^2 {\spb2.5}^2
          \over {\spb1.2} {\spb2.3} {\spb3.4} {\spa4.5} {\spb5.1}}
   -{i\over 3} {{\spa1.2} {\spa4.1}^2 {\spb2.4}^3
          \over {\spa4.5} {\spa5.1} {\spa2.4} {\spb2.3} {\spb3.4} s_{51}}
            \nn \\
&& \null
   +{i\over 3} {{\spa3.2} {\spa5.3}^2 {\spb2.5}^3
          \over {\spa5.4} {\spa4.3} {\spa2.5} {\spb2.1} {\spb1.5} s_{34}}
   +{i\over 6}
    {{\spa1.3}^2 \spb2.4\spb2.5 \over s_{34} \spa4.5 s_{51}}
 \, . \label{5target}
\ea
Here the contribution of an $\NeqOne$ chiral multiplet is
\ba
A^{\NeqOne}_{5;1}(1^-,2^+,3^-,4^+,5^+) &=& \nn\\
&&\outdent
  \cg A_5^\tree(1^-,2^+,3^-,4^+,5^+)\Biggl[
   {1\over2 \e} \biggl\{  \biggl({\mu^2\over -s_{34}}\biggr)^{\e}
                         +\biggl({\mu^2\over -s_{51}}\biggr)^\e  \biggr\} + 2
                 \Biggr]
                  \label{Vf5m} \\
&&\outdent \null
  +i\, \cg \Biggl[
         {{\spa1.3}^2 {\spa4.1} {\spb2.4}^2
         \over {\spa4.5} {\spa5.1}}
           {\Ls_1
   \bigl( {-s_{23}\over -s_{51}},\,{-s_{34}\over -s_{51}} \bigr)
            \over s_{51}^2}
      -{{\spa1.3}^2 {\spa5.3} {\spb2.5}^2
         \over {\spa3.4} {\spa4.5}}
          {\Ls_1 \bigl({-s_{12}\over -s_{34}},\,{-s_{51}\over -s_{34}}
\bigr)
           \over s_{34}^2} \nn \\
&& \hskip 15mm \null
      +{1\over2} {{\spa1.3}^3
          (\spa1.5 \spb5.2 \spa2.3-\spa3.4 \spb4.2 \spa2.1)
             \over \spa1.2\spa2.3\spa3.4\spa4.5\spa5.1}
           {\Ll_0 \bigl({-s_{34}\over -s_{51}}\bigr)\over s_{51}}
\Biggr]
\,. \nn
\ea
We have introduced the function,
\ba
\Ls_1(r_1,r_2) & \equiv & {1 \over (1-r_1-r_2)^2} \Bigl[
\Li_2(1-r_1) + \Li_2(1-r_2) + \ln r_1 \, \ln r_2 - {\pi^2 \over 6}
\nn \\
& & \qquad \qquad \qquad \,\,
+ \,(1-r_1-r_2) \biggl( \Ll_0 (r_1) + \Ll_0 (r_2) \biggr) \Bigr]\, ,
\ea
as well as the $\Ll_i$-functions,
\begin{eqnarray}
\Ll_0(r) &=&
 {\ln(r)\over 1-r}
\,, \nn \\
\Ll_1(r) &=&
{\Ll_0(r)+1\over 1-r}\,, \nn \\
\Ll_2(r) &=&
{\ln(r)-(r-1/r)/2\over (1-r)^3} \,.
\label{Lsdef}
\end{eqnarray}
These functions are free of spurious singularities as
$r \rightarrow 1$, by design.

The task of this subsection is to start with the cut
completion~(\ref{A5Neq0m}), and use the recursive bootstrap to derive
the remaining rational terms~(\ref{5target}).  This example has been
considered already in ref.~\cite{Bootstrap}, but the detailed
construction was not given there.  We recall the example here because
the general all-multiplicity solution will follow a similar
construction.  (For $n$-point amplitudes, however, we will use a cut
completion which differs from the above one at $n=5$, but follows
somewhat more naturally from the form of the cut parts obtained in
ref.~\cite{BBSTQCD}.)

We use a $\Shift{1}{3}$ shift, in the notation of \eqn{SpinorShift}.
The first thing we have to ensure is that the cut
completion $\Cuth_5$ in~\eqn{A5Neq0m} is a satisfactory one,
with respect to this shift.  It should not produce any
$z$-dependent denominator factors that vanish at spurious
(unphysical) values of $z$.  Notice that $\Cuth_5$ contains
factors of $\spa2.4$ and $\spa2.5$ in denominators.  These
spurious singularities are not a problem for the $\Shift{1}{3}$ shift,
because they do not acquire any dependence on $z$.
(The double poles in $\spa2.4$ and $\spa2.5$ are actually cancelled
by the behavior of the $\Ls_1$-containing functions multiplying them.)
Spurious denominator factors of the form $(s_{23}-s_{51})$,
$(s_{12}-s_{34})$, and $(s_{34}-s_{51})$ are dealt with via the $\Ll_2$
functions.

The next step is to examine the large-parameter behavior of $\Cuth_5(z)$,
which can be obtained straightforwardly by applying the $\Shift{1}{3}$
shift to \eqn{A5Neq0m}. It turns out that in this case,
\be
\InfPart{\Shift13}{\Cuth_5} =
0
\,.
\label{inf513}
\ee
The large-$z$ contributions cancel between the series expansion of
the (poly)logarithmic terms and the rational terms of the
cut-completion.  This feature is not true in general.  Specifically,
the cut-completion chosen for six or more gluons, as given in
appendix~\ref{CutAppendix}, has a large-$z$ contribution (see
\eqn{eq:definition_of_InfC}).

As mentioned in the introduction, we `assume' that the full amplitude
has vanishing large-$z$ behavior under the $\Shift{1}{3}$
shift,
\be
\InfPart{\Shift13}{A_{5;1}^{\NeqZero}} = 0 \,.
\label{InfA513}
\ee
Of course, in the five-gluon case it is not an assumption, because we know
the full amplitude in advance.

\begin{figure}[t]
  \centerline{\epsfxsize 3.5 truein\epsfbox{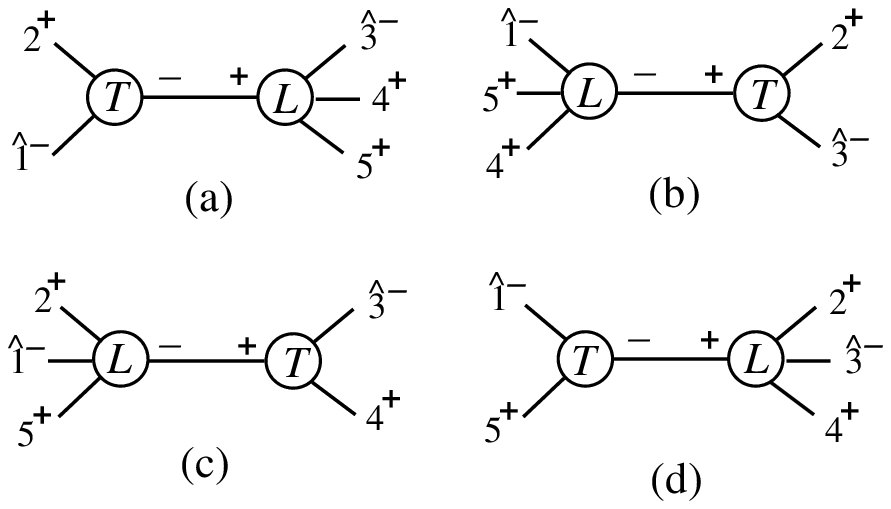}} \caption{
    Non-vanishing recursive diagrams for the amplitude
    $A^{\NeqZero}_{5;1}(1^-,2^+,3^-,4^+,5^+)$, using a $\Shift{1}{3}$ shift,
    as given in eqs.~(\ref{rec513a})-(\ref{rec513d}).
  }
\label{rec513fig}
\end{figure}

With a $\Shift13$ shift, following \eqn{RationalRecursion},
four recursive diagrams give non-vanishing contributions,
as illustrated in \fig{rec513fig},
\begin{equation}
R_5^D = \DiagrammaticRational_5^{\rm (a)} +
\DiagrammaticRational_5^{\rm (b)}
+
\DiagrammaticRational_5^{\rm (c)}
+ \DiagrammaticRational_5^{\rm (d)}\,.
\end{equation}
As mentioned above, the necessary lower-point amplitudes and vertices
are listed in the appendix of our companion paper~\cite{Genhel}, except for
$R_4(1^-,2^+,3^-,4^+)$, which is given by,
\be
R_4(1^-,2^+, 3^-, 4^+) =
\biggl({1\over 3\eps}  + {8\over 9} \biggr) A_4^\tree(1^-,2^+, 3^-, 4^+)
- i { {\spa{1}.{3} }^2 \spb{1}.{2} \spb{2}.{3} \over \spa{3}.{4}
\spa{4}.{1} {\spb{1}.{3}}^2 } \, .
\label{R4mpmp}
\ee
(The alert reader will note that
$\InfPart{\Shift13}{R_4} = i\, {\sandmm{1}.{3}.2}^2/{\sandmm{4}.{1}.3}^2 \neq 0$,
due to the last term in \eqn{R4mpmp}.  However, this behavior
is cancelled by that of the logarithmic terms in $A_{4;1}^{\NeqZero}$,
so that $\InfPart{\Shift13}{A_{4;1}^{\NeqZero}} = 0$.)

We obtain for the recursive diagrams,
\bea
\DiagrammaticRational_5^{\rm (a)} & = &
 A_3^{\tree} (-\Kh_{12}^-,\hat{1}^-,2^+) {i \over s_{12}}
R_4(\hat{3}^-,4^+,5^+,\Kh_{12}^+) =
 - {i \over 3} \, { \spb3.5 {\spb2.4}^3 \over \spa4.5 \spb1.2
              \spb1.3 {\spb3.4}^2}
 \, ,
 \label{rec513a} \\
\DiagrammaticRational_5^{\rm (b)} & = &
 R_4(\hat{1}^-,\Kh_{23}^-,4^+,5^+)
{i \over s_{23} } A_3^{\tree} (\hat{3}^-,-\Kh_{23}^+,2^+) =
 i \left( {1 \over 3 \e } + {8 \over 9} \right)
{ {\spa1.3}^3 \over \spa5.1 \spa2.3 \spa2.4 \spa4.5 }
 \, , \hskip 1 cm
\\
\DiagrammaticRational_5^{\rm (c)} & = &
R_4 (\hat{1}^-,2^+,\Kh_{34}^-, 5^+)
{i \over s_{34} } A_3^{\tree}(\hat{3}^-,4^+,-\Kh_{34}^+) \nn \\
& = &
 i \left( {1 \over 3 \e } + {8 \over 9} \right)
{ {\spa1.3}^3 \spa1.4 \over \spa5.1 \spa1.2 \spa3.4 \spa4.5 \spa2.4 }
+ i { {\spa1.3}^3 \over \spa1.4 \spa3.4 {\spa2.5}^2 }
 \, ,
\\
\DiagrammaticRational_5^{\rm (d)} & = &
A_3^{\tree} (\hat{1}^-,-\Kh_{51}^-,5^+) {i \over s_{51}}
R_4(\hat{3}^-,4^+,\Kh_{51}^+,2^+) =
 {i \over 3}\,
{{\spb3.5}^3 {\spb2.4}^3 \over \spa2.4 \spb1.3 {\spb2.3}^2
                {\spb3.4}^2 \spb1.5}
\,. \label{rec513d}
\eea
As remarked earlier, there is no factorization-function
contribution for the case of MHV amplitudes.

\begin{figure}[t]
\centerline{\epsfxsize 3.5 truein\epsfbox{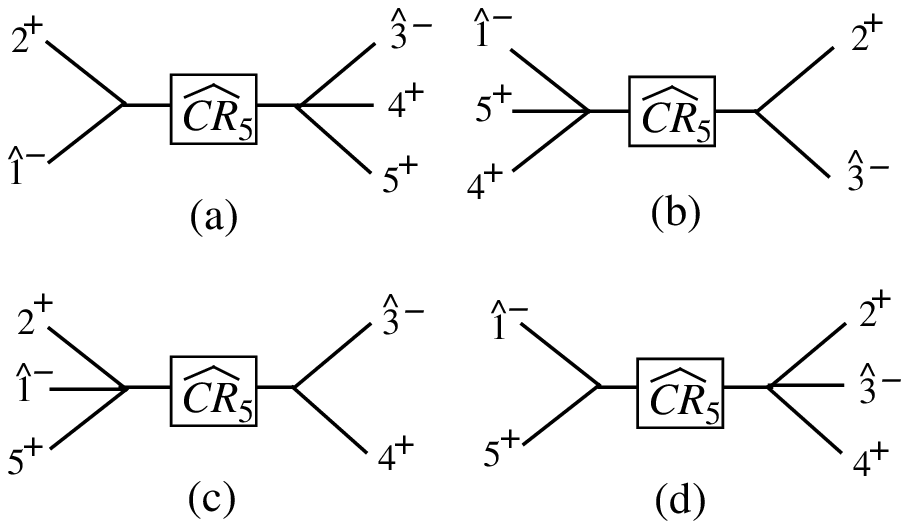}}
\caption{
Channels giving overlap contributions to
$A^{\NeqZero}_{5;1}(1^-,2^+,3^-,4^+,5^+)$ with a $\Shift{1}{3}$
shift, at the values of $z$ given in~\eqn{olap513}.}
\label{overlap513fig}
\end{figure}

In addition, there are four channels that can potentially contribute
overlap terms,
\bea
z^{\rm (a)} =  - { \spb{1}.{2} \over
\spb{2}.{3} } \, , & \qquad  &
z^{\rm (b)}  = {  \spa{2}.{3} \over \spa{1}.{2} } \, , \nn \\
z^{\rm (c)}  = - {  \spa{3}.{4} \over \spa{1}.{4} } \, , & \qquad  &
z^{\rm (d)}  =  { \spb{1}.{5} \over \spb{3}.{5} } \, .
\label{olap513}
\eea
The overlap contributions are illustrated in \fig{overlap513fig}.
For our choice of cut completion, \eqn{A5Neq0m}, they are given by,
\bea
\Overlap^{\rm (a)} & = &
 -{i \over 3} {\spa1.2 {\spb2.4}^3 \over \spb1.2 {\spb3.4}^2 \spa4.5 \spa2.5 }
\label{olap513a} \, ,\\
\Overlap^{\rm (b)} & = &
-  {i\over 3}\left( {1 \over \e } + 2 \right)
{ {\spa1.3}^3 \over \spa5.1 \spa2.3 \spa2.4 \spa4.5 }
\, , \\
\Overlap^{\rm (c)} & = &
- {i\over 3} \left( {1 \over \e } + 2 \right)
{ {\spa1.3}^3 \spa1.4 \over \spa5.1 \spa1.2 \spa3.4 \spa4.5 \spa2.4 }
+
{i \over 6} {{\spa1.3}^2 \spb2.4 \spa1.4 \over s_{34} \spa4.5 \spa1.5 \spa2.4 }
- i\, {4 \over 3}
{ {\spa1.3}^2 \spb2.5 \spa1.5 \over s_{34} \spa4.5 \spa1.4 \spa2.5 } \nn \\
& & \null - i\, {2 \over 3} { {\spa1.3}^2 \spb4.5 \over s_{34} \spa2.4 \spa2.5 }
+ {i \over 3} { {\spa1.3}^3 \sandmm{4}.{(3-5)}.{2} \over
s_{34} \spa3.4 \spa4.5 \spa1.4 \spa2.5 }
- i { {\spa1.2}^2 {\spa1.3}^3 \spb2.3 \spa4.5 \over
s_{34} \spa1.5 {\spa1.4}^2 \spa2.4 {\spa2.5}^2 } \nn \\
& & \null
+ 2\, i\, { {\spa1.3}^2 {\spa1.5}^2 \spa2.4 \spb2.5 \over
s_{34} \spa4.5 {\spa1.4}^2 {\spa2.5}^2 }
- i\, { \spa1.2 {\spa1.3}^2 \spb2.5 \left(
\spa1.2 \spa4.5 - \spa1.5 \spa2.4 \right) \over
s_{34} {\spa1.4}^2 \spa2.4 {\spa2.5}^2 } \nn \\
& & \null
- i\, { {\spa1.3}^3 \spa1.5 \spb3.5 \over s_{34} {\spa1.4}^2 {\spa2.5}^2 }
- i\, { {\spa1.3}^3 \over \spa1.5 \spa2.4 \spa3.4 \spa2.5 }
\, , \\
\Overlap^{\rm (d)} & = &
{i \over 6} { \spa1.3 \spb2.4 \spb3.5 \sandmm{1}.{(2-5)}.{4} \over
{\spb3.4}^2 \spa4.5 \spa1.5 \spb1.5 \spa2.4 }
- {i \over 3} { {\spa1.4}^2 \spa1.2 {\spb2.4}^3 \over \spb2.3
   \spb3.4 \spa4.5 \spb1.5 {\spa1.5}^2 \spa2.4 }
\nn \\
& & \null
+ {i \over 3} { {\spa1.4} {\spb2.4}^3 {\spb3.5}^2 \over {\spb2.3}^2 {\spb3.4}^2
\spa4.5 \spb1.5 \spa2.4 }
+ {i \over 3} { \spa1.5 {\spb4.5}^2 (s_{15} + s_{35}) \over
{\spb3.4}^2 \spa4.5 \spb1.5 \spa2.4 \spa2.5}
 \,  .
\label{olap513d}
\eea

The final result is the sum of all contributions, eqs.~(\ref{inf513}),
(\ref{rec513a})-(\ref{rec513d}), and (\ref{olap513a})-(\ref{olap513d}),
\be
\Remaining_5(1^-,2^+,3^-,4^+,5^+) = -\InfPart{\Shift13}{\Cuth_5}
+ {\DiagrammaticRational_5}^{\rm (a)} +
{\DiagrammaticRational_5}^{\rm (b)} +
{\DiagrammaticRational_5}^{\rm (c)} +
{\DiagrammaticRational_5}^{\rm (d)} +
\Overlap_5^{\rm (a)} + \Overlap_5^{\rm (b)} +
\Overlap_5^{\rm (c)} + \Overlap_5^{\rm (d)} \, .
\ee
The result agrees with \eqn{5target}, after appropriate simplification.
The complete rational part $\Rational_5(1^-,2^+,3^-,4^+,5^+)$,
which is needed in the recursive construction of higher-point
amplitudes in the next subsection, is found
by adding the rational terms of $\Cuth_5$ and $\Remaining_5$.
It can be read off from \eqns{A5Neq0m}{5target},
\be
\Rational_5(1^-, 2^+, 3^-, 4^+, 5^+) =
{1 \over \cg} A^{\NeqZero}_{5;1}(1^-,2^+,3^-,4^+,5^+)
\Bigr|_{\ln,\Li_2,\pi^2 = 0} \, .
\label{R5mpmpp}
\ee

\subsection{Six-point amplitudes with two negative helicities}

Without loss of generality, we label one of the negative-helicity
gluons as gluon 1 and the other as gluon $m$.  Other configurations
can be obtained by cyclicly permuting the labels of the gluons.
We introduce a shorthand for the argument list in the $n$-gluon case,
\be
A_{n;1}^{\NeqZero}(1,m)
\equiv A_{n;1}^{\NeqZero}(1^-,2^+,\ldots,(m-1)^+,m^-,(m+1)^+,\ldots,n^+) \,.
\label{shortargs}
\ee
The amplitude where the other negative-helicity gluon is adjacent,
$A_{n;1}^{\NeqZero}(1,2)$, can be found in
refs.~\cite{Bootstrap,FordeKosower}.  At six points, there are two
other independent configurations with two negative-helicity
gluons, $A^{\NeqZero}_{6;1}(1,3)$ and $A^{\NeqZero}_{6;1}(1,4)$.
We choose a $[1,m\rangle$ shift, in the notation of \eqn{SpinorShift},
with $m=3,4$ in the respective cases.

The cut parts were obtained in ref.~\cite{BBSTQCD}.
As explained above, we
complete these cut-containing terms to remove $z$-dependent spurious
singularities that arise when performing the $\Shift{1}{m}$ shift we've chosen.
The result is given in \eqn{BBST}. Taking $n=6$ in this expression gives the
completed-cut parts required here, $\Cuth_6(0) \equiv \Cuth_6(1,m)$.
It is straightforward to compute $\Inf \Cuth_6(1,m)$
by first shifting $\Cuth_6(0)$ by the $[1,m\rangle$ shift,
to obtain $\Cuth_6(z)$, and then series expanding
around $w = 1/z$, keeping the $w^0$ term.  (The result is given
by \eqn{eq:definition_of_InfC}, with $n=6$.)

For the case of $A^{\NeqZero}_{6;1}(1,3)$ with a $[1,3\rangle$ shift,
the recursive contribution is given by,
\bea
R^D_6(1,3) & = &
A_3^{\tree} (-\Kh_{12}^-,\hat{1}^-,2^+) {i \over s_{12}}
R_5(\hat{3}^-,4^+,5^+,6^+,\Kh_{12}^+) \nn \\
& + & R_5(\hat{1}^-,\Kh_{23}^-,4^+,5^+,6^+)
{i \over s_{23} } A_3^{\tree} (\hat{3}^-,-\Kh_{23}^+,2^+) \nn \\
& + & A_4^{\tree} (\hat{1}^-,\Kh_{234}^-,5^+,6^+)
{i \over s_{234} } R_4 (\hat{3}^-,4^+,-\Kh_{234}^+,2^+) \nn \\
& + & R_4(\hat{1}^-,\Kh_{234}^+,5^+,6^+)
{i \over s_{234} } A_4^{\tree} (\hat{3}^-,4^+,-\Kh_{234}^-,2^+) \nn \\
& + & R_5 (\hat{1}^-,2^+,\Kh_{34}^-, 5^+, 6^+)
{i \over s_{34} } A_3^{\tree}(\hat{3}^-,4^+,-\Kh_{34}^+) \nn \\
& + & A_4^{\tree} (\hat{1}^-,2^+,\Kh_{345}^-,6^+)
{i \over s_{345} } R_4(\hat{3}^-,4^+,5^+,-\Kh_{345}^+) \nn \\
& + & R_4 (\hat{1}^-,2^+,\Kh_{345}^+,6^+)
{i \over s_{345} } A_4^{\tree}(\hat{3}^-,4^+,5^+,-\Kh_{345}^-) \nn \\
& + & A_3^{\tree} (\hat{1}^-,-\Kh_{61}^-,6^+) {i \over s_{61}}
R_5(\hat{3}^-,4^+,5^+,\Kh_{61}^+,2^+) \, ,
\label{rec613}
\eea
where we have omitted terms that vanish.  As mentioned in
\sect{RecursionReviewSection}, there is no
factorization-function contribution for MHV amplitudes.
\Eqn{rec613} is illustrated in \fig{rec613fig}. The amplitudes
entering \eqn{rec613} are listed in the appendix of ref.~\cite{Genhel}
and in section~\ref{5ptSection}.
Note that the lower-point amplitudes have at most two negative
helicities. This general feature of the one-loop MHV amplitudes
allows us to solve the recursion relation for all $n$, as shown in
\sect{AllnSection}.

\begin{figure}[t]
  \centerline{\epsfxsize 5 truein\epsfbox{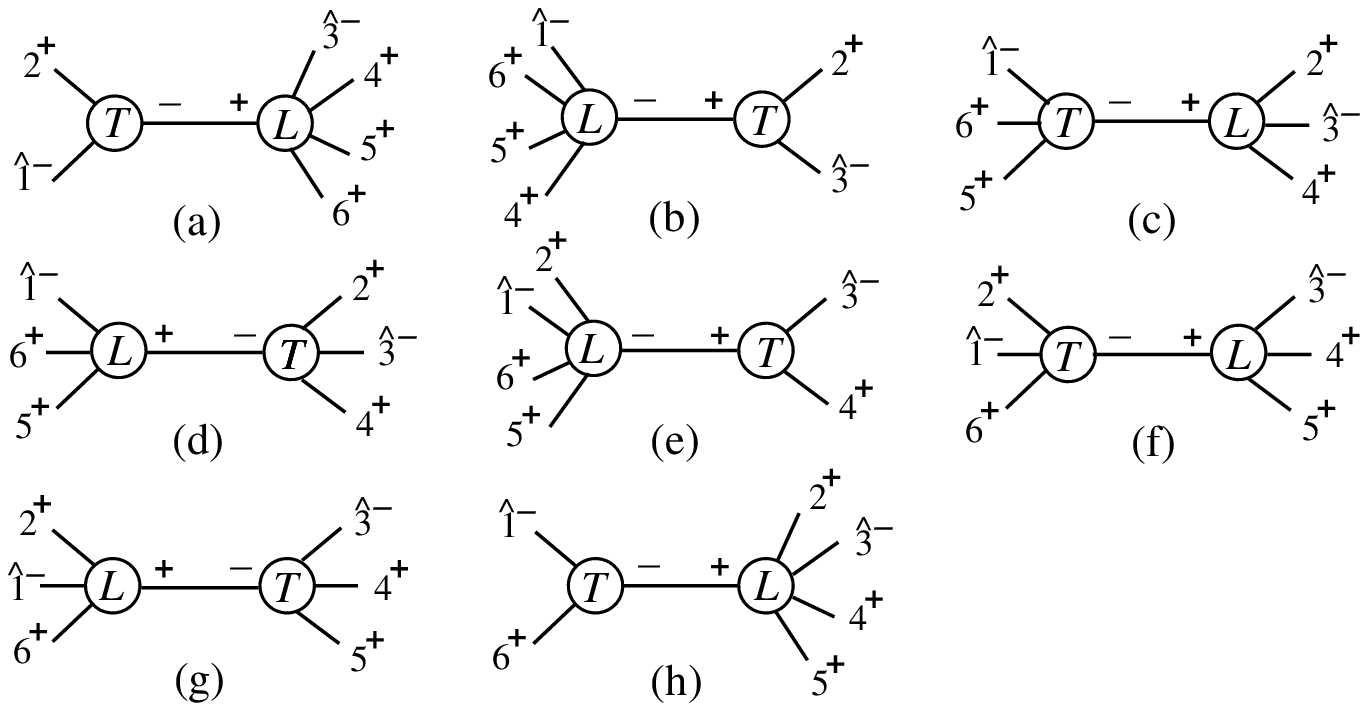}} \caption{
    Non-vanishing recursive diagrams for the amplitude
    $A^{\NeqZero}_{6;1}(1,3)$, using a $\Shift{1}{3}$ shift,
    as given in \eqn{rec613}. }
\label{rec613fig}
\end{figure}

\begin{figure}[t]
\centerline{\epsfxsize 5 truein\epsfbox{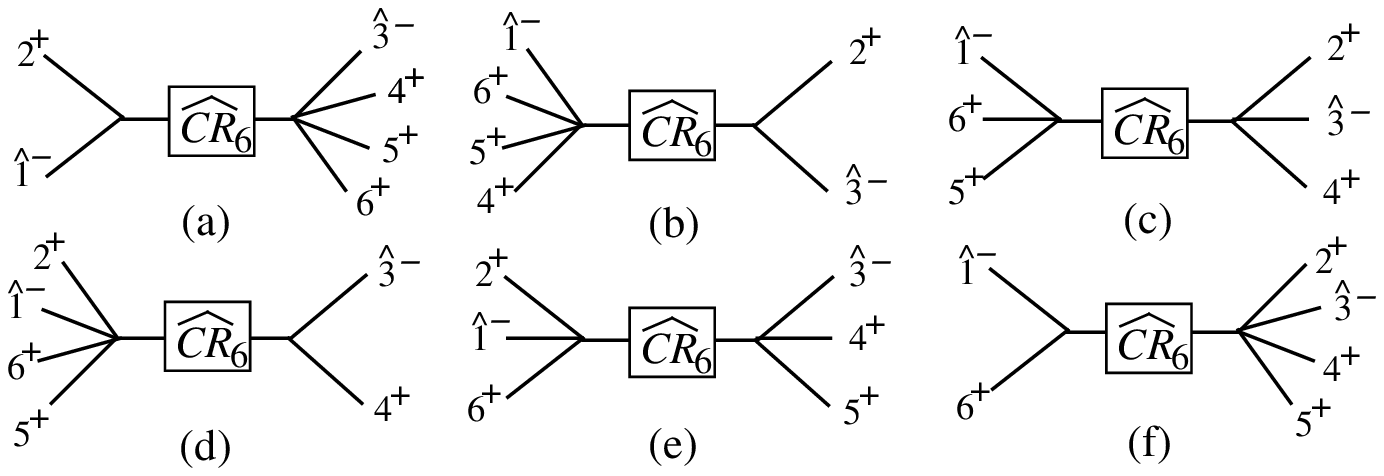}}
\caption{
Channels that can give overlap contributions to
$A^{\NeqZero}_{6;1}(1,3)$ with a $\Shift{1}{3}$
shift, at the values of $z$ given in \eqn{olap613}. }
\label{overlap613fig}
\end{figure}

Once a cut completion is chosen, we can apply the shift
$[1,3\rangle$ to $\CuthRat_6$ to obtain the overlap contributions
$\Overlap_6$. For $A^{\NeqZero}_{6;1}(1,3)$ there
are six possible channels which can give poles, illustrated in
\fig{overlap613fig},
\bea
z^{\rm (a)} =  - { \spb{1}.{2} \over
\spb{2}.{3} } \, , & \qquad  &
z^{\rm (b)}  = {  \spa{2}.{3} \over \spa{1}.{2} } \, , \nn \\
z^{\rm (c)}  =  - { s_{234} \over \spab{1}.{(2+4)}.{3} } \, , &\qquad
&
z^{\rm (d)}  = - {  \spa{3}.{4} \over \spa{1}.{4} } \, , \nn \\
z^{\rm (e)}  =  - {  s_{345} \over \spab{1}.{(4+5)}.{3} } \, , &
\qquad & z^{\rm (f)}  =  { \spb{1}.{6} \over \spb{3}.{6} } \, .
\label{olap613}
\eea
Depending on the specific cut completion, some channels may give
vanishing overlap contributions.

After assembling and simplifying the recursive and overlap diagrams,
we obtain the result presented in \eqn{totR6hat13} of
appendix~\ref{SixpointAppendix}
for $\Remaining_6(1,3)$, for the specific cut completion~(\ref{BBST}).
We have checked numerically that our result is
symmetric under the flip $(123456) \leftrightarrow (321654)$,
even though the recursion relation~(\ref{rec613}) does not keep the flip
symmetry manifest. Furthermore, our result displays the correct
multi-particle and collinear factorization limits, as well as cancellation
of all spurious singularities.  We give some numerical values
in \sect{NumericsSection}.

\begin{figure}[t]
\centerline{\epsfxsize 5 truein\epsfbox{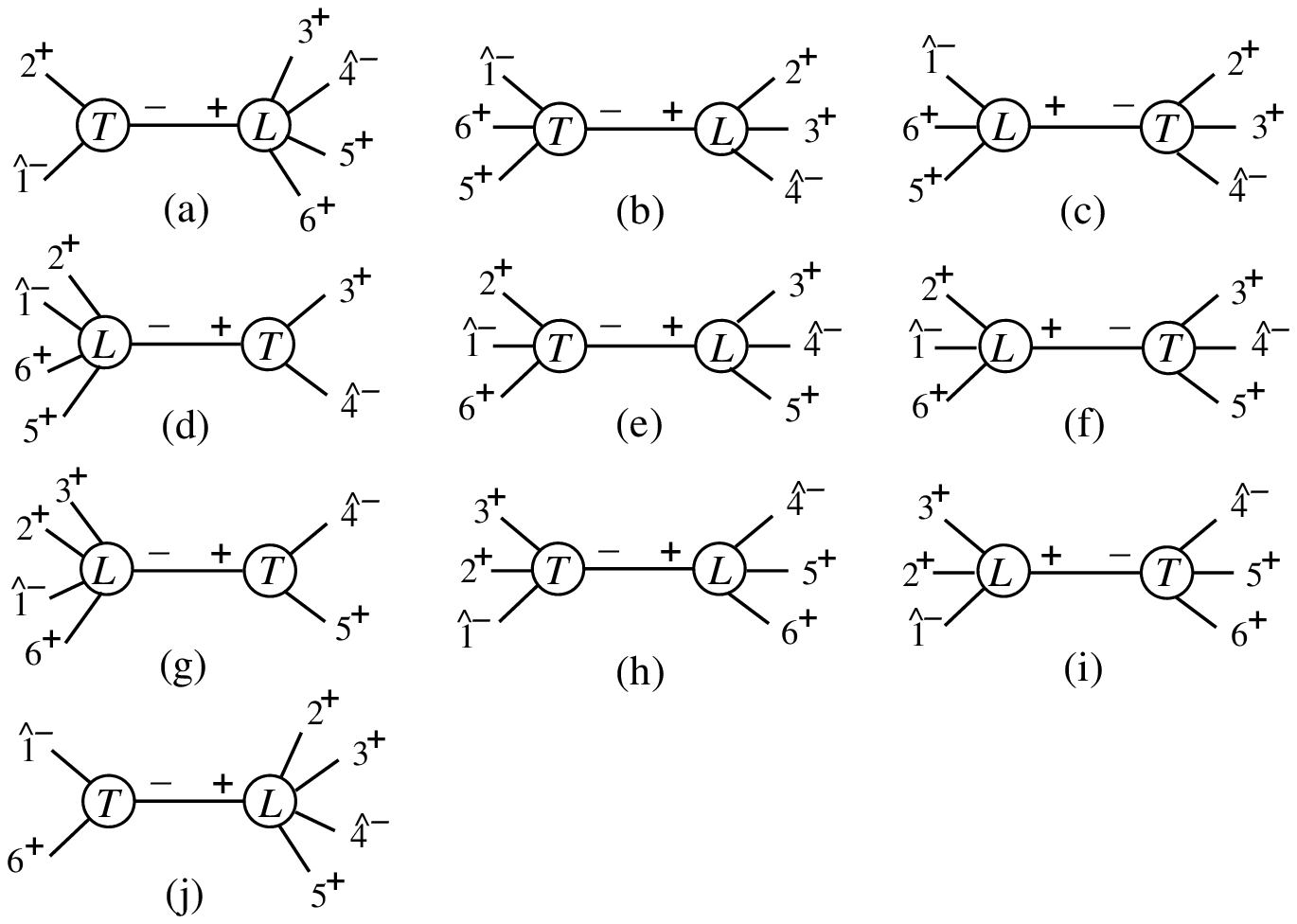}}
\caption{
  Non-vanishing recursive diagrams for the amplitude
  $A^{\NeqZero}_{6;1}(1,4)$, using a $\Shift{1}{4}$ shift,
  as given in \eqn{rec614}. }
\label{rec614fig}
\end{figure}

The computation of $A^{\NeqZero}_{6;1}(1,4)$ uses the $[1,4\rangle$ shift.
The recursive contribution is given by,
\bea
R^D_6(1,4) & = &
A_3^{\tree} (\hat{1}^-,2^+,-\Kh_{12}^-) {i \over s_{12}}
R_5(\hat{4}^-,5^+,6^+,\Kh_{12}^+,3^+) \nn \\
& + & A_4^{\tree} (\hat{1}^-,\Kh_{234}^-,5^+,6^+)
{i \over s_{234} } R_4 (\hat{4}^-,-\Kh_{234}^+,2^+,3^+) \nn \\
& + & R_4(\hat{1}^-,\Kh_{234}^+,5^+,6^+)
{i \over s_{234} } A_4^{\tree} (\hat{4}^-,-\Kh_{234}^-,2^+,3^+) \nn \\
& + & R_5 (\hat{1}^-,2^+,\Kh_{34}^-, 5^+, 6^+)
{i \over s_{34} } A_3^{\tree}(\hat{4}^-,-\Kh_{34}^+,3^+) \nn \\
& + & A_4^{\tree} (\hat{1}^-,2^+,\Kh_{345}^-,6^+)
{i \over s_{345} } R_4(\hat{4}^-,5^+,-\Kh_{345}^+,3^+) \nn \\
& + & R_4 (\hat{1}^-,2^+,\Kh_{345}^+,6^+)
{i \over s_{345} } A_4^{\tree}(\hat{4}^-,5^+,-\Kh_{345}^-,3^+) \nn \\
& + & R_5 (\hat{1}^-,2^+,3^+,\Kh_{45}^-, 6^+)
{i \over s_{45} } A_3^{\tree}(\hat{4}^-,5^+,-\Kh_{45}^+) \nn \\
& + & A_4^{\tree} (\hat{1}^-,2^+,3^+,\Kh_{456}^-)
{i \over s_{456} } R_4(\hat{4}^-,5^+,6^+,-\Kh_{456}^+) \nn \\
& + & R_4 (\hat{1}^-,2^+,3^+,\Kh_{456}^+)
{i \over s_{456} } A_4^{\tree}(\hat{4}^-,5^+,6^+,-\Kh_{456}^-) \nn \\
& + & A_3^{\tree} (\hat{1}^-,-\Kh_{61}^-,6^+) {i \over s_{61}}
R_5(\hat{4}^-,5^+,\Kh_{61}^+,2^+,3^+) \, ,
\label{rec614}
\eea
as illustrated in \fig{rec614fig}. Note that the amplitude
$A^{\NeqZero}_{6;1}(1,4)$ has a flip symmetry
under the simultaneous exchange of gluons $2 \leftrightarrow 6$
and $3 \leftrightarrow 5$.  This flip symmetry remains manifest
in the presence of the $\Shift{1}{4}$ shift.
(A second flip symmetry, $(123456) \leftrightarrow (432165)$,
is {\it not} kept manifest.)
Therefore we only need to compute six recursive diagrams in~\eqn{rec614};
the other four diagrams can be obtained by the flip symmetry.
Diagram (a) can be
obtained from diagram (j) by the flip symmetry, likewise diagram (b)
from (h), (c) from (i), and (d) from (g).  Diagrams (e) and
(f) are invariant under the symmetry.

\begin{figure}[t]
\centerline{\epsfxsize 5 truein\epsfbox{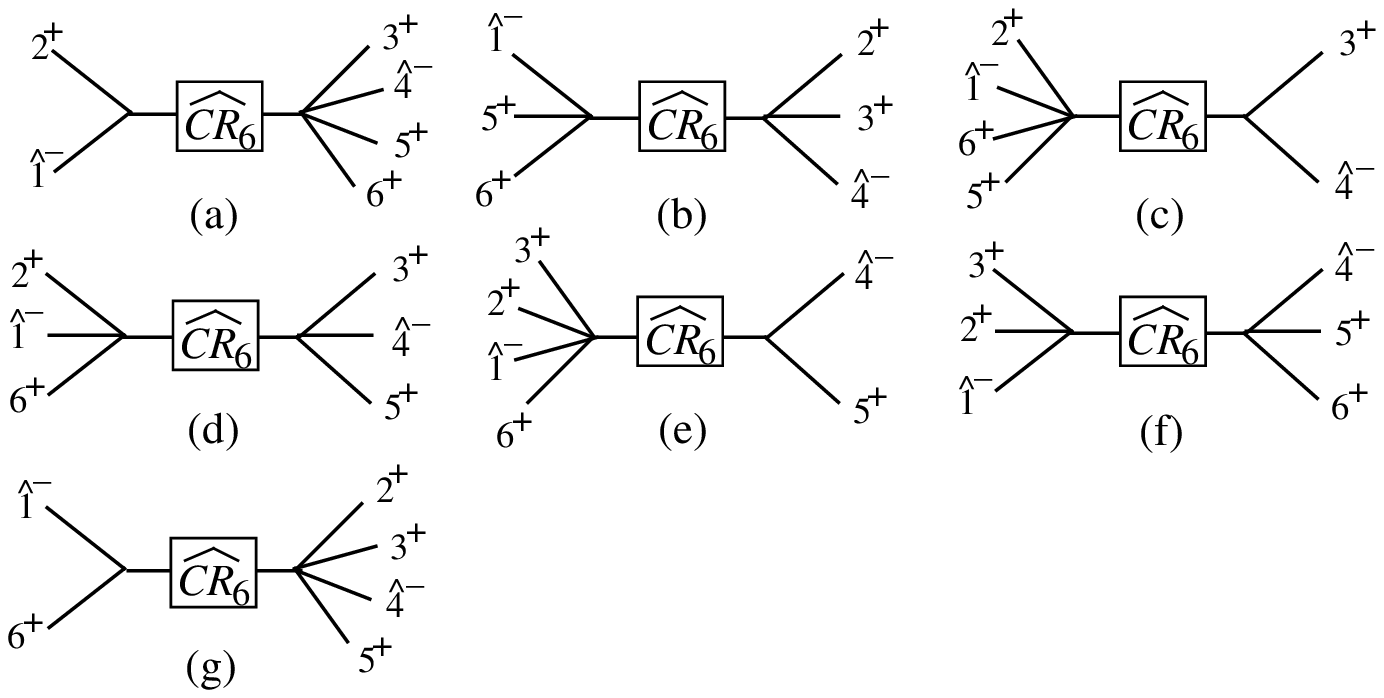}}
\caption{
Channels giving overlap contributions to
$A^{\NeqZero}_{6;1}(1,4)$ with a $\Shift{1}{4}$
shift, at the values of $z$ given in~\eqn{olap614}. }
\label{overlap614fig}
\end{figure}

With a $[1,4\rangle$ shift there are seven channels that can give
contributions to the overlap $\Overlap_6(1,4)$
as illustrated in \fig{overlap614fig},
\bea
z^{\rm (a)}  = - {
\spb{1}.{2} \over \spb{2}.{4} } \, , & \qquad &
z^{\rm (b)}  = - {  s_{234} \over \spab{1}.{(2+3)}.{4} } \, , \nn \\
z^{\rm (c)}  =  {  \spa{3}.{4} \over \spa{1}.{3} } \, , & \qquad &
z^{\rm (d)}  = - {  s_{345} \over \spab{1}.{(3+5)}.{4} } \, , \nn \\
z^{\rm (e)} = - {\spa{4}.{5} \over \spa{1}.{5} } \, , & \qquad &
z^{\rm (f)}  = - {  s_{456} \over \spab{1}.{(5+6)}.{4} } \, , \nn \\
z^{\rm (g)}  =  { \spb{1}.{6} \over \spb{4}.{6} } \, .
\label{olap614}
\eea
Again, we can make use of the aforementioned flip symmetry,
$2 \leftrightarrow 6$ and $3 \leftrightarrow 5$, because our cut
completion, $\CuthRat_6(1,4)$ in \eqn{eq:CR_hat_unshifted}, was chosen
to be flip symmetric. (This property relies in part on the symmetry
and antisymmetry under $r\to1/r$ of the functions $\Ll_2(r)$ and $\Lnl_1(r)$
defined in \eqns{Lsdef}{L1hat}, respectively.)
The flip symmetry relates overlaps $\Overlap^{\rm (a)}$
and $\Overlap^{\rm (g)}$, $\Overlap^{\rm (b)}$ and $\Overlap^{\rm (f)}$, and
$\Overlap^{\rm (c)}$ and $\Overlap^{\rm (e)}$, whereas $\Overlap^{\rm (d)}$ is
invariant.

The result for the rational part
$\Remaining_6(1,4)$, complementing the completed-cut terms
given in \eqn{BBST}, is listed in appendix~\ref{SixpointAppendix}.
We have checked numerically that the other, non-manifest, flip symmetry
$(123456) \leftrightarrow (432165)$
of this amplitude holds, and that the proper factorization limits
are obtained in all channels.  We give numerical results for the amplitude
in the next subsection.

\subsection{Numerical results for various six-, seven- and
eight-point amplitudes} \label{NumericsSection}

All-$n$ analytical results were previously found for the $\NeqZero$
one-loop amplitudes with all gluons of positive
helicity~\cite{AllPlus}, and all but one gluon of positive
helicity~\cite{Mahlon}. These results have been recomputed via a
simplified version of the bootstrap
approach~\cite{OnShellRecurrenceI,Qpap}.  (It is simpler because these
amplitudes contain no cuts.)  The full bootstrap, as described
in~\sect{RecursionReviewSection}, has been used to compute the
$n$-gluon split-helicity next-to-MHV amplitudes~\cite{Genhel}, using
cut-containing terms from ref.~\cite{RecurCoeff}.  As described in
more detail in the next section, we have now obtained all-$n$ results
for all MHV configurations, by combining the cut parts~\cite{BBSTQCD}
with the rational terms from \sect{AllnSection} of this paper.  The
case of MHV amplitudes with two nearest neighboring negative-helicity
legs in the color ordering was worked out in ref.~\cite{FordeKosower}.

Here we list some numerical values of the MHV amplitudes for a
particular phase-space point, defined below, for six, seven, and eight
gluons.  In the six-point case,
we compare to results~\cite{EGZ06} that were obtained with
a semi-numerical approach~\cite{EGZ}, and more recently,
analytically~\cite{XYZ6}.
For the seven- and eight-gluon cases, we have checked numerically
that our results have the correct flip symmetries (in fact this
was checked up to $n=11$), as well the correct limits in all
multi-particle and collinear factorization channels.

Specifically, for the six-point case we use the same phase-space
point as in refs.~\cite{EGZ06,Genhel}, with the six momenta $k_i$
chosen as follows,
\ba
\label{SixPointKinematics}
k_1 & = & \frac{\mu}{2} (-1,
+\sin\theta, +\cos\theta \sin\phi, +\cos\theta \cos\phi ),
 \nonumber \\
k_2 & = & \frac{\mu}{2} (-1,  -\sin\theta, -\cos\theta \sin\phi,
-\cos\theta \cos\phi ),
 \nonumber \\
k_3 & = & \frac{\mu}{3} (1,1,0,0), \nonumber \\
k_4 & = & \frac{\mu}{7} (1,\cos\beta,\sin\beta,0), \nonumber \\
k_5 & = & \frac{\mu}{6} (1,\cos\alpha \cos\beta, \cos\alpha
\sin\beta,
\sin\alpha), \nonumber \\
k_6 & = & -k_1-k_2-k_3-k_4-k_5\, ,
\ea
where
\ba
&& \theta = {\pi\over 4}\,, \hskip 1 cm \phi = {\pi\over 6}
\,, \hskip 1 cm \alpha = {\pi \over 3} \,, \hskip 1 cm \cos \beta
= - {7\over 19} \,.
\ea
Note that the energies of $k_1$ and $k_2$ are negative and
$k_i^2=0$. In order to have energies of $O(1)$, the authors of
ref.~\cite{EGZ06} chose the scale $\mu=n=6$~[GeV].  As usual $\mu$
also denotes the scale which is used to carry the dimensionality
of the $D$-dimensional integrals.

For the seven-point case we choose the same kinematic
 point as in ref.~\cite{Genhel},
\ba
k_1 & = & \frac{\mu}{2} (-1,  +\sin\theta, +\cos\theta
\sin\phi, +\cos\theta \cos\phi ),
 \nonumber \\
k_2 & = & \frac{\mu}{2} (-1,  -\sin\theta, -\cos\theta \sin\phi,
-\cos\theta \cos\phi ),
 \nonumber \\
k_3 & = & \frac{\mu}{3} (1,1,0,0), \nonumber \\
k_4 & = & \frac{\mu}{8} (1,\cos\beta, \sin\beta,0), \nonumber \\
k_5 & = & \frac{\mu}{10} (1,\cos\alpha \cos\beta, \cos\alpha
\sin\beta,
\sin\alpha), \nonumber \\
k_6 & = & \frac{\mu}{12} (1, \cos\beta \cos\gamma, \sin\beta
\cos\gamma,
\sin\gamma), \nonumber \\
k_7 & = & -k_1-k_2-k_3-k_4-k_5-k_6\, ,
\label{SevenPointKinematics}
\ea
where
\ba
 \theta = {\pi\over 4}\,,\hskip 1 cm
 \phi   =  {\pi\over 6}\,,\hskip 1 cm
 \alpha = {\pi \over 3}\,,\hskip 1 cm
 \gamma = {2 \pi \over 3} \,, \hskip 1 cm
 \cos \beta = - {37\over 128} \,,
\ea
and $\mu = n = 7$~GeV.


\begin{table}
\caption{\label{Neq4Table} Numerical results for the non-vanishing
$\NeqFour$ six-, seven-, and eight-point amplitudes with two
negative-helicity legs in the FDH scheme. The kinematic points are
given in eqs.~(\ref{SixPointKinematics}),
(\ref{SevenPointKinematics}) and (\ref{EightPointKinematics}).
The analytic expressions used for
this table are from ref.~\cite{Neq4Oneloop}.  }
\begin{tabular}{||c||c|c||}
\hline \hline
Helicity& $1/\e$ & $\e^0$ \\
\hline \hline ${-}{-}{+}{+}{+}{+}$ &  $\;
       448.1350970  + i \, 288.8591589
$ & $ \;
        231.6837670 + i \, 1219.687214  \hskip .1 cm
$ \\
\hline ${-}{+}{-}{+}{+}{+}$ &  $\;
   145.1068197 + i \, 93.53303095
$ & $ \;
   75.01955289 + i \, 394.9365577
$ \\
\hline
${-}{+}{+}{-}{+}{+}$ &  $\;
  7.064109769 - i\, 10.20934744
$ & $ \;
  28.56235909 - i\,  4.462571113
$ \\
\hline \hline
${-}{-}{+}{+}{+}{+}{+}$ & $
2923.502435 + i \, 683.4723607
$ &  $
 5112.775012 + i \,  6035.881921
$ \\
\hline ${-}{+}{-}{+}{+}{+}{+}$ & $
946.6344939 + i\, 221.3093804
$ &  $
1655.524253 + i \, 1954.427662
$ \\
\hline ${-}{+}{+}{-}{+}{+}{+}$ & $
21.02279700 - i\,  56.55151527
$ &  $
133.2644743  - i\,  86.65176251
$ \\
\hline \hline
${-}{-}{+}{+}{+}{+}{+}{+}$ & $
62.40652480 + i \, 54.16810878
$ &  $
\; 42.55501770 + i \, 181.2843342
$ \\
\hline ${-}{+}{-}{+}{+}{+}{+}{+}$ & $
7.274637638 + i \, 6.314297490
$ &  $
4.960576389 + i \, 21.13205060
$ \\
\hline ${-}{+}{+}{-}{+}{+}{+}{+}$ & $
\; 0.06051809107 + i \, 0.01883973611 \;
$ &  $
\; 0.08570866540 + i \, 0.1142513680 \;
$ \\
\hline ${-}{+}{+}{+}{-}{+}{+}{+}$ & $
1.592056562 - i \, 1.878665237
$ &  $
5.386800498 - i \, 1.331984344
$ \\
\hline \hline
\end{tabular}
\end{table}

\begin{table}
\caption{\label{Neq1Table} Numerical results for the non-vanishing
$\NeqOne$ chiral contributions to six-, seven-, and eight-point
MHV amplitudes. The kinematic points are given in
eqs.~(\ref{SixPointKinematics}), (\ref{SevenPointKinematics}), and
(\ref{EightPointKinematics}). The analytical expressions were
obtained from ref.~\cite{Neq1Oneloop}.}
\begin{tabular}{||c||c|c||}
\hline \hline
Helicity& $1/\e$ & $\e^0$ \\
\hline \hline
${-}{-}{+}{+}{+}{+}$ & $
- \,28.11035867 + i \, 4.643367883
$ & $ \hskip .5 mm
-\, 108.9419206 + i \, 35.02980993
\hskip .5 mm $ \\
\hline ${-}{+}{-}{+}{+}{+}$ & $
- \,9.102176499 + i\,  1.503529518
$ & $
- \,35.86914908 + i \, 5.750500896
$ \\
\hline ${-}{+}{+}{-}{+}{+}$ & $ \hskip .5 mm
0.08664490662 + i \; 0.6577514371
\hskip .5 mm $ & $
- \, 2.226022769 + i \, 3.230760457
$ \\
\hline \hline
${-}{-}{+}{+}{+}{+}{+}$ & $
-\,104.5611840 + i \, 45.34709475
$ & $
-\,429.6932951 + i \, 209.0560823
$ \\
\hline ${-}{+}{-}{+}{+}{+}{+}$ & $
-\, 33.85706894 + i \, 14.68345762
$ & $
-\, 162.2581945 + i \,  46.94912355
$ \\
\hline ${-}{+}{+}{-}{+}{+}{+}$ & $
0.6394298622  + i \, 2.199205804
$ & $
-\, 9.326523676 + i \, 12.89705727
$ \\
\hline \hline
${-}{-}{+}{+}{+}{+}{+}{+}$ & $
-\, 3.088708769 + i \, 0.2144746165
$ &  $
-\, 7.455275518 + i \, 0.7776396411
$ \\
\hline ${-}{+}{-}{+}{+}{+}{+}{+}$ & $
-\, 0.3600462794 + i \, 0.02500099345
$ &  $
-\, 0.7753751687 - i \, 0.06138368438
$ \\
\hline ${-}{+}{+}{-}{+}{+}{+}{+}$ & $
\; -\, 0.002103801109 + i \, 0.001101616598 \;
$ &  $
\; -\, 0.008916187143 - i \, 0.003055492662 \;
$ \\
\hline ${-}{+}{+}{+}{-}{+}{+}{+}$ & $
0.007480422983 + i \, 0.09196001806
$ &  $
-0.2405797542 + i \, 0.3716908859
$ \\
\hline \hline
\end{tabular}
\end{table}

\begin{table}
\caption{\label{67Table} Numerical results for six-, seven-, and
eight-point $\NeqZero$ amplitudes as described in the text, evaluated
 at the specific phase space points in
 eqs.~(\ref{SixPointKinematics}), (\ref{SevenPointKinematics}), and
 (\ref{EightPointKinematics}).}.
\begin{tabular}{||c||c|c||}
\hline \hline
Helicity& $1/\epsilon$ & $\epsilon^0$ \\
\hline \hline
${-}{-}{+}{+}{+}{+}$ & $
-\,9.370119558 + i \, 1.547789294
$ & $
-\, 45.80779561 + i \, 13.03695870
$ \\
\hline
${-}{+}{-}{+}{+}{+}$ & $
-\, 3.034058833 + i \, 0.5011765059
$ & $
-\, 15.20562226 + i \, 1.709378044
$ \\
\hline
${-}{+}{+}{-}{+}{+}$  & $ \hskip .5 mm
0.02888163554 + i\, 0.2192504790
\hskip .5 mm $ & $ \hskip .5 mm
-\, 1.449837907 + i\, 0.1763294054
\hskip .5 mm $ \\
\hline\hline
${-}{-}{+}{+}{+}{+}{+}$  & $
-\,34.85372799 + i \, 15.11569825
$ & $
-\,176.2169235 + i \,87.93931019
$ \\
\hline ${-}{+}{-}{+}{+}{+}{+}$  & $
-\, 11.28568965 + i \, 4.894485872
$ & $
-\, 58.27730664 + i\, 13.67438826
$ \\
\hline ${-}{+}{+}{-}{+}{+}{+} $ & $
0.2131432874  + i\, 0.7330686014
$ & $
-\,6.238940131 - i \, 0.4283898751
$ \\
\hline \hline
${-}{-}{+}{+}{+}{+}{+}{+}$ & $
-\,1.029569590 + i \, 0.07149153884
$ & $
-\,4.244770988 + i \, 0.3284878412
$ \\
%
\hline ${-}{+}{-}{+}{+}{+}{+}{+}$ & $
-\,0.1200154265 + i\, 0.008333664483
$ &  $
-\,0.5328645055 - i\, 0.1225617739
$ \\
\hline ${-}{+}{+}{-}{+}{+}{+}{+}$ & $ \;
-\,0.0007012670363 + i\, 0.0003672055326
\; $ &  $ \;
0.007966818918 - i\, 0.05081471136
\; $ \\
\hline ${-}{+}{+}{+}{-}{+}{+}{+}$ & $
0.002493474328 + i\, 0.03065333935
$ &  $
-\, 0.1173717691 + i\, 0.1867703328
$ \\
\hline \hline
\end{tabular}
\end{table}

At eight points, it turns out that the choice in
ref.~\cite{Genhel} sits on top of a possible spurious singularity
(where $\langle 7^\pm|(1+2)|5^\pm \rangle = 0$), so here we choose
a different reference kinematic point,
\ba
k_1 &=& {\mu \over 2} (-1,\, 1/\sqrt{2},\, -1/2,\,  1/2) \,, \nn\\
k_2 &=& {\mu\over 2} (-1,\, -  1/\sqrt{2},\, 1/2,\, -1/2)\,,\nn\\
k_3 &=& {\mu \over 5} (1, 1, 0, 0)\,,\nn\\
k_4 &=& {\mu \over 5} (1, \cos\beta, \sin\beta, 0) \,,\nn\\
k_5 &=& {\mu \over 6}  (1, \cos\alpha \cos\beta,
           \cos\alpha \sin\beta, \sin\alpha) \,, \nn\\
k_6 &=& {\mu \over 7} (1,  \cos\gamma \cos\beta,
             \cos\gamma \sin\beta,  \sin\gamma) \,, \nn \\
k_7 &=& {\mu \over 8} (1, \cos\delta \cos\beta,
                       \cos\delta \sin\beta, \sin\delta) \,, \nn\\
k_8 &=& -k_1-k_2-k_3-k_4-k_5-k_6-k_7 \,,
\label{EightPointKinematics}
\ea
where
\ba
\cos\alpha = {8\over 17} \,, \hskip 1 cm \cos\beta =
-{7193\over 8258} \,, \hskip 1 cm \cos\delta = -{3\over 5} \,,
\hskip 1 cm \cos\gamma =  {3\over 5} \,,
\ea
and $\mu = n = 8$~GeV.

Our results for the $\NeqZero$ MHV amplitudes are presented in
table~\ref{67Table}. The full QCD amplitudes can be reconstructed
from the $\NeqZero$ amplitudes and the $\NeqFour$ and $\NeqOne$
supersymmetric parts via \eqn{AnQCD}. The corresponding
supersymmetric amplitudes are listed in tables~\ref{Neq4Table} and
\ref{Neq1Table}, using results of
refs.~\cite{Neq4Oneloop,Neq1Oneloop}.
Note that we have extracted an overall factor of $i\, \cg$ from the
numerical values presented in the tables.
We do not include the
coefficients of the leading $1/\e^2$ singularities of the
$\NeqFour$ amplitudes in the table, as these are easily extracted
from the values of tree amplitudes, for any helicity
configuration,
\be
A^{\NeqFour}_{n;1} \Bigr|_{1/\e^2} =  -  {n\cg \over \e^2}
A^{\tree}_n \, .
\ee
The numerical values of the tree amplitudes may be read off from
the values of the $1/\e$ singularities of either the $\NeqOne$ or
the $\NeqZero$ loop amplitudes,
\be
A^{\NeqOne}_{n;1}
\Bigr|_{1/\e} =   {\cg\over \e} A^{\tree}_n \,, \hskip 2cm
A^{\NeqZero}_{n;1} \Bigr|_{1/\e} =   {\cg\over 3\e} A^{\tree}_n
\,,
\ee
given in tables~\ref{Neq1Table} and \ref{67Table}. Our results in
table~\ref{67Table} at six points match those of
ref.~\cite{EGZ06}, taking into account differing phase
conventions.  The rational terms in our results agree completely
with formulae provided by the authors of ref.~\cite{XYZ6},
to high numerical accuracy and for a large number of
phase-space points.


\section{One-Loop $n$-Gluon $\NeqZero$
Amplitudes with Two Negative Helicities}
\label{AllnSection}

\def\indentA{\hskip 7mm}
\def\indentB{\hskip 5mm}

We now turn to the analytic construction of the all-$n$ amplitude.
The complete amplitude is decomposed according to,
\be
\Al_{n;1}(1,m) = c_{\Gamma} \bigg[ \hat{C}_n(1,m) + \hat{R}_n(1,m) \bigg] \,,
\label{eq:the_comple_all_n_amplitude}
\ee
where $1$ and $m$ label the negative-helicity legs.
Here $\hat{C}_n$ is the appropriate cut completion of $C_n$,
given in \eqn{BBST}.  To calculate $\hat{R}_n(1,m)$ we will, as in the
examples of the previous section, choose a $\Shift 1m$ shift.  The
remaining rational terms are given by \eqn{BasicFormula}. Using the
assumption, extrapolated from the four- and five-gluon cases,
that $\Inf A_n$ vanishes for a $\Shift 1m$ shift\footnote{As
 mentioned earlier, the assumption that $\Inf A_n$ vanishes in this
case can be checked by requiring consistency of the final result with all
factorization limits.}, we have,
\begin{eqnarray}
\hat{R}_n(1,m)= R^D_n(1,m)+O_n(1,m) - \Inf\hat{C}_n(1,m)
\,.
\end{eqnarray}
Our first step is to determine the large-$z$ behavior of the
cut-completed terms $\hat{C}_n(1,m)$.
The result of this exercise is given in \eqn{eq:definition_of_InfC}
of \app{CutAppendix}.

Next we turn our attention to the overlap terms, $\Overlap_n$.
These terms are calculated by taking the residues
of the poles in $z$ generated by shifting $\CuthRat_n$ of
\eqn{eq:CR_hat_unshifted} by $\Shift{1}{m}$. In $\Overlap_n$ we do not
include the residues of poles coming from the tree amplitude
which is an overall prefactor of $\CuthRat_n(1,m)$ in
\eqn{eq:CR_hat_unshifted}. Such residues would cancel against
corresponding terms produced by the insertion of $\CuthRat_n$
into~\eqn{eq:unwinding_def} below.  Hence we shall drop both types
of terms from the following discussion.  We refer to these
modified overlap terms as $\tilde{O}_n$.
The final remaining piece
necessary for calculating the complete rational term comes from the
recursive rational diagrams, $R^D_n(1,m)$. These
terms are computed using an on-shell recursion relation for $\hat{R}_n$
generated by our chosen $\Shift 1 m$ shift. The non-zero terms of the
recursion relation coming from this shift are,
\begin{eqnarray}
&&\hspace*{-0.8cm}\hat{R}_n(1,m) \nonumber\\
&=&\hR_{n-1}(\hat{1}^-,\ldots,(m-2)^+,\hK^-_{(m-1)m},(m+1)^+,\ldots,n^+)
\frac{i}{s_{(m-1)m}} \nonumber\\
&&\hphantom{\hR_{n-1}((m+1)^+,\ldots,\hat{1}^-,\ldots,(m-2)^+}
\times\At_3(-\hK^+_{(m-1)m},(m-1)^+,\hat{m}^-) \nonumber\\
&&\null
+\hR_{n-1}(\hat{1}^-,\ldots,(m-1)^+,\hK^-_{m(m+1)},(m+2)^+,\ldots,n^+)
\frac{i}{s_{m(m+1)}} \nonumber\\
&&\hphantom{\hR_{n-1}((m+2)^+,\ldots,\hat{1}^-,\ldots,(m-1)^+}
\times\At_3(-\hK^+_{m(m+1)},\hat{m}^-,(m+1)^+) \nonumber\\
&&\null
+\Asdef(1^-,2^+,\ldots,m^-,\ldots,n^+) + \tilde{O}_n(1,m)
- \Inf\hat{C}_n(1,m) \,.
\label{eq:unwinding_def}
\end{eqnarray}
Here we have combined all the pieces containing known all-$n$
amplitudes~\cite{Mahlon,Qpap} into $\Asdef$, which is defined by,
\begin{eqnarray}
&&\hspace*{-0.8cm}\Asdef(1^-,2^+,\ldots ,m^-,\ldots ,n^+)
\nonumber\\
&=&\sum_{j_1=2}^{m}\sum_{j_2=\max(m,j_1+2)}^{\min(n,j_1+n-3)}
   \Bigg(\At_{n-j_2+j_1}(\hat{1}^-,2^+,\ldots ,(j_1-1)^+,
\hK^-_{j_1\ldots j_2},(j_2+1)^+,\ldots ,n^+)\frac{i}{s_{j_1\ldots j_2}}
\nonumber\\
&&\hphantom{\sum_{j_1=}\sum_{j_2=}\Bigg(\At(}\times
R_{j_2-j_1+2}(-\hK^+_{j_1\ldots j_2},j_1^+,\ldots ,\hat{m}^-,\ldots ,j_2^+)
\nonumber\\
&& \null
+ R_{n-j_2+j_1}(\hat{1}^-,2^+,\ldots ,(j_1-1)^+,
\hK^+_{j_1\ldots j_2},(j_2+1)^+,\ldots ,n^+)\frac{i}{s_{j_1\ldots j_2}}
\nonumber\\
&&\hphantom{\sum_{j_1=}\sum_{j_2=}\Bigg(\At(}
 \times\At_{j_2-j_1+2}
 (-\hK^-_{j_1\ldots j_2},j_1^+,\ldots ,\hat{m}^-,\ldots ,j_2^+)
\Bigg) \,.
\label{eq:asdef}
\end{eqnarray}
As mentioned above, the reason $\hat{R}$
instead of the full rational term $R$ appears on the right-hand
side of \eqn{eq:unwinding_def},
is that the additional $\CuthRat_n$ terms have been cancelled
against certain overlap contributions.  The three-point loop
vertices appearing in \eqn{eq:asdef} vanish,
$R_3(\hat{1}^-,2^+,\hK^+_{3\ldots n})
= R_3(\hat{1}^-,\hK^+_{2\ldots(n-1)},n^+) = 0$.

The recursive rational contributions naturally split themselves into
two classes.  The first class consists of the
one-loop amplitudes with one negative-helicity leg (which are
completely rational) and the tree amplitudes that multiply them.
These terms are fully known and are contained in $\Asdef$.
The second class of recursive contributions
is, however, more difficult.  They are given by the first two terms
of \eqn{eq:unwinding_def} and consist of the rational parts of
lower-point one-loop amplitudes with two negative-helicity legs, along
with the tree amplitudes that multiply them.  Specifically, these
unknown amplitudes are exactly those we are trying to determine,
having the same number of negative-helicity legs, but with fewer
positive-helicity legs.  The key for obtaining an all-$n$
expression is to solve recursively for this class of terms.


\subsection{Solving the recursion relation}

To solve \eqn{eq:unwinding_def} recursively for
$\hat{R}_n(1,m)$, we follow the same
general plan as given in ref.~\cite{FMassive} for the on-shell recursion
relations for massive tree-level
amplitudes~\cite{GloverMassive,Massive}.  (Other examples of
systematic solutions to on-shell recursion relations may be found in
refs.~\cite{TreeRecurResults,RecurCoeff,FordeKosower}.)

Our basic tactic is to insert the left-hand side of
\eqn{eq:unwinding_def} into the right-hand side of
\eqn{eq:unwinding_def} repeatedly. At each insertion we find that our
desired amplitude $\hat{R}_n(1,m)$ appears on the right-hand side with
one fewer positive-helicity leg, and multiplied by one more three-point
gluon vertex. This `unwinding' of the amplitude continues until we
have reduced the right-hand side of $\hat{R}_n$
(\eqn{eq:unwinding_def}) down to, in this case, $\hat{R}_4$ and a sum
of terms that contain only known quantities ($\Asdef$, overlap terms
$\tilde{O}$ and large-$z$ subtraction terms $\Inf \Cuth$) multiplied by
strings of $\At_3$ vertices.

The presence of two terms on the right-hand side of
\eqn{eq:unwinding_def} containing $\hat{R}_{n-1}$ means that we remove
a leg from either side of $m$ at each step of this unwinding. For
example, our solution contains the following four terms involving
$\Asdef$ at the second unwinding step:
\begin{eqnarray}
&&\hspace*{-0.8cm}
\Asdef(\hat{1}^-,\ldots ,(m-3)^+,\hK^-_{(m-2)(\hK_{(m-1)m})},(m+1)^+,\ldots ,n^+)
\frac{i}{s_{(m-2)(\hK_{(m-1)m})}}\label{eq:example_4_terms_unwinding}
\\
&&\times\At_3(-\hK^+_{(m-2)(\hK_{(m-1)m})},(m-2)^+,\hK^-_{(m-1)m})\frac{i}{s_{(m-1)m}}
\At_3(-\hK^+_{(m-1)m},(m-1)^+,\hat{m}^-) \,,\nonumber
\\
&&\hspace*{-0.8cm}
\Asdef(\hat{1}^-,\ldots ,(m-2)^+,\hK^-_{(\hK_{(m-1)m})(m+1)},(m+2)^+,\ldots ,n^+)
\frac{i}{s_{(\hK_{(m-1)m})(m+1)}}\nonumber
\\
&&\times\At_3(-\hK^+_{(\hK_{(m-1)m})(m+1)},\hK^-_{(m-1)m},(m+1)^+)
\frac{i}{s_{(m-1)m}}\At_3(-\hK^+_{(m-1)m},(m-1)^+,\hat{m}^-) \,,
\nonumber
\\
&&\hspace*{-0.8cm}
\Asdef(\hat{1}^-,\ldots ,(m-2)^+,\hK^-_{(m-1)(\hK_{m(m+1)})},(m+2)^+,\ldots ,n^+)
\frac{i}{s_{(m-1)(\hK_{m(m+1)})}}\nonumber
\\
&&\times\At_3(-\hK^+_{(m-1)(\hK_{m(m+1)})},(m-1)^+,\hK^-_{m(m+1)})
\frac{i}{s_{m(m+1)}}
\At_3(-\hK^+_{m(m+1)},\hat{m}^-,(m+1)^+)\,,
\nonumber
\\
&&\hspace*{-0.8cm}
\Asdef(\hat{1}^-,\ldots ,(m-1)^+,\hK^-_{(\hK_{m(m+1)})(m+2)},(m+3)^+,\ldots ,n^+)
\frac{i}{s_{(\hK_{m(m+1)})(m+2)}}\nonumber
\\
&&\times\At_3(-\hK^+_{(\hK_{m(m+1)})(m+2)},\hK^-_{m(m+1)},(m+2)^+)
\frac{i}{s_{m(m+1)}}
\At_3(-\hK^+_{m(m+1)},\hat{m}^-,(m+1)^+) \,.
\nonumber
\end{eqnarray}
As can be seen, we get a term for each possible order we can
extract two legs from around the leg $m$ in the amplitude. Therefore,
after extracting $l$ legs ({\it i.e.}, performing $l$ unwinding
steps) we obtain $2^l$ terms. Each term corresponds to a
particular ordering of the $l$ legs that we extract from
around $m$.

At each step of the unwinding we must choose new shifted momenta.
We always choose to shift the two negative-helicity
legs of $\hat{R}$. For example, after the first step we choose
$\Shift {\hat 1} {\hK_{m(m+1)}} $ or $\Shift {\hat1} { \hK_{(m-1)m}}$ as
the shifted legs, depending upon which of the two terms on the right-hand 
side of \eqn{eq:unwinding_def} we insert $\hR_{n-1}$ into.
Similarly, when we perform a
second insertion, of $\hR_{n-2}$, we choose the intermediate $\hK$
momentum leg of the last shift and the previously shifted $\hat{1}$
leg.

\begin{figure}[t]
\centerline{\epsfxsize 3.5 truein\epsfbox{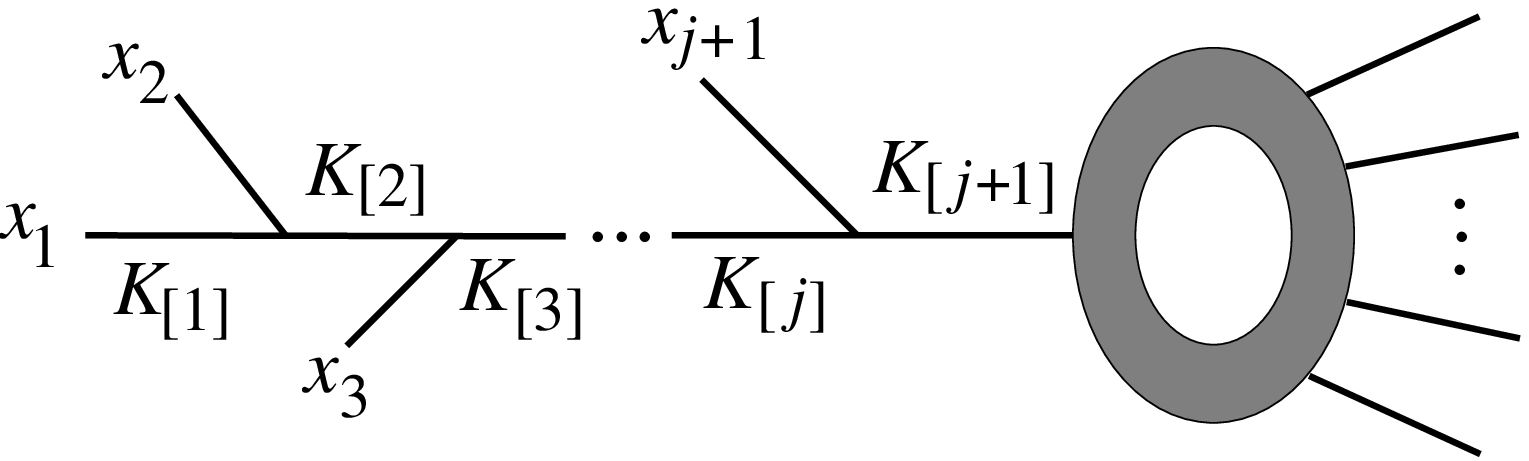}}
\caption{The propagators corresponding to the sequence of $K_{[i]}$.}
\label{UnwindFigure}
\end{figure}

At each step we extract one leg from $\hR_i$ and create one extra
three-point gluon vertex. The order of extraction can be thought of as
a path of extracted legs denoted by the ordered set of momenta
$\{x_1,x_2,\ldots,x_{j+1}\}$, starting with $x_1=k_m$. Then $x_2$
denotes the momentum of the first extracted leg after $m$, $x_3$ is
the momentum of the leg extracted after $x_2$, and so on, until we
reach the last extracted leg $x_{j+1}$ (if $j$ unwinding steps are
performed).  We need to define two sequences of momentum sums related
to the chain of intermediate momenta appearing as arguments of the
three-point gluon vertices.  We denote these sequences by $\{ K_{[i]}
\}$, $i=1,\ldots,j+1$ and $\{ \hK_{[i]} \}$, $i=1,\ldots,j$. 
We also define a sequence $\{\hat k_{[i]}\}$ of
corresponding shifts of $k_1$, for 
 $i=0,\ldots,j$.  All these momenta are on shell,
even $\{ K_{[i]} \}$.  The momentum $K_{[i]}$ is defined recursively
by adding the external momentum $x_i$ to $K_{[i-1]}$ and performing a
shift in order to bring the sum back on shell.  The shift also alters
the propagator leg of the previous unwinding step from $K_{[i-1]}$ to
$\hK_{[i-1]}$, keeping it on shell.  (See \fig{UnwindFigure}.)
The sequence of shifted $\{ \hat{k}_{[i]} \}$ is dictated by momentum
conservation.

The elements of the sequences involved in the first unwinding step are,
for the case $x_2=m-1$,
\begin{eqnarray}
K^{\mu}_{[1]}&=&k_m^{\mu} \,,
\nonumber\\
K^{\mu}_{[2]}&=&k_{m-1}^{\mu}+k_m^{\mu}-\frac{K^2_{(m-1)m}
\Bmp{1^-|\gamma^{\mu}|k^-_m}}{2\Bmp{1^-|\s k_{m-1}|k^-_m}}\equiv\hK_{(m-1)m} \,,
\nonumber\\
\hK^{\mu}_{[1]}&=&k_m^{\mu}-\frac{K^2_{(m-1)m}
\Bmp{1^-|\gamma^{\mu}|k^-_m}}{2\Bmp{1^-|\s k_{m-1}|k^-_m}}\equiv \hat{m}\,,
\nonumber\\
\hat{k}^{\mu}_{[0]}&=&k^{\mu}_1 \,,
\nonumber\\
\hat{k}^{\mu}_{[1]}&=&k_1^{\mu}+\frac{K^2_{(m-1)m}
\Bmp{1^-|\gamma^{\mu}|k^-_m}}{2\Bmp{1^-|\s k_{m-1}|k^-_m}}\equiv \hat{1} \,.
\label{KKksequences}
\end{eqnarray}
In this notation, the expression obtained at the first step
of the unwinding process is,
\begin{eqnarray}
&&\hspace*{-0.8cm}
\frac{i\At_3(-K^+_{[2]},(m-1)^+,\hK^-_{[1]})}{K_{(m-1)m}^2}
\Asdef(\hat{k}^-_{[1]},2^+,\ldots,(m-2)^+,K^-_{[2]},(m+1)^+,\ldots,n^+) \,.
\label{eq:just_before_the_3_poin_id_examp}
\end{eqnarray}

More generally, after $j$ unwinding steps, we have
the following product of $\At_3$ vertices multiplied by an $\Asdef$
term,
\begin{eqnarray}
&&\hspace*{-0.8cm}(-1)^{i_2-m} \Biggl[ \prod_{r=2}^{j+1}
\frac{i\At_3(-K^+_{[r]},x_{r}^+,\hK^-_{[r-1]})}
{(x_{r}+K_{[r-1]})^2}
\Biggr]
\Asdef(\hat{k}^-_{[j]},2^+,\ldots,
(i_1-1)^+,K^-_{[j+1]},(i_2+1)^+,\ldots,n^+) \,,
\nonumber\\
\label{eq:just_before_the_3_poin_id}
\end{eqnarray}
where the propagators corresponding to the $K_{[i]}$ are
depicted in \fig{UnwindFigure}.  Here $j$ legs have been extracted, 
$(m-i_1)$ legs from the left of $m$ and $(i_2-m)$ legs from the right of $m$. 
The factor of $(-1)^{i_2-m}$, with $(i_2-m)$ the number of legs $x_i$ that
are after $m$ in the external ordering, comes from reversing the order
of the arguments of the three-point vertices of legs extracted in
positions above $m$ in the external ordering, {\it i.e.},
\begin{eqnarray}
\At_3(-K^+_{[r]},\hK^-_{[r-1]},x_{r}^+)
=(-1) \At_3(-K^+_{[r]},x_{r}^+,\hK^-_{[r-1]}) \,.
\end{eqnarray}
This reordering is needed so that all extracted three-point vertices
have the same order as given in \eqn{eq:just_before_the_3_poin_id}.

The recursive definitions of the sequences $\{ K_{[i]} \}$,
$\{ \hK_{[i]} \}$ and $\{ \hat{k}_{[i]} \}$, which satisfy
the on-shell requirements, are
\begin{eqnarray}
K^{\mu}_{[1]} &=& k_m^{\mu} \,,
\nonumber\\
K^{\mu}_{[i]} &=& x_{i}^{\mu}+K_{[i-1]}^{\mu}+w^{\mu}_{[i]} \,,
\nonumber\\
\hK^{\mu}_{[i]} &=& K_{[i]}^{\mu}+w^{\mu}_{[i+1]}\,,
\nonumber\\
\hat{k}^{\mu}_{[0]} &=& k^{\mu}_1,\;
\nonumber\\
\hat{k}^{\mu}_{[i]} &=& \hat{k}^{\mu}_{[i-1]}-w^{\mu}_{[i+1]} \,,
\label{eq:def_of_shifted_4_momentum}
\end{eqnarray}
where
\begin{equation}
w^{\mu}_{[i]} = - \frac{(x_{i}+K_{[i-1]})^2
\Bmp{\hat{k}_{[i-2]}^-|\gamma^{\mu}|K^-_{[i-1]}}}
{2\Bmp{\hat{k}_{[i-2]}^-|\s x_{i}|K^-_{[i-1]}}}
= - \frac{(x_{i}+K_{[i-1]})^2
\Bmp{1^-|\gamma^{\mu}|K^-_{[i-1]}}}{2\Bmp{1^-|\s x_{i}|K^-_{[i-1]}}} \,.
\label{wshiftdef}
\end{equation}
We can make the identification on the right-hand side of \eqn{wshiftdef} 
because the choice of shifts is such that the unshifted spinors at step $i$ are
given by,
\begin{eqnarray}
\langle \hat{k}^-_{[i]}|=\langle \hat{k}^-_{[i-1]}|= \cdots =\langle
1^-|\,,&& \hskip 1 cm \langle \hK_{[i]}^+|=\langle \hK_{[i-1]}^+|=
\cdots =\langle m^+| \,.
\label{eq:unwinding_shift_simp}
\end{eqnarray}
The shifted spinors at step $i$ are given by
\begin{eqnarray}
\langle \hat{k}_{[i]}^+|=\langle \hat{k}_{[i-1]}^+|+\frac{(x_{i+1}+K_{[i]})^2}{\Bmp{1^-|\s
    x_{i+1}|K^-_{[i]}}}\langle K^+_{[i]}| \,,&& \hskip .5 cm
\langle \hK_{[i]}^-|=
 \langle K_{[i]}^-| - \frac{(x_{i+1}+K_{[i]})^2}{\Bmp{1^-|\s
    x_{i+1}|K^-_{[i]}}}\langle K^-_{[i]}| \,. \hskip .8 cm
\label{eq:spinor_shift_simp}
\end{eqnarray}
From \eqn{eq:def_of_shifted_4_momentum} we can see that momentum conservation is
satisfied,
\begin{eqnarray}
\hat{k}^{\mu}_{[i-1]}+K^{\mu}_{[i]}=\hat{k}^{\mu}_{[i-2]}+x_{i}^{\mu}+K_{[i-1]}^{\mu}
= \cdots 
= k^{\mu}_1+\sum_{\ell=1}^{i}x^{\mu}_\ell \,.
\label{eq:unwinding_mom_simp}
\end{eqnarray}

Using the representation of the one-loop $n$-gluon amplitudes
with one negative helicity in ref.~\cite{Qpap}, one can see
that the left-handed spinor $\tilde\lambda_1$
for the negative-helicity leg (leg 1) never appears.
From \eqn{eq:unwinding_shift_simp}, the right-handed spinor
$\lambda_1$ that does appear is unshifted,
$\lambda_{\hat{k}_{[i]}}=\langle\hat{k}_{[i]}^-| = \langle1^-|$.

Next, using the identity,
\begin{eqnarray}
&&\hspace*{-0.8cm}\prod_{r=2}^{j+1}
\frac{i \At_3(-K^+_{[r]},x_{r}^+,\hK^-_{[r-1]})}{(x_{r}+K_{[r-1]})^2}
=  i \frac{\At_{j+2}(1^-,x^+_{j+1},x_j^+,\ldots,x^+_2,m^-)}{\Bmp{1 K_{[j+1]}}^2} \,,
\label{eq:the_key_unwinding_id}
\end{eqnarray}
we rewrite the general term~(\ref{eq:just_before_the_3_poin_id}) as
\begin{eqnarray}
&&\hspace*{-0.8cm}(-1)^{(i_2-m)}\At_{j+2}(1^-,x_{j+1}^+,x_j^+,\ldots,x_2^+,m^-)
\frac{i}{\spash{1}.{K_{[j+1]}}^2}
\nonumber\\
&&\times\Asdef(1^-,2^+,\ldots,(i_1-1)^+,K_{[j+1]}^-,(i_2+1)^+,\ldots,n^+)
\,.
\label{eq:gen_term_unwinding_1}
\end{eqnarray}
If $\Asdef$ contains a $\lambda_m$ spinor (where $m$ is the
negative-helicity leg of $\Asdef$, other than leg $1$) then 
we rewrite the general form of this term as
\begin{eqnarray}
  &&\hspace*{-0.8cm}(-1)^{(i_2-m)}\At_{j+2}(1^-,x_{j+1}^+,x_j^+,\ldots,x_2^+,m^-)
\frac{i}{\spash{1}.{K_{[j+1]}}^2}
  \nonumber\\
  &&\times\Asdefp(1^-,2^+,\ldots,(i_1-1)^+,K_{[j+1]}^-,(i_2+1)^+,\ldots,n^+)f(j+1) \,.
\label{eq:isolated_gen_term}
\end{eqnarray}
Here $\Asdefp$ is defined to contain no $\lambda_m$ spinors; we construct
$f(j+1)$ such that it contains all such factors. The argument $(j+1)$
denotes the dependence of this function on the momentum of the last
leg that was extracted for this term in the unwinding, $x_{j+1}$. This
step is necessitated by our desire to combine together, into a single 
simple form, all terms containing the same set of extracted legs.
The terms involved have to be independent of the order of
extraction of these legs, for it to be possible to combine them.

Looking at \eqns{eq:gen_term_unwinding_1}{eq:isolated_gen_term} 
we see immediately that $K_{[j]}$
depends upon the order of the path of extracted legs.
Because of the homogeneity of the spinors in the amplitude, the
denominator factor of $\Bmp{1K_{[j]}}^2$ must combine with a
complementary factor in the quantity $\Asdef$ multiplying it, 
to generate spinor strings of the form
$\langle 1^-|\s K_{[j]}\ldots $. 
This string can be rewritten using
$\langle 1^-|\s K_{(\hK_{a\ldots b})c}=\langle 1^-|\s K_{a\ldots b c}$ 
to get 
$\langle 1^-| \s {\hat{K}}_{i_1\ldots i_2}\ldots 
= \langle 1^-| \s K_{i_1\ldots i_2}\ldots$
(because the spinor $\lambda_1$ in the shift always contracts with
$\langle 1^-|$ to give $\spa1.1=0$).
Now, $K_{i_1\ldots i_2}$ is just
the consecutive sum of momenta between the left-most extracted leg
$i_1$ and the right-most extracted leg $i_2$.  So these terms become
independent of the order of the extracted legs in the unwinding, and
henceforth we will simply replace any such appearance of $K_{[j]}$
with $\hat{K}_{i_1\ldots i_2}$.

However, this is not the only source of a path dependence. The presence
of a $\lambda_m$ spinor in $\Asdef$ also leads to such a
dependence, for example a $\Bmp{y m}$ term in $\Asdef$ becomes
$\Bmp{y K_{[j+1]}}$ during the unwinding. We find that we
cannot completely eliminate the path dependence coming from a $\Bmp{y m}$
term.  We can only reduce it to that of the last extracted leg. 
To see this, consider again $\Bmp{y K_{[j+1]}}$ as it will appear 
in the amplitude combined with $\Bmp{1K_{[j+1]}}$\,,
\begin{eqnarray}
\frac{\Bmp{y K_{[j+1]}}[K_{[j+1]}K_{[j]}]}{\Bmp{1K_{[j+1]}}[K_{[j+1]}K_{[j]}]}
\equiv
\frac{\Bmp{y x_{j+1}}[x_{j+1} K_{[j]}]}{\Bmp{1x_{j+1}}[x_{j+1} K_{[j]}]}
=\frac{\Bmp{y x_{j+1}}}{\Bmp{1x_{j+1}}} \,.
\label{xsimpl}
\end{eqnarray}
The factor of $\Bmp{1K_{[j+1]}}$ in the denominator here is always
guaranteed to be present due to the homogeneity of the spinors.  From
this example we see that, after removing any dependence upon the
shifted external momentum from the amplitude, we can distinguish
different paths of extracted legs \textit{only} by the last leg
extracted from $\hR$. For the amplitudes we consider here we find
that such $\lambda_m$ spinors are always present, so to proceed we
must first isolate all $\lambda_m$ as in \eqn{eq:isolated_gen_term}.
Once they are isolated, we can straightforwardly reduce the
path-dependence to that of the final extracted leg only.

With this simplified dependence upon the extracted path we can now
proceed to combine together all terms containing the same set of extracted
legs $\{i_1,\ldots,m, \ldots, i_2\}$.  All terms that correspond
to each possible path between two legs are combined.  For example,
considering the case when $i_1=m-1$ and $i_2=m+1$, we
combine the middle two terms of \eqn{eq:example_4_terms_unwinding}, as
these correspond to the two possible ways of extracting the set of
legs $\{m-1,m,m+1\}$. In general for the set of extracted legs
$\{i_1,\ldots, i_2\}$ we see that the last leg extracted from such a
contributing term can only be either $i_1$ or $i_2$. Therefore we
write the sum of all possible extraction paths of the legs
$\{i_1,\ldots,m, \ldots, i_2\}$ as two sums, one for
$x_{j+1}=k_{i_1}$ and one for $x_{j+1}=k_{i_2}$, to get,
\begin{eqnarray}
&&\hspace*{-0.0cm}\Bigg(
f(i_1) \, (-1)^{n_{\beta_1}}
\sum_{\sigma\in {\rm OP}(\alpha_1,\beta_1)}
\At_{i_2-i_1+2}(1^-,i_1^+,\sigma(\alpha_1,\beta_1),m^-)
\nonumber\\
&& - f(i_2) \, (-1)^{(n_{\beta_1}-1)}
\sum_{\sigma\in {\rm OP}(\alpha_2,\beta_2)}
\At_{i_2-i_1+2}(1^-,i_2^+,\sigma(\alpha_2,\beta_2),m^-) \Bigg)
\nonumber\\
&&\hspace*{0.3cm}
\times\frac{i}{\spash{1}.{\hK_{i_1\ldots i_2}}^2}
\Asdef'(1^-,2^+,\ldots,(i_1-1)^+,\hK_{i_1\ldots i_2}^-,(i_2+1)^+,\ldots,n^+)
\nonumber\\
&=&\Bigg(
f(i_1) \,
\frac{\spa{1}.{m}^3}{\spa{1}.{i_1}\spa{i_1}.{m}^3} \, (-1)^{n_{\beta_1}}
\sum_{\sigma\in {\rm OP}(\alpha_1,\beta_1)}
\At_{i_2-i_1+1}(i_1^-,\sigma(\alpha_1,\beta_1),m^-)
\nonumber\\
&&-
f(i_2) \,
\frac{\spa{1}.{m}^3}{\spa{1}.{i_2}\spa{i_2}.{m}^3} \, (-1)^{n_{\beta_2}}
\sum_{\sigma\in {\rm OP}(\alpha_2,\beta_2)}
\At_{i_2-i_1+1}(i_2^-,\sigma(\alpha_2,\beta_2),m^-)\Bigg)
\nonumber\\
&&\hspace*{0.3cm}
\times\frac{i}{\spa{1}.{\hK_{i_1\ldots i_2}}^2}
\Asdef'(1^-,2^+,\ldots,(i_1-1)^+,
        \hK_{i_1\ldots i_2}^-,(i_2+1)^+,\ldots,n^+) \,,
\label{eq:simplifying_the_unwinding_1}
\end{eqnarray}
where $\sigma(\alpha_i,\beta_i)$ indicates one of the possible
`ordered permutations (OP)' or `mergings' of the ordered sets
$\alpha_i$ and $\beta_i$, with
$\alpha_1=\{i_1+1,\ldots,m-1\}$,
$\beta_1=\{i_2,\ldots,m+1\}$, $\alpha_2=\{i_1,\ldots,m-1\}$ and
$\beta_2=\{i_2-1,\ldots,m+1\}$.  The elements of each merging are
the union of the two sets,
$\alpha_i \cup \beta_i \equiv \{x_j,\ldots,x_2\}$.
The merged ordering must preserve the order of the elements
$\alpha_i$ and $\beta_i$ individually, but any relative ordering
of elements of $\alpha_i$ with respect to those of $\beta_i$
is permitted.  Also, $n_{\beta_i}$ is the
number of elements in the set $\beta_i$. We can then rewrite
\eqn{eq:simplifying_the_unwinding_1} using the Kleiss-Kuijf
relation~\cite{KleissKuijf} between tree amplitudes
(for which a simple proof was given in ref.~\cite{LanceColor}).
The Kleiss-Kuijf relation reads, in our notation,
\begin{equation}
A_n^{\tree}(1,\alpha,n,\beta^T)
= (-1)^{n_\beta} \sum_{\sigma\in{\rm OP}(\alpha,\beta)}
 A_n^{\tree}(1,\sigma(\alpha,\beta) ,n) \,,
\label{eq:KKrelation}
\end{equation}
where $\beta^T$ is $\beta$ written in the reversed order.
Applying it to \eqn{eq:simplifying_the_unwinding_1} and
rearranging the resulting MHV tree amplitudes to restore gluon 1,
we obtain,
\begin{eqnarray}
&&\hspace*{-0.8cm}\frac{\spa{i_2}.{1}\spa{i_1}.{m}f(i_1)
-\spa{i_1}.{1}\spa{i_2}.{m}f(i_2)}{\spa{1}.{m}\spa{i_2}.{i_1}}
\At_{i_2-i_1+2}(1^-,i_1^+,\ldots,m^-,\ldots,i_2^+)\nonumber\\
&&\times\frac{i}{\spa{1}.{\hK_{i_1\ldots i_2}}^2}
\Asdef'(1^-,2^+,\ldots,(i_1-1)^+,\hK_{i_1\ldots i_2}^-,(i_2+1)^+,\ldots,n^+) \,.
\label{eq:gen_m_rep_id_for_unwinding}
\end{eqnarray}
The final result is then given by summing
\eqn{eq:gen_m_rep_id_for_unwinding} over all possible sets of
extracted legs $\{i_1,\ldots,m,\ldots,i_2\}$, which is equivalent to
summing over all possible factorization channels of the amplitude.

In the simplest case, $\Asdef$ contains no $\lambda_m$ spinors and
$f(i)=1$.  In this case, using the Schouten identity, 
\eqn{eq:gen_m_rep_id_for_unwinding} reduces to
\begin{eqnarray}
&&\At_{i_2-i_1+2}(1^-,i_1^+,\ldots,m^-,\ldots,i_2^+) \nn \\
&&\hskip1.3cm
\times \frac{i}{\spa{1}.{\hK_{i_1\ldots i_2}}^2}
\Asdef(1^-,2^+,\ldots,(i_1-1)^+,\hK_{i_1\ldots i_2}^-,(i_2+1)^+,\ldots,n^+)\,.
\end{eqnarray}
The case of the amplitude considered here is more complicated.  
We always have to take into account a $\lambda_m$ spinor contained in the
denominator factor of $s_{j_1\ldots \hat{m}\ldots j_2}$ in the $\Asdef$
of \eqn{eq:asdef}. During the unwinding, 
$s_{j_1\ldots \hat{m}\ldots j_2}$ becomes 
$s_{j_1\ldots K_{[j+1]}\ldots j_2}$,
which, using \eqn{xsimpl}, can be rewritten as
\begin{eqnarray}
s_{j_1\ldots K_{[j+1]}\ldots j_2}&=&\Bmp{K_{[j+1]}^-|\s
K_{j_1\ldots (i_1-1),(i_2+1)\ldots j_2}|K_{[j+1]}^-}+s_{j_1\ldots
(i_1-1)(i_2+1)\ldots j_2} \nonumber\\
&=&\frac{\Bmp{x_{j+1}^-|\Gdefd{j_1,i_1,i_2,j_2}|1^+}}{\Bmp{x_{j+1}1}} 
\,,
\end{eqnarray}
where we define,
\begin{eqnarray}
\Gdef{a,b,c,d}&=&\s K_{a\ldots d} \s K_{a\ldots (b-1),(c+1)\ldots d}\,,\nn\\
\Gdefd{a,b,c,d}&=&\s K_{a\ldots (b-1),(c+1)\ldots d}\s K_{a\ldots d} \,.
\label{GdefA}
\end{eqnarray}
and again $x_{j+1}$ is the last leg extracted during the unwinding.

Hence, if we isolate a factor of $1/s_{j_1\ldots \hat{m}\ldots j_2}$ 
from the remainder of $\Asdef$ (along the lines of
\eqn{eq:isolated_gen_term}), then we see that $f(j+1)$ for
the final extracted leg $x_{j+1}$ is given by
\begin{eqnarray}
&&\hspace*{-0.8cm}f(j+1)
=\frac{\spa{x_{j+1}}.{1}}{\Bmp{x_{j+1}^-|\Gdefd{j_1,i_1,i_2,j_2}|1^+}} \,.
\label{eq:def_of_f_below}
\end{eqnarray}
Inserting this result for $f(j+1)$ into
\eqn{eq:gen_m_rep_id_for_unwinding} and taking into account the cases
when $i_1=m$ or $i_2=m$ we get the result for a specific set of
extracted legs $\{i_1,\ldots,m,\ldots,i_2\}$,
\begin{eqnarray}
&&\hspace*{-0.8cm}
-\sum_{j_1=2}^{i_1} \sum_{j_2=\max(i_2,i_2+2-i_1+j_1)}^{\min(n,n-3+j_1)}
\Kdef{j_1,j_2,i_1,i_2,m}
\At_{i_2-i_1+2}(1^-,i_1^+,\ldots,m^-,\ldots ,i_2^+)\frac{1}{\spa{1}.{\hK_{i_1\ldots i_2}}^2}
\nonumber\\
&&\hphantom{s_{a\ldots b}\spa{1}.{i_1}\spa{i_2}.{1}}
\times
\Asdef^{'\,j_1j_2}(1^-,2^+,\ldots ,(i_1-1)^+,\hK_{i_1\ldots i_2}^-,(i_2+1)^+,\ldots ,n^+) \,,
\end{eqnarray}
with
\begin{eqnarray}
\Kdef{a,b,i_1,i_2,m}=\left\{
\begin{array}{ll}
i_1 = m \textrm{ and } i_2 = m:&1/s_{a\ldots b} \comma\\
i_1 = m :&
\spa{i_2}.{1}/\sandmp{i_2}.{\Gdefd{a,i_1,i_2,b}}.1 \comma
\\
i_2 = m :&
\spa{1}.{i_1}/\sandmp{1}.{\Gdef{a,i_1,i_2,b}}.{i_1} \comma
\\
\textrm{otherwise}:&
\displaystyle
 \frac{\sandmp{m}.{\Gdefd{a,i_1,i_2,b}}.1\spa{1}.{i_1}\spa{i_2}.{1}}
{\spa{m}.{1}\sandmp{1}.{\Gdef{a,i_1,i_2,b}}.{i_1}
\sandmp{i_2}.{\Gdefd{a,i_1,i_2,b}}.{1}}
\,,
\end{array}
\right.
\end{eqnarray}
and
\begin{eqnarray}
&&\hspace*{-0.8cm}\Asdef^{'\,j_1j_2}(1^-,2^+,\ldots ,m^-,\ldots ,n^+)
\nonumber\\
&\equiv&
\At_{n-j_2+j_1}(\hat{1}^-,2^+,\ldots ,(j_1-1)^+,
\hK^-_{j_1\ldots j_2},(j_2+1)^+,\ldots ,n^+)
\nonumber\\
&&\hphantom{\sum_{j_1=}\sum_{j_2=}\Bigg(\At(}\times
R_{j_2-j_1+2}(-\hK^+_{j_1\ldots j_2},j_1^+,\ldots ,\hat{m}^-,\ldots ,j_2^+)
\nonumber\\
&& \null
+  R_{n-j_2+j_1}(\hat{1}^-,2^+,\ldots ,(j_1-1)^+,
\hK^+_{j_1\ldots j_2},(j_2+1)^+,\ldots ,n^+)
\nonumber\\
&&\hphantom{\sum_{j_1=}\sum_{j_2=}\Bigg(\At(}
 \times\At_{j_2-j_1+2}
 (-\hK^-_{j_1\ldots j_2},j_1^+,\ldots ,\hat{m}^-,\ldots ,j_2^+) \,,
\label{eq:asdefp}
\end{eqnarray}
which are the terms of $\Asdef$ with a factor of $i$ and 
the denominator factors of
$s_{j_1\ldots (\hK_{i_1\ldots i_2})\ldots j_2}$ removed.

To obtain the final result we must include all possible factorization
channels.  We also must include the overlap and $\Inf \hat{C}_n$ terms.
Finally, there is a contribution from the terminal step of the unwinding,
because $\hR_4$ is non-zero for this amplitude.  The complete result for
$\hat{R}_n$ is then given by,
\begin{eqnarray}
&&\hspace*{-0.8cm}\hat{R}_n(1,m)
\nonumber\\
&=& - \sum_{i_1=2}^{m}}{\sum_{i_2=m}^{\min(n,n+i_1-5)}
\At_{i_2-i_1+2}(1^-,i_1^+,\ldots,m^-,\ldots ,i_2^+)\frac{1}{\spa{1}.{\hK_{i_1\ldots i_2}}^2}
\Bigg[
\nonumber\\
&&\hskip1cm \null
 i \Inf \hat{C}_{n-i_2+i_1}(1^-,2^+,\ldots ,(i_1-1)^+,\hK_{i_1\ldots
  i_2}^-,(i_2+1)^+,\ldots ,n^+)
\nonumber\\
&&\hskip1cm
+ \sum_{j_1=2}^{i_1} \sum_{j_2=\max(i_2,i_2+2-i_1+j_1)}^{\min(n,n-3+j_1)}
\Kdef{j_1,j_2,i_1,i_2,m}
\nonumber\\
&&\hskip2cm
\times\Big(
A_s^{'\,j_1j_2}
(1^-,2^+,\ldots ,(i_1-1)^+,\hK_{i_1\ldots i_2}^-,(i_2+1)^+,\ldots ,n^+)
\nonumber\\
&&\hskip2cm \null
+ \tilde{\Overlap}^{'\,j_1j_2}_{n-i_2+i_1}
(1^-,2^+,\ldots ,(i_1-1)^+,\hK_{i_1\ldots
  i_2}^-,(i_2+1)^+,\ldots ,n^+)\Big)
\Bigg]
\nonumber\\
&& - \, \At_{n-2}(1^-,3^+,\ldots,(m-1)^+,m^-,(m+1)^+,\ldots,(n-1)^+)
\frac{1}{\Bmp{n2}^2}
\nonumber\\
&& + \, {2\over9} \At_n(1,m)
\,.
\label{eq:complete_unwinding_result}
\end{eqnarray}
Here $A_s^{'\,j_1j_2}$ is as given in \eqn{eq:asdefp} and in a
corresponding treatment $\tilde{\Overlap}^{'\,j_1j_2}$ represents the
terms of $\tilde{\Overlap}$ containing denominator factors of
$s_{j_1\ldots (\hK_{i_1\ldots i_2})\ldots j_2}$, but with those
factors extracted; as in $A_s^{'\,j_1j_2}$ we also extract an overall factor 
of $i$. Furthermore, as mentioned above,
$\tilde{\Overlap}^{'\,j_1j_2}$ does not include terms that stem from
the residues of poles of the overall tree term in $\CuthRat_n$,
because they are precisely cancelled by $\CuthRat_n$.  For this reason,
$\CuthRat_n$ is not included in \eqn{eq:complete_unwinding_result}.
Finally, note that for $i_1=i_2=m$, the `tree amplitude' in
\eqn{eq:complete_unwinding_result} is to be evaluated as
$\At_2(1^-,m^-) = i{\spa1.{m}}^4/(\spa{1}.{m}\spa{m}.{1}) = - i
{\spa1.{m}}^2$.


\subsection{The final result}

After combining all the necessary terms into
\eqn{eq:complete_unwinding_result} we find that the final result for
$\hat{R}_n$ is given by,
\begin{eqnarray}
&&\hspace*{-0.8cm}\hat{R}_n(1,m)
\nonumber\\
&=&
\At_{n}(1,m)
\nonumber\\
&&\times\Bigg\{
\sum_{i_1=2}^{m} \sum_{i_2=m}^{\min(n,n+i_1-5)}
 \Bigg( \sum_{j_1=2}^{i_1} \sum_{j_2=\max(i_2,i_2+2-i_1+j_1)}^{\min(n,n-3+j_1)}
\frac{\Kdef{j_1,j_2,i_1,i_2,m}}{3}
\frac{\spa{(i_1-1)}.{i_1}\spa{i_2}.{(i_2+1)}}{\spa{1}.{i_1}\spa{i_2}.{1}}
\nonumber\\
&&\times\frac{{h^{i_1i_2}_{1;j_1j_2}(n)} {h^{i_1i_2}_{2;j_1j_2}(n)}
  (T^{i_1i_2}_{1;j_1j_2}(n)+T^{i_1i_2}_{2;j_1j_2}(n)
  +s^4_{j_1\ldots(i_1-1),(i_2+1)\ldots j_2}
   (T^{i_1i_2}_{3;j_1j_2}(n)+T^{i_1i_2}_{4;j_1j_2}(n)))}
 {\sandmp{(j_1-1)}.{\Gbdef{j_1,i_1,i_2,j_2}}.1
  \sandmp{1}.{\Gbdefd{j_1,i_1,i_2,j_2}}.{(j_2+1)} }
\nonumber\\&& \null
+T^{i_1i_2}_6(m,n)+T^{i_1i_2}_7(n)\Bigg)
+T_5(m,n)
\Bigg\}\,,
\end{eqnarray}
where we have introduced
\begin{eqnarray}
\Gbdef{a,b,c,d}&=&\Ksl_{a\ldots (b-1),(c+1)\ldots d}
\Ksl_{b\ldots c} \,,
\nonumber\\
\Gbdefd{a,b,c,d}&=&\Ksl_{b\ldots c} \Ksl_{a\ldots (b-1),(c+1)\ldots d} \,,
\end{eqnarray}
in addition to the definitions in \eqn{GdefA}.
This solution then depends on the functions,
\begin{eqnarray}
h^{i_1i_2}_{1;j_1j_2}(n)&=&
\left\{\begin{array}{ll}
j_1 = i_1: &1/s_{(i_2+1)\ldots j_2} \comma
\\
j_1<i_1: &\displaystyle
-\frac{\spa{(j_1-1)}.{j_1}\sandmp{1}.{\Gbdef{}}.1}
{\sandmp{1}.{\Gbdefd{}}.{j_1}\sandmp{(i_1-1)}.{\Gdefd{}}.1}\comma
\end{array}\right.
\end{eqnarray}
\begin{eqnarray}
h^{i_1i_2}_{2;j_1j_2}(n)&=&\left\{\begin{array}{ll}
j_2 = i_2:&1/s_{j_1\ldots(i_1-1)}\comma
\\
j_2>i_2:&\displaystyle \frac{\spa{j_2}.{(j_2+1)}\sandmp{1}.{\Gbdef{}}.1}
{\sandmp{j_2}.{\Gbdef{}}.1\sandmp{1}.{\Gdef{}}.{(i_2+1)}}\comma
\end{array}\right.
\end{eqnarray}
where above and throughout this section $\Gdef{}\equiv
\Gdef{j_1,i_1,i_2,j_2}$, $\Gdefd{}\equiv \Gdefd{j_1,i_1,i_2,j_2}$,
$\Gbdef{}\equiv \Gbdef{j_1,i_1,i_2,j_2}$ and $\Gbdefd{}\equiv
\Gbdefd{j_1,i_1,i_2,j_2}$. The function $T_1$ is given by,
\begin{eqnarray}
T^{i_1i_2}_{1;j_1j_2}(n)=
\sum_{l=i_2+1}^{j_2}
 f^{i_1i_2}_{1;j_1j_2}(l,n)
+ \sum_{l=j_1-1}^{i_1-2}
 f^{i_1i_2}_{1;j_1j_2}(l,n)
\,,
\nonumber\\
\end{eqnarray}
with
\begin{eqnarray}
f^{i_1i_2}_{1;j_1j_2}(l,n)=\left\{\begin{array}{l}
j_1\leq l\leq i_1-2 :
\\ \indentA\displaystyle
\frac{\sandmp{1}.{\Gdef{}\Ksl_{l{}(l+1)}\Ksl_{(l+1)\ldots(i_1-1)}\Gdefd{}}.1\sandmp{1}.{\Gdef{}}.l
\sandmp{1}.{\Gdef{}}.{(l+1)}}{\spa{l}.{(l+1)}} \comma
\\
l = j_1-1\textrm{ and }j_1 = i_1:\;\;0 \comma
\\
l = j_1-1\textrm{ and }j_1<i_1:
\\
\displaystyle
\indentA -s_{j_1\ldots (i_1-1),(i_2+1)\ldots j_2}
 \sandmp{1}.{\Gdef{}\Ksl_{(j_1+1)\ldots(i_1-1),(i_2+1)\ldots j_2}\Ksl_{j_1\ldots(i_1-1)}\Gdefd{}}.1
\\\indentA \indentB\displaystyle\times
 \frac{\sandmp{1}.{\Ksl_{j_1\ldots j_2}\Ksl_{i_1\ldots i_2}}.1
\sandmp{1}.{\Gdef{}}.{j_1}}{\sandmp{1}.{\Gbdefd{}}.{j_1}} \comma
\\
l = j_2 :
\\ {\indentA\displaystyle}
s_{j_1\ldots (i_1-1),(i_2+1)\ldots j_2}
  \sandmp{1}.{\Gdef{}\Ksl_{j_1\ldots(i_1-1),(i_2+1)\ldots(j_2-1)}\Ksl_{(i_2+1)\ldots j_2}\Gdefd{}}.1
\\ \indentA\indentB\displaystyle\times
\frac{\sandmp{1}.{\Gdef{}}.{j_2}
\sandmp{1}.{\Ksl_{j_1\ldots j_2}\Ksl_{i_1\ldots i_2}}.1}
{\sandmp{j_2}.{\Gbdef{}}.1} \comma
\\
i_2+1\leq l < j_2 :
\\ \indentA\displaystyle
-\frac{\sandmp{1}.{\Gdef{}\Ksl_{l{}(l+1)}\Ksl_{(i_2+1)\ldots l}\Gdefd{}}.1
\sandmp{1}.{\Gdef{}}.{l}
\sandmp{1}.{\Gdef{}}.{(l+1)}}{\spa{l}.{(l+1)}} \fstop
\end{array}\right.
\end{eqnarray}
Next, the $T_2$ term reads,
\begin{eqnarray}
&&\hspace*{-0.8cm}T^{i_1i_2}_{2;j_1j_2}(n)=
\sum_{l=i_2+2}^{j_2}
f^{i_1i_2}_{2;j_1j_2}(l,n)
+ \sum_{l=j_1-1}^{i_1-3}
f^{i_1i_2}_{2;j_1j_2}(l,n)
\,,
\end{eqnarray}
\newpage
with
\begin{eqnarray}
&&\hspace*{-0.8cm}f^{i_1i_2}_{2;j_1j_2}(l,n)=
\nonumber\\
&&\hspace*{-0.4cm}\left\{
\begin{array}{l}
 j_1< l\leq (i_1-1) : \\
\indentA\displaystyle  - \sum_{p=l+1}^{i_1-2}
 \frac{\spa{(l-1)}.{l}\sandmp{1}.{\Gdef{}\Ksl_{l\ldots p}\Ksl_{(p+1)\ldots(i_1-1)}\Gdefd{}}.{1}^3}
{\sandmp{1}.{\Gdef{}\Ksl_{(p+1)\ldots(i_1-1)}\Ksl_{l\ldots p}}.{(l-1)}
\sandmp{1}.{\Gdef{}\Ksl_{(p+1)\ldots (i_1-1)}\Ksl_{l\ldots p}}.{l}}
\\
\indentA\indentB\displaystyle\times\frac{\spa{p}.{(p+1)}
\sandmp{1}.{\Gdef{}\Ksl_{l\ldots (i_1-1)}[\Fc(l,p)]^2\Ksl_{(p+1)\ldots (i_1-1)}\Gdefd{}}.{1}}{s_{l\ldots p}
\sandmp{1}.{\Gdef{}\Ksl_{l\ldots (i_1-1)}\Ksl_{l\ldots p}}.{p}
\sandmp{1}.{\Gdef{}\Ksl_{l\ldots (i_1-1)}\Ksl_{l\ldots p}}.{(p+1)}}\comma
\\
l = j_1 :
\\ \indentA\displaystyle - \sum_{p=j_1+1}^{i_1-2}
\frac{\sandmp{1}.{\Gbdefd{}}.{j_1}
\sandmp{1}.{\Gdef{}\Ksl_{j_1\ldots p}\Ksl_{(p+1)\ldots (i_1-1)}\Gdefd{}}.{1}^3}
{\sandmp{1}.{\Gdef{}\Ksl_{(p+1)\ldots (i_1-1)}\Ksl_{j_1\ldots p}\Gbdef{}}.{1}
\sandmp{1}.{\Gdef{}\Ksl_{(p+1)\ldots (i_1-1)}\Ksl_{j_1\ldots p}}.{j_1}}
\\ \indentA\indentB\displaystyle
\times\frac{\spa{p}.{(p+1)}
\sandmp{1}.{\Gdef{}\Ksl_{j_1\ldots (i_1-1)}[\Fc(j_1,p)]^2\Ksl_{(p+1)\ldots (i_1-1)}\Gdefd{}}.{1}}{s_{j_1\ldots p}
\sandmp{1}.{\Gdef{}\Ksl_{j_1\ldots (i_1-1)}\Ksl_{j_1\ldots p}}.{p}\sandmp{1}.{ \Gdef{}\Ksl_{j_1\ldots (i_1-1)}\Ksl_{j_1\ldots p}}.{(p+1)}}\comma
\\
l = j_1-1 \textrm{ and }(j_1\geq i_1-1\textrm{ or }j_2 = i_2):0 \comma
\\
l = j_1-1 \textrm{ and }(j_1<i_1-1\textrm{ and }j_2>i_2):
\\ \indentA\displaystyle - \sum_{p=j_1}^{i_1-2}
\frac{\sandmp{1}.{\Ksl_{j_1\ldots j_2}\Ksl_{i_1\ldots i_2}}.{1}
\sandmp{j_2}.{\Ksl_{j_1\ldots (i_1-1),(i_2+1)\ldots j_2}\Ksl_{i_1\ldots i_2}}.{1}}{\sandmp{1}.{\Gdef{j_1,p+1,j_2,j_2}\Gbdef{p+1,i_1,i_2,j_2}}.{1}}
\\
\indentA\indentB\displaystyle
\times\frac{%
\sandmp{1}.{\Gdef{}\Ksl_{(p+1)\ldots (i_1-1),(i_2+1)\ldots j_2}\Ksl_{(p+1)\ldots (i_1-1)} \Gdefd{}}.{1}^3}
{\sandmp{1}.{\Gdef{} \Ksl_{(p+1)\ldots (i_1-1)}\Uc{i_1i_2}{j_1j_2}(p,j_2+1)}.{j_2}\sandmp{1}.{\Gdef{}\Ksl_{(p+1)\ldots (i_1-1)}\Ksl_{j_1\ldots p} \Gbdef{}}.{1}}
\\
\indentA\indentB\displaystyle
\times\frac{\spa{p}.{(p+1)}\sandmp{1}.{\Gdef{}\Ksl_{(i_2+1)\ldots j_2}[\Ft{i_1i_2}{-;j_1j_2}(j_1,p)]^2\Ksl_{(p+1)\ldots (i_1-1)}\Gdefd{}}.{1}}
{\sandmp{1}.{\Gdef{}\Ksl_{(i_2+1)\ldots j_2}\Uc{i_1i_2}{j_1j_2}(p,j_2+1)}.{p}\sandmp{1}.{\Gdef{}\Ksl_{(i_2+1)\ldots j_2}\Uc{i_1i_2}{j_1j_2}(p,j_2+1)}.{(p+1)}}
\comma
\\
i_2+1\leq l\leq j_2 :
\\ \indentA\displaystyle
\sum_{p=l+1}^{j_2}
\fb^{i_1i_2}_{2;j_1j_2}(l,p,n)
 + \sum_{p=j_1-1}^{i_1-2}
\fb^{i_1i_2}_{2;j_1j_2}(l,p,n) \comma
\end{array}
\right.
\end{eqnarray}
\newpage
and
\begin{eqnarray}
&&\hspace*{-0.8cm}\fb^{i_1i_2}_{2;j_1j_2}(l,p,n)=
\nonumber\\
&&\hspace*{-0.4cm}\left\{
\begin{array}{l}
j_1-1<p\leq(i_1-1) :
\\ \indentA\displaystyle
\frac{\sandmp{1}.{\Ksl_{j_1\ldots j_2}\Ksl_{i_1\ldots i_2}}.{1}}
{\sandmp{1}.{\Gdef{j_1,p+1,l-1,j_2}\Gbdef{p+1,i_1,i_2,l-1}}.{1}}
\\ \indentA\indentB\displaystyle
\times\frac{\spa{(l-1)}.{l}\sandmp{1}.{ \Gdef{}\Ksl_{(p+1)\ldots(i_1-1),(i_2+1)\ldots (l-1)}\Ksl_{(p+1)\ldots (i_1-1)} \Gdefd{}}.{1}^3}{\sandmp{1}.{ \Gdef{}\Ksl_{(p+1)\ldots (i_1-1)}\Uc{i_1i_2}{j_1j_2}(p,l)}.{(l-1)}\sandmp{1}.{ \Gdef{}\Ksl_{(p+1)\ldots (i_1-1)}\Uc{i_1i_2}{j_1j_2}(p,l)}.{l}}
\\ \indentA\indentB\displaystyle
\times\frac{\spa{p}.{(p+1)}\sandmp{1}.{ \Gdef{}\Ksl_{(i_2+1)\ldots (l-1)}[\Ft{i_1i_2}{-;j_1j_2}(l,j_2;j_1,p)]^2\Ksl_{(p+1)\ldots (i_1-1)}\Gdefd{}}.{1}}{\sandmp{1}.{ \Gdef{}\Ksl_{(i_2+1)\ldots (l-1)}\Uc{i_1i_2}{j_1j_2}(p,l)}.{p}\sandmp{1}.{ \Gdef{}\Ksl_{(i_2+1)\ldots (l-1)}\Uc{i_1i_2}{j_1j_2}(p,l)}.{(p+1)}} \comma
\\
p = j_1-1  \textrm{ and }j_1 = i_1: \;\;0 \comma
\\
p = j_1-1  \textrm{ and }j_1<i_1:
\\  \indentA\displaystyle
-\frac{\sandmp{1}.{\Ksl_{j_1\ldots j_2}\Ksl_{i_1\ldots i_2}}.{1}
}{\sandmp{1}.{\Gdef{j_1,j_1,l-1,j_2}\Gbdef{j_1,i_1,i_2,l-1}}.{1}
  \sandmp{1}.{ \Gdef{}\Ksl_{j_1\ldots (i_1-1)}\Uc{i_1i_2}{j_1j_2}(j_1-1,l)}.{(l-1)}}
\\ \indentA\indentB\displaystyle
\times\frac{\spa{(l-1)}.{l}
 \sandmp{1}.{ \Gdef{}\Ksl_{j_1\ldots(i_1-1),(i_2+1)\ldots (l-1)}\Ksl_{j_1\ldots (i_1-1)} \Gdefd{}}.{1}^3}
{\sandmp{1}.{ \Gdef{}\Ksl_{j_1\ldots (i_1-1)}\Uc{i_1i_2}{j_1j_2}(j_1-1,l)}.{l}}
\\  \indentA\indentB\displaystyle
\times\frac{\sandmp{1}.{\Gbdefd{}}.{j_1}\sandmp{1}.{ \Gdef{}\Ksl_{(i_2+1)\ldots (l-1)}[\Fbt{i_1i_2}{-;j_1j_2}(l,j_2)]^2\Ksl_{j_1\ldots (i_1-1)}\Gdefd{}}.{1}}{\sandmp{1}.{ \Gdef{}\Ksl_{(i_2+1)\ldots (l-1)}\Ksl_{l\ldots j_2}\Gbdef{}}.{1}\sandmp{1}.{ \Gdef{}\Ksl_{(i_2+1)\ldots (l-1)}\Uc{i_1i_2}{j_1j_2}(j_1-1,l)}.{j_1}} \comma
\\
(l+1)\leq p <j_2 :
\\  \indentA\displaystyle
\frac{\spa{(l-1)}.{l}\sandmp{1}.{ \Gdef{}\Ksl_{l\ldots p}\Ksl_{(i_2+1)\ldots p} \Gdefd{}}.{1}^3}{\sandmp{1}.{ \Gdef{}\Ksl_{(i_2+1)\ldots p}\Ksl_{l\ldots p}}.{(l-1)}\sandmp{1}.{ \Gdef{}\Ksl_{(i_2+1)\ldots p}\Ksl_{l\ldots p}}.{l}}
\\ \indentA\indentB\displaystyle
\times\frac{\spa{p}.{(p+1)}\sandmp{1}.{ \Gdef{}\Ksl_{(i_2+1)\ldots (l-1)}[\Fc(l,p)]^2\Ksl_{(i_2+1)\ldots p} \Gdefd{}}.{1}}{s_{l\ldots p}\sandmp{1}.{ \Gdef{}\Ksl_{(i_2+1)\ldots (l-1)}\Ksl_{l\ldots p}}.{p}\sandmp{1}.{ \Gdef{}\Ksl_{(i_2+1)\ldots (l-1)}\Ksl_{l\ldots p}}.{(p+1)}} \comma
\\
p = j_2 :
\\  \indentA\displaystyle
\frac{\spa{(l-1)}.{l}\sandmp{1}.{ \Gdef{}\Ksl_{l\ldots j_2}\Ksl_{(i_2+1)\ldots j_2} \Gdefd{}}.{1}^3}{\sandmp{1}.{ \Gdef{}\Ksl_{(i_2+1)\ldots j_2}\Ksl_{l\ldots j_2}}.{(l-1)}\sandmp{1}.{ \Gdef{}\Ksl_{(i_2+1)\ldots j_2}\Ksl_{l\ldots j_2}}.{l}}
\\ \indentA\indentB\displaystyle
\times\frac{\sandmp{j_2}.{\Gbdef{}}.{1}\sandmp{1}.{ \Gdef{}\Ksl_{(i_2+1)\ldots (l-1)}[\Fc(l,j_2)]^2\Ksl_{(i_2+1)\ldots j_2} \Gdefd{}}.{1}}{s_{l\ldots j_2}\sandmp{1}.{ \Gdef{}\Ksl_{(i_2+1)\ldots (l-1)}\Ksl_{l\ldots j_2}}.{j_2}\sandmp{1}.{ \Gdef{}\Ksl_{(i_2+1)\ldots (l-1)}\Ksl_{l\ldots j_2}\Gbdef{}}.{1}} \fstop
\end{array}
\right.
\end{eqnarray}
The $T_3$ term is given by,
\begin{eqnarray}
T^{i_1i_2}_{3;j_1j_2}(n)&=&
\sum_{l=2}^{j_1}
f^{i_1i_2}_{3;j_1j_2}(l,n)
+ \sum_{l=j_2+1}^{n-1}
f^{i_1i_2}_{3;j_1j_2}(l,n)
\,,
\end{eqnarray}
which depends on
\begin{eqnarray}
f^{i_1i_2}_{3;j_1j_2}(l,n)=\left\{
\begin{array}{l}
2\leq l<j_1-1 :
\displaystyle -\frac{\sandmp{1}.{\Ksl_{l(l+1)}\Ksl_{1\ldots l}}.{1}\spa{1}.{l}\spa{1}.{(l+1)}}{\spa{l}.{(l+1)}} \comma
\\
l = j_1-1 :
\displaystyle -\frac{\sandmp{1}.{\Ksl_{(j_1-1)\ldots j_2}\Ksl_{1\ldots (j_1-1)}}.{1}\spa{1}.{(j_1-1)}\sandmp{1}.{\Gbdef{}}.{1}}{\sandmp{(j_1-1)}.{\Gbdef{}}.{1}} \comma
\\
l = j_1 \textrm{ and } j_2 = n:\;\;0 \comma
\\
l = j_1 \textrm{ and } 
j_2<n:
\\ \indentA\displaystyle
\frac{\sandmp{1}.{\Ksl_{j_1\ldots (j_2+1)}\Ksl_{(j_2+1)\ldots n}}.{1}\sandmp{1}.{\Gbdef{}}.{1}\spa{1}.{(j_2+1)}}{\sandmp{1}.{\Gbdefd{}}.{(j_2+1)}} \comma
\\
j_2+1\leq l \leq n-1 :
\displaystyle \frac{\sandmp{1}.{\Ksl_{l(l+1)}\Ksl_{(l+1)\ldots n}}.{1}\spa{1}.{l}\spa{1}.{(l+1)}}{\spa{l}.{(l+1)}} \fstop
\end{array}
\right.
\end{eqnarray}
Following on, the $T_4$ term reads,
\begin{eqnarray}
T^{i_1i_2}_{4;j_1j_2}(n)=
\sum_{l=3}^{j_1}
 f^{i_1i_2}_{4;j_1j_2}(l,n)
 + \sum_{l=j_2+1}^{n-2}
f^{i_1i_2}_{4;j_1j_2}(l,n)
\,,
\end{eqnarray}
with
\begin{eqnarray}
f^{i_1i_2}_{4;j_1j_2}(l,n)=\left\{
\begin{array}{l}
j_2+1 < l\leq (n-2):
\\ \indentA \displaystyle - \sum_{p=l+1}^{n-1}
\frac{\spa{(l-1)}.{l}\sandmp{1}.{\Ksl_{l\ldots p}\Ksl_{(p+1)\ldots n}}.{1}^3}{\sandmp{1}.{\Ksl_{(p+1)\ldots n}\Ksl_{l\ldots p}}.{(l-1)}\sandmp{1}.{\Ksl_{(p+1)\ldots n}\Ksl_{l\ldots p}}.{l}}
\\ \indentA\indentB\displaystyle
\times\frac{\spa{p}.{(p+1)}\sandmp{1}.{\Ksl_{l\ldots n}[\Fc(l,p)]^2\Ksl_{(p+1)\ldots n}}.{1}}
{s_{l\ldots p}\sandmp{1}.{\Ksl_{l\ldots n}\Ksl_{l\ldots p}}.{p}
 \sandmp{1}.{\Ksl_{l\ldots n}\Ksl_{l\ldots p}}.{(p+1)}} \comma
\\
l = j_2+1:
\\ \indentA\displaystyle - \sum_{p=j_2+2}^{n-1}
\frac{\sandmp{1}.{\Gbdefd{}}.{(j_2+1)}
  \sandmp{1}.{\Ksl_{j_2+1\ldots p}\Ksl_{(p+1)\ldots n}}.{1}^3}
     {\sandmp{1}.{\Ksl_{(p+1)\ldots n}\Ksl_{j_2+1\ldots p}\Gbdef{}}.{1}
      \sandmp{1}.{\Ksl_{(p+1)\ldots n}\Ksl_{j_2+1\ldots p}}.{(j_2+1)}}
\\ \indentA\indentB\displaystyle
\times\frac{\spa{p}.{(p+1)}\sandmp{1}.{\Ksl_{j_2+1\ldots n}[\Fc(j_2+1,p)]^2\Ksl_{(p+1)\ldots n}}.{1}}
{s_{j_2+1\ldots p}\sandmp{1}.{\Ksl_{j_2+1\ldots n}\Ksl_{j_2+1\ldots p}}.{p}
\sandmp{1}.{\Ksl_{j_2+1\ldots n}\Ksl_{j_2+1\ldots p}}.{(p+1)}} \comma
\\
l = j_1 \textrm{ and } (j_2\geq n-1 \textrm{ or } j_1 = 2):\;\; 0 \comma
\\
l = j_1 \textrm{ and } (j_2<n-1 \textrm{ and } j_1>2):
\\ \indentA\displaystyle \sum_{p=j_2+1}^{n-1}
\frac{\sandmp{1}.{\Ksl_{j_1\ldots j_2}\Ksl_{i_1\ldots i_2}}.{1}}
{\sandmp{1}.{\Ksl_{j_1\ldots j_2}\Ksl_{(j_2+1)\ldots p}\Gbdef{j_1,i_1,i_2,p}}.{1}}
\\ \indentA\indentB\displaystyle
\times\frac{1}{\sandmp{1}.{ \Ksl_{(p+1)\ldots n} \Uc{i_1i_2}{j_1j_2}(j_1-1,p+1)}.{(j_1-1)}}
\\ \indentA\indentB\displaystyle
\times\frac{\sandmp{(j_1-1)}.{\Gbdef{}}.{1}
  \sandmp{1}.{\Ksl_{j_1\ldots p}\Ksl_{(p+1)\ldots n}}.{1}^3}
  {
 \sandmp{1}.{\Ksl_{(p+1)\ldots n}\Ksl_{(j_2+1)\ldots p}\,\Gbdef{}}.{1}}
\\ \indentA\indentB\displaystyle
\times\frac{\spa{p}.{(p+1)}}
{\sandmp{1}.{\Ksl_{1\ldots (j_1-1)} \Uc{i_1i_2}{j_1j_2}(j_1-1,p)}.{p}}
\\ \indentA\indentB\displaystyle
\times\frac{\sandmp{1}.{\Ksl_{1\ldots (j_1-1)}[\Ft{i_1i_2}{+;j_1j_2}(j_2+1,p)]^2\Ksl_{(p+1)\ldots n}}.{1}}
{\sandmp{1}.{\Ksl_{1\ldots (j_1-1)} \Uc{i_1i_2}{j_1j_2}(j_1-1,p+1)}.{(p+1)}}
 \comma
\\
3\leq l< j_1:
\displaystyle \sum_{p=l+1}^{j_1}
\fb^{i_1i_2}_{4;j_1j_2}(l,p,n)
+ \sum_{p=j_2+1}^{n-1}
\fb^{i_1i_2}_{4;j_1j_2}(l,p,n)  \comma
\end{array}
\right.
\end{eqnarray}
and
\begin{eqnarray}
\fb^{i_1i_2}_{4;j_1j_2}(l,p,n)=\left\{
\begin{array}{l}
j_2+1 \leq p\leq n-1 :
\\ \indentA\displaystyle
 \frac{\sandmp{1}.{\Ksl_{j_1\ldots j_2}\Ksl_{i_1\ldots i_2}}.{1}}
{\sandmp{1}.{\Gbdefd{l,j_1,j_2,p}\Gbdef{l,i_1,i_2,p}}.{1}}
\frac{\spa{(l-1)}.{l}}{\sandmp{1}.{ \Ksl_{(p+1)\ldots n}\Uc{i_1i_2}{j_1j_2}(l-1,p+1)}.{(l-1)}}
\\ \indentA\indentB\displaystyle
\times\frac{\sandmp{1}.{\Ksl_{l\ldots p}\Ksl_{(p+1)\ldots n}}.{1}^3}
{\sandmp{1}.{\Ksl_{(p+1)\ldots n}\Uc{i_1i_2}{j_1j_2}(l-1,p+1)}.{l}}
\\ \indentA\indentB\displaystyle
\times\frac{\spa{p}.{(p+1)}}
{\sandmp{1}.{\Ksl_{1\ldots (l-1)}\Uc{i_1i_2}{j_1j_2}(l-1,p+1)}.{p}}
\\ \indentA\indentB\displaystyle
\times\frac{%
\sandmp{1}.{\Ksl_{1\ldots (l-1)}[\Ft{i_1i_2}{+;j_1j_2}(l,j_1-1;j_2+1,p)]^2\Ksl_{(p+1)\ldots n}}.{1}}
{\sandmp{1}.{\Ksl_{1\ldots (l-1)}\Uc{i_1i_2}{j_1j_2}(l-1,p+1)}.{(p+1)}} \comma
\\
p = j_1 \textrm{ and }j_2 = n :\;\;0 \comma
\\
p = j_1 \textrm{ and }j_2<n :
\\ \indentA\displaystyle
\frac{\sandmp{1}.{\Ksl_{j_1\ldots j_2}\Ksl_{i_1\ldots i_2}}.{1}}{\sandmp{1}.{\Ksl_{j_1\ldots j_2}\Ksl_{l\ldots (j_1-1)}\Gbdef{l,i_1,i_2,j_2}}.{1}}
\\ \indentA\indentB\displaystyle
\times\frac{\spa{(l-1)}.{l}}{\sandmp{1}.{\Ksl_{(j_2+1)\ldots n}\Uc{i_1i_2}{j_1j_2}(l-1,j_2+1)}.{(l-1)}}
\\ \indentA\indentB\displaystyle
\times\frac{\sandmp{1}.{\Ksl_{l\ldots j_2}\Ksl_{(j_2+1)\ldots n}}.{1}^3}{\sandmp{1}.{\Ksl_{(j_2+1)\ldots n} \Uc{i_1i_2}{j_1j_2}(l-1,j_2+1)}.{l}}
\\ \indentA\indentB\displaystyle
\times\frac{%
 \sandmp{1}.{\Ksl_{1\ldots (l-1)}[\Fbt{i_1i_2}{+;j_1j_2}(l,j_1-1)]^2\Ksl_{(j_2+1)\ldots n}}.{1}}
{\sandmp{1}.{\Ksl_{1\ldots (l-1)}\Ksl_{l\ldots (j_1-1)}\Gbdef{}}.{1}}
\\ \indentA\indentB\displaystyle
\times\frac{\sandmp{1}.{\Gbdefd{}}.{(j_2+1)}}
{\sandmp{1}.{\Ksl_{1\ldots (l-1)}\Uc{i_1i_2}{j_1j_2}(l-1,j_2+1)}.{(j_2+1)}}
\comma
\\
l+1\leq p< j_1-1 :
\\ \indentA\displaystyle
\frac{\spa{(l-1)}.{l}\sandmp{1}.{\Ksl_{l\ldots p}\Ksl_{1\ldots p}}.{1}^3}
{\sandmp{1}.{\Ksl_{1\ldots p}\Ksl_{l\ldots p}}.{(l-1)}
 \sandmp{1}.{\Ksl_{1\ldots p}\Ksl_{l\ldots p}}.{l}}
\\ \indentA\indentB\displaystyle
\times\frac{\spa{p}.{(p+1)}
\sandmp{1}.{\Ksl_{1\ldots (l-1)}[\Fc(l,p)]^2\Ksl_{1\ldots p}}.{1}}
{s_{l\ldots p}\sandmp{1}.{\Ksl_{1\ldots (l-1)}\Ksl_{l\ldots p}}.{p}
 \sandmp{1}.{\Ksl_{1\ldots (l-1)}\Ksl_{l\ldots p}}.{(p+1)}}
\comma
\\
p = j_1-1 :
\\ \indentA\displaystyle
 \frac{\spa{(l-1)}.{l}
      \sandmp{1}.{\Ksl_{l\ldots j_1-1}\Ksl_{1\ldots j_1-1}}.{1}^3}
 {s_{l\ldots j_1-1}
  \sandmp{1}.{\Ksl_{1\ldots j_1-1}\Ksl_{l\ldots j_1-1}}.{(l-1)}
  \sandmp{1}.{\Ksl_{1\ldots j_1-1}\Ksl_{l\ldots j_1-1}}.{l}}
\\ \indentA\indentB\displaystyle
\times\frac{\sandmp{(j_1-1)}.{\Gbdef{}}.{1}
   \sandmp{1}.{\Ksl_{1\ldots (l-1)}[\Fc(l,j_1-1)]^2\Ksl_{1\ldots j_1-1}}.{1}}
{\sandmp{1}.{\Ksl_{1\ldots (l-1)}\Ksl_{l\ldots j_1-1}}.{(j_1-1)}
\sandmp{1}.{\Ksl_{1\ldots (l-1)}\Ksl_{l\ldots j_1-1}\Gbdef{}}.{1}}
\fstop
\end{array}
\right.
\end{eqnarray}
Above we have used the following abbreviations,
\begin{eqnarray}
U^{i_1i_2}_{j_1j_2}(p,l)&=&
\Ksl_{(p+1)\ldots(i_1-1),(i_2+1)\ldots (l-1)}
 -\frac{\Ksl_{i_1\ldots i_2}{|1^+\rangle\langle1^-|\Gdef{}}}
       {\sandmp{1}.{\Ksl_{j_1\ldots j_2}\Ksl_{i_1\ldots i_2}}.{1}}
\,,
\end{eqnarray}
as well as
\begin{eqnarray}
\Ft{i_1i_2}{\pm;j_1j_2}(a,b)&=&
\Fc(a,b)
\mp\frac{\Ksl_{j_1\ldots j_2}|1^+\rangle\langle 1^-|\Gbdefd{}}
 {\sandmp{1}.{\Ksl_{j_1\ldots j_2}\Ksl_{i_1\ldots i_2}}.{1}}\Ksl_{a\ldots b}
\comma \nonumber\\
\Fbt{i_1i_2}{\pm;j_1j_2}(a,b)&=&
\Fc(a,b)\pm\Ksl_{a\ldots b}
 \frac{\Gbdef{}|1^+\rangle\langle 1^-|\Ksl_{j_1\ldots j_2}}
{\sandmp{1}.{\Ksl_{j_1\ldots j_2}\Ksl_{i_1\ldots i_2}}.{1}} \comma
\nonumber\\
\Ft{i_1i_2}{\pm;j_1j_2}(a,b;c,d)&=&
\Fbt{i_1i_2}{\pm;j_1j_2}(a,b)
+\Ft{i_1i_2}{\pm;j_1j_2}(c,d)+\Ksl_{a\ldots b}\Ksl_{c\ldots d} \comma
\nonumber\\
\Fc(a,b)&=&\sum_{i=a}^{b-1}\sum_{m=i+1}^b\s k_i\s k_m
\,.
\end{eqnarray}
Next we have,
\begin{eqnarray}
T_5(m,n)=\left\{\begin{array}{ll}
m = 2 \textrm{ or } m = n: & \hskip .3cm \displaystyle \frac{2}{9} \comma
\\
\textrm{otherwise} :&
\displaystyle \vphantom{\Biggl(nn \Biggr)}\frac{2}{9} -\frac{\spa{1}.{2}\spa{2}.{3}\spa{(n-1)}.{n}\spa{n}.{1}}{\spa{2}.{n}^2\spa{1}.{3}\spa{(n-1)}.{1}}
\, .
\end{array}\right.
\end{eqnarray}
The unwinding of \eqn{eq:CR_hat_unshifted} gives the following contribution,
\begin{eqnarray}
&&\hspace*{-0.5cm}T_6^{i_1i_2}(m,n)=
\frac{1}{2} \! \sum_{j_1=1}^{i_1-1}
\sum_{j_2=\max(i_2,i_2+2-i_1+j_1)}^{\min(n,n+j_1-2)}
\sum_{r=2}^3
\frac{\Kdef{j_1+1,j_2,i_1,i_2,m}\spa{(i_1-1)}.{i_1}\spa{i_2}.{(i_2+1)}}
{\spa{1}.{i_1}\spa{i_2}.{1}} \hskip -.2 cm \nn \\
&& \hskip -.4 cm
\times\left(\Hdef^{i_1i_2}_{+;j_1j_2}(r,j_1,j_2+1)
+\Hdef^{i_1i_2}_{-;j_1j_2}(r,j_1+1,j_2+1)- \Hdef^{i_1i_2}_{+;j_1j_2}(r,j_1+1,j_2)-
\Hdef^{i_1i_2}_{-;j_1j_2}(r,j_1,j_2) \right)
\! .
\nonumber\\
\label{eq:CRn_overall_def_a}
\end{eqnarray}
The terms entering this expression are given by,
\begin{eqnarray}
\Hdef^{i_1i_2}_{\pm;j_1j_2}(r,a,b)=
H^{i_1i_2}_{j_1j_2}(r,a,b)\pm
\Hb^{i_1i_2}_{j_1j_2}(r,b,a)
\,,
\end{eqnarray}
where
$\Hdef^{i_1i_2}_{\pm;(i_1-1)j_2}(r,j_1+1,y)
=\Hdef^{i_1i_2}_{\pm;(i_1-1)j_2}(r,i_2,y)$.
We have introduced the following functions,
\begin{eqnarray}
\Hb^{i_1i_2}_{j_1j_2}(3,a,b)&=&
-\frac{1}{3}\frac{\Cc^{i_1i_2}_{j_1j_2}(a,b,a)}
{\sandmp{a}.{\Gbdef{j_1+1,i_1,i_2,j_2}}.{1}^2}\,
\,, \hskip .5 cm
\Hb^{i_1i_2}_{j_1j_2}(2,a,b)=
-\frac{1}{2}
 \frac{\Sc^{i_1i_2}_{j_1j_2}(a,b,a)}{\sandmp{a}.{\Gbdef{j_1+1,i_1,i_2,j_2}}.{1}}
\,, \hskip .4 cm
\nonumber\\
H^{i_1i_2}_{j_1j_2}(3,a,b)&=&
\frac{1}{3}\frac{
\Cc^{i_1i_2}_{j_1j_2}(a,a,b)}{\sandmp{a}.{\Gbdef{j_1+1,i_1,i_2,j_2}}.{1}^2}
\,, \hskip .5 cm
H^{i_1i_2}_{j_1j_2}(2,a,b)=
-\frac{1}{2}
\frac{\Sc^{i_1i_2}_{j_1j_2}(a,a,b)}{\sandmp{a}.{\Gbdef{j_1+1,i_1,i_2,j_2}}.{1}}
\,,~~~~~~
\end{eqnarray}
with the constraints
$H^{i_1i_2}_{j_1j_2}(r,i_2,x)=0$, $\Hb^{i_1i_2}_{j_1j_2}(r,i_2,x)=0$,
$\Hb^{i_1i_2}_{j_1j_2}(r,x,i_2)=0$
and
$H^{i_1i_2}_{j_1j_2}(2,x,i_2)=0$ as well as
$H^{i_1i_2}_{j_1j_2}(r,x,y)=0=\Hb^{i_1i_2}_{j_1j_2}(r,x,y)$ when
$K_{(j_1+1)\ldots (i_1-1),(i_2+1)\ldots j_2}=k_{x}$. We also need
\begin{eqnarray}
\mathcal{C}^{i_1i_2}_{j_1j_2}(j,a,b)&=&
-f^{i_1i_2}_{\mathcal{C};j_1j_2}(b,a,b)f^{i_1i_2}_{\mathcal{D};j_1j_2}(j)
\sandmp{1}.{\Gdef{j_1+1,i_1,i_2,j_2}}.{j}
\label{eg:def_of_cC}\\
&&\times
\left(\sandmp{1}.{\Gdef{j_1+1,i_1,i_2,j_2}}.{j}
+ s_{(j_1+1)\ldots(i_1-1),(i_2+1)\ldots j_2}\spa{1}.{j}\right)\nonumber
\,,
\end{eqnarray}
\begin{eqnarray}
\mathcal{S}^{i_1i_2}_{j_1j_2}(j,a,b)
&=&2f^{i_1i_2}_{\mathcal{D};j_1j_2}(j)\frac{\spa{1}.{a}\spa{1}.{b}
\sandmp{1}.{\Gdef{j_1+1,i_1,i_2, j_2}}.{a}
\sandmp{1}.{\Gdef{j_1+1,i_1,i_2, j_2}}.{b}}
{\spa{a}.{b}^2\sandmp{1}.{\Ksl_{(j_1+1)\ldots j_2} \Ksl_{i_1\ldots i_2}}.{1}}
\label{eg:def_of_cS}\\
&&\times
\sandmp{1}.{\Gdef{j_1+1,i_1,i_2, j_2}}.{j}
\,,
\nonumber
\end{eqnarray}
and
\begin{eqnarray}
f^{i_1i_2}_{\mathcal{C};j_1j_2}(j,a,b)&=&\left\{\begin{array}{ll}
j = i_2: &
\displaystyle -\sandmp{1}.{\Gbdef{j_1+1,i_1,i_2,j_2}}.{1} \comma \\
\textrm{otherwise}: &
\displaystyle \spa{1}.{b}\sandmp{1}.{\Gdef{j_1+1,i_1,i_2,j_2}}.{a}/
\spa{a}.{b} \comma
\end{array}
\right.
\\
f^{i_1i_2}_{\mathcal{D};j_1j_2}(j)&=&\left\{\begin{array}{ll}
j_1 = i_1-1\textrm{ and }j_2 = i_2+1: &
\displaystyle\frac{\sand{1}.{\ksl_{j}}.{(i_2+1)}}{\sandmp{(i_1-1)}.{\Gdefd{j_1+1,i_1,i_2,j_2}}.{1}} \comma
\\
j_1 = i_1-2\textrm{ and }j_2 = i_2: &
\displaystyle -\frac{\sand1.{\ksl_j}.{(i_1-1)}}{\sandmp{1}.{\Gdef{j_1+1,i_1,i_2,j_2}}.{(i_2+1)}} \comma
\\
\textrm{otherwise}: & \frac{s_{(j_1+1)\ldots (i_1-1),(i_2+1)\ldots j_2}
\sandmp1.{\Ksl_{(j_1+1)\ldots j_2} \ksl_j}.{1}}{\sandmp{(i_1-1)}.{\Gdefd{j_1+1,i_1,i_2,j_2}}.{1}\sandmp{1}.{\Gdef{j_1+1,i_1,i_2,j_2}}.{(i_2+1)}}
\,.
\end{array}
\right.
\end{eqnarray}
Finally, the large-$z$ contribution, \eqn{eq:definition_of_InfC},
results in,
\begin{eqnarray}
&&\hspace*{-0.8cm}
T^{i_1i_2}_7(n)=
\label{eq:Bndry_overall_def_a} \\
&& \sum_{j_1=2}^{i_1-1} \sum_{j_2=i_2+1}^{n}
\frac{\spa{(i_1-1)}.{i_1}\spa{i_2}.{(i_2+1)}}
     {\spa{1}.{(i_2+1)}\spa{1}.{i_1}\spa{i_2}.{1}\spa{(i_1-1)}.{1}}
\Bigg(
\frac{\spa{1}.{j_1}^3\spa{1}.{j_2}^3\spb{j_2}.{j_1}^2}{\spa{j_1}.{j_2}^2
\sandpp{j_1}.{\Ksl_{i_1\ldots i_2}}.{1}\sandpp{j_2}.{\Ksl_{i_1\ldots i_2}}.{1}}
\nonumber\\
&&
-R^{i_1i_2}_{+;j_1j_2}(j_1,j_2)+R^{i_1i_2}_{+;j_1j_2}(j_1+1,j_2-1)+R^{i_1i_2}_{-;j_1j_2}(j_1,j_2-1)-R^{i_1i_2}_{-;j_1j_2}(j_1+1,j_2)\Bigg) \,,
\nonumber
\end{eqnarray}
where we set $R^{i_1i_2}_{\pm;j_1j_2}(a,b)=0$ if $K^2_{a\ldots b}=0$
or if $K_{a\ldots b}=K_{i_1\ldots i_2}$; otherwise
$R^{i_1i_2}_{\pm;j_1j_2}(a,b)$ is given by
\begin{eqnarray}
R^{i_1i_2}_{\pm;j_1,j_2}(a,b)&=&
\frac{\spa{1}.{j_1}^2\spa{1}.{j_2}^2}{2
\spa{j_1}.{j_2}^2\sandmp{1}.{\Ksl_{a\ldots b}\Ksl_{i_1\ldots i_2}}.{1}}
\Bigg(\frac{\sandmp{1}.{\Ksl_{a\ldots b} \ksl_{j_1}}.{1}^2}
           {\sandmp{1}.{\ksl_{j_1}\Ksl_{i_1\ldots i_2}}.{1}}
  \pm\frac{\sandmp{1}.{\Ksl_{a\ldots b} \ksl_{j_2}}.{1}^2}
          {\sandmp{1}.{\ksl_{j_2}\Ksl_{i_1\ldots i_2}}.{1}}
\nonumber\\
&&\null
\hskip 3 cm +\sandmp{1}.{\Ksl_{a\ldots b} \ksl_{j_1}}.{1}
   \pm\sandmp{1}.{\Ksl_{a\ldots b}\ksl_{j_2}}.{1}\Bigg) \,.
\label{Rcoeffaformula}
\end{eqnarray}


\section{Conclusions}
\label{ConclusionSection}

At the LHC, events involving large numbers of jets will play a central
role in investigations and measurements of new physics.  On the
theoretical side, a proper understanding of such events will require
NLO calculations.  These calculations in turn require one-loop amplitudes with
large numbers of hard colored final-state particles.

In this paper we have provided new and non-trivial examples of
one-loop QCD amplitudes with an arbitrary number of external gluons.  We
have computed these amplitudes using the on-shell bootstrap
method~\cite{Bootstrap,Genhel}.  Previously, $n$-gluon amplitudes with
special helicity configurations, with two or three color-adjacent
negative-helicity legs, have been computed~\cite{FordeKosower,Genhel}.
In the computation of three color-adjacent negative-helicity legs,
several theoretical issues mentioned in the introduction ---
non-standard channels and large shift-parameter behavior --- arose and
were resolved in a general manner.  The calculation presented here is
in these respects simpler, in that our choice of shift eliminates
these issues.  Here we have extended the results to cover all one-loop
corrections to the celebrated MHV amplitudes of Parke and
Taylor~\cite{ParkeTaylor}.  These are amplitudes with the two negative
helicities in arbitrary positions in the color ordering.  These
amplitudes are most conveniently expressed in a supersymmetric
decomposition~\cite{GGGGG,Neq1Oneloop}, for which many of the
ingredients had been computed previously in
refs.~\cite{Neq4Oneloop,Neq1Oneloop,BBSTQCD}.  The ensemble of terms,
including the rational parts of the scalar-loop contributions given
here, provide an expression for this class of amplitudes that is
compact when compared to expectations based on a brute-force
diagrammatic calculation.

These results will be of direct use in studies of multi-jet physics at the
LHC.  They also allow us to confirm the relatively mild increase in
complexity of our methods as the number of external legs increases.  For
many years, it has been widely believed that the rapid growth in the
complexity of gauge theory calculations is an intrinsic part of the
perturbative expansion.  Our construction illustrates in a non-trivial
context that this is not so for the amplitudes presented here.

In constructing the amplitudes we made use of a number of
empirically-observed properties.  In the calculations described in
this paper, we have confirmed the validity of these assumptions using
the stringent requirements of proper symmetries and factorization in
all channels under real momenta.  For six gluons, we also confirmed
agreement with the numerical results of Ellis, Giele and
Zanderighi~\cite{EGZ06} and also with the analytic results of
Xiao, Yang and Zhu~\cite{XYZ6}.

Although analytic results for the rational parts of the two remaining,
$\NeqZero$ non-MHV six-gluon amplitudes have been obtained
recently~\cite{XYZ6}, it is still of interest to use the recursive
bootstrap to construct them from the known cut-containing
parts~\cite{BFM}.  One reason to do so is in order to rearrange the
rational parts, so that their spurious singularities are more
manifestly cancelled against those of the cut parts. A second reason
would be to see whether a more compact form can be obtained, in the
interest of faster numerical evaluation of the amplitude.

We may contrast the speed of numerical evaluation using our
expressions with that of the semi-numerical approach of
refs.~\cite{GieleGloverNumerical,EGZ}. The semi-numerical computation
of the complete six-gluon amplitude takes 9 seconds to evaluate one
$\NeqZero$ helicity configuration on a 2.8 GHz Pentium
processor~\cite{EGZ06}.  There are 64 helicity configurations in
total.  (Subsequent helicity configurations will not take as long as 9
seconds, though.)  We have implemented the $\NeqFour$, $\NeqOne$ and
$\NeqZero$ components of the six-gluon amplitude in C++, except for
the $\NeqZero$ components of the non-MHV helicity configurations ---
$({-}{+}{-}{+}{-}{+})$, $({-}{-}{+}{-}{+}{+})$ and permutations (14 in
all, out of 64).  The terms we have implemented so far require 30
milliseconds in total to evaluate to 9 significant digits on a 2 GHz
Xeon processor, for all 64 helicity configurations, which is
considerably faster than the semi-numerical approach.

Besides the issue of speed, there is also the question of
numerical instability due to round-off error,
near physical and unphysical kinematic singularities.  
Singularities that cancel within the $\Ll_i$
and $\Ls_i$ functions can easily be patched, using Taylor
expansions, in the numerical implementation of those functions.
Experience with NLO programs for multi-jet production indicates
that the relatively mild spurious singularities remaining in 
our amplitudes will not cause significant numerical difficulties.
As described in ref.~\cite{CGM}, 
the sizes of regions of numerical instability depend 
heavily on the powers to which singular denominator factors are raised.
To assess the potential for numerical instabilities arising 
near the remaining spurious singularities, we consider 
the examples of $e^+e^-\rightarrow 4$ jets and $pp \rightarrow W,Z+2$ 
jets.  The relevant one-loop 
amplitudes, in the form presented in ref.~\cite{ZFourPartons}, 
have similarly mild spurious singularities, having been obtained 
with an early version of the on-shell bootstrap.
In ref.~\cite{ZFourPartonsNLODS}, these amplitudes were implemented 
in an NLO program for $e^+e^-$ annihilation into four jets.  No numerical
difficulties were encountered because of the tiny size of the unstable
regions.  Similarly, no numerical difficulties have 
arisen~\cite{CampbellPrivate} in the 
implementation of these amplitudes in the more general 
crossed kinematics arising for $pp \rightarrow W,Z+2$ jet 
production~\cite{MCFM}.
These results suggest that jet programs using the amplitudes in this paper
will also be free of significant complications arising from round-off
error.

There are a number of open issues that would be important to address.
For example, it would be useful to have a first-principles derivation
of the complex factorization properties, as well as of the behavior of
loop amplitudes at large values of the shift parameter.  In this
regard, recent papers~\cite{VamanYao} linking tree-level on-shell
recursion with gauge-theory Lagrangians in particular gauges may prove
useful.  The unitarity method with $D$-dimensional
cuts~\cite{BernMorgan} may also assist in the formal understanding of
properties of gauge-theory amplitudes.
To apply our techniques to non-MHV amplitudes, one must find shifts
that avoid non-standard factorization channels~\cite{Genhel}.
Also, one should construct a satisfactory cut-completion, 
free of spurious singularities in the shift parameter.  This construction
is more intricate than for the MHV amplitudes presented here.

It would also be very important to apply the on-shell bootstrap
to processes involving external vector bosons and quarks.
(For it to be applicable to cases with massive quarks, one would need
to first extend the methods to allow massive particles in the loop.)
Such processes are of crucial importance
for understanding backgrounds to new physics in supersymmetric
and other extensions of the Standard Model.

With a set of one-loop multi-parton
matrix elements in hand, one can proceed to construct a numerical program for
NLO differential cross sections.  Although the construction of such programs is
non-trivial, it is well-understood, and very general formalisms are
known~\cite{NLOAlgorithms}.

Another interesting open problem concerns the twistor-space
properties~\cite{WittenTopologicalString} of the loop amplitudes.
There have already been some studies of these
properties~\cite{TwistorLoop,BenaLoopTwistor,HolomorphicAnomaly}. In
particular, the coefficients of box integrals have a surprisingly
simple twistor-space structure, exhibiting delta-function support on
intersecting lines, as described in some detail for the case of
$\NeqFour$ super-Yang-Mills
theory~\cite{NeqFourSevenPoint,BCFUnitarity,NeqFourNMHV,BCFCoplanarity}.
It would be interesting to map out the twistor-space properties of the
complete amplitudes, especially in non-supersymmetric theories, using
the all-$n$ expressions obtained here and
elsewhere~\cite{AllPlus,Mahlon,Neq1Oneloop,BBSTQCD,Qpap,Bootstrap,Genhel}.
Using the twistor-space structure as a guide, it may, for example, be
possible to construct a set of loop-level MHV vertices, incorporating
both rational and cut terms, in analogy with the tree-level
construction of amplitudes with generic helicities using MHV
vertices~\cite{CSW}.

It might also be possible to use the one-loop MHV amplitudes to obtain
insight into the size of NLO corrections to hadron collider processes
with a large number of jets.  At tree level, one of the early
applications of the MHV (Parke-Taylor)
amplitudes~\cite{ParkeTaylor,BGSix,MPX} was as the basis of schemes
for estimating multi-jet rates at hadron colliders, in advance of the
availability of exact matrix elements for all helicity configurations.
One approach was to simply assume that the non-MHV amplitudes were the
same as the MHV amplitudes, and multiply the MHV terms in the cross
section by a simple combinatoric factor~\cite{KunsztStirling}.
Subsequently, a procedure known as `infrared reduction', designed to
match the known collinear behavior, was applied to the case
of four-jet production, and gave results quite similar to the exact
matrix elements~\cite{Maxwell}.  It would be interesting to see
whether similar procedures could be applied sensibly to NLO
computations as well, making use of the MHV loop amplitudes reported
here, as well as similar ones containing quarks, should they become
available.

We anticipate that the on-shell unitarity-bootstrap approach will have
widespread applications to computing the higher-multiplicity
amplitudes required for next-to-leading order computations of
phenomenological interest at the Large Hadron Collider.

\section*{Acknowledgments}

We thank Xun Su, Zhi-Guang Xiao, Gang Yang and Chuan-Jie Zhu for
sending us their results prior to publication~\cite{XYZgen5,XYZ6}, and
for discussions and assistance in comparing our six-point results with
theirs.  L.D. thanks the Aspen Center for Physics, where part of this
work was performed, for hospitality.  We thank Academic Technology
Services at UCLA for computer support.  The figures were generated
using Jaxodraw~\cite{Jaxo}, based on Axodraw~\cite{Axo}.

\appendix


\section{Completed-Cut Terms of $\NeqZero$ MHV Amplitudes}
\label{CutAppendix}

In this appendix we collect the cut parts for the $\NeqZero$ one-loop
amplitudes with two negative-helicity gluons, which were obtained
in ref.~\cite{BBSTQCD}, and complete them in a convenient way.  We
also give the rational parts of these completed-cut terms in
a form convenient for computing the overlap terms.
Converting the expressions in ref.~\cite{BBSTQCD} to a notation
similar to that used for the $\NeqOne$ amplitudes in
ref.~\cite{Neq1Oneloop}, and adding suitable rational terms
via $\Ll_i$ functions, we have the completed-cut parts of
the MHV amplitude,
%
\ba
&& \hskip -.4 cm
\Cuth^{\NeqZero}_n(1,m) =
  \nn \\
&&  \hskip .7 cm \null
  \At_n(1,m)   \nn\\
&& \hskip 1.1cm \null
 \times\Biggl\{ \sum_{j_1=2}^{m-1} \sum_{j_2=m+1}^n
   - {1\over2} [b^m_{j_1, j_2}]^2
\Mz \Bigl( s_{(j_1+1) \ldots j_2}, s_{j_1\ldots(j_2 -1) };
     s_{(j_1+1)\ldots (j_2 -1)}, s_{(j_2+1)\ldots (j_1-1)} \Bigr) \nn\\
&& \null \hskip 1.5 cm
+ \sum_{ 2 \le j < m }\;
  \sum_{ a \in \SelectSetb}
 \Biggl[
   d^m_{j,a}{ \Ll_2 \biggl( { -s_{(j+1) \ldots a}
                   \over - s_{j\ldots a}} \biggr)   \over  s_{j\ldots a}^3}
    + e^m_{j,a}{\Lnl_1 \biggl({ -s_{(j+1) \ldots a}
                  \over - s_{j\ldots a}} \biggr) \over s_{j\ldots a}^2}
          \nn  \\
&& \hskip 4. cm \null
    + \biggl({1\over 6}\, c^m_{j,a} +  f^m_{j,a} \biggl)
    { \Ll_0 \biggl( {-s_{(j+1)\ldots a} \over  - s_{j \ldots a}}
                              \biggr) \over s_{j\ldots a}}
   - {1\over 4} (b^m_{j,a} - b^m_{j,a+1})
             \ln\biggl( {-s_{(j+1)\ldots a} \over -s_{j\ldots a} } \biggr)
\Biggr] \nn \\
&&  \null \hskip 1.5 cm
+ \sum_{m < j \leq n} \;
  \sum_{a \in \SelectSeta}
\Biggl[  d^m_{j,a}
  { \Ll_2 \biggl({-s_{(a+1)\ldots j} \over -s_{(a+1)\ldots (j-1)}} \biggr)
                   \over s_{(a+1)\ldots(j-1)}^3}
  +e^m_{j,a} \,
  { \Lnl_1 \biggl({-s_{(a+1)\ldots j} \over -s_{(a+1)\ldots (j-1)}} \biggr)
                   \over s_{(a+1)\ldots(j-1)}^2}    \nn  \\
&& \null \hskip 4. cm
+ \biggl({1\over 6}\, c^m_{j,a} +  f^m_{j,a} \biggl)
            { \Ll_0 \biggl({- s_{(a+1)\ldots j} \over
                                       - s_{(a+1)\ldots(j-1)}} \biggr)
                        \over s_{(a+1)\ldots (j-1)} }
    \nn  \\
&& \hskip 4. cm \null
  - {1\over 4} (b^m_{j,a} - b^m_{j,a+1})
             \ln\biggl( {-s_{(a+1)\ldots j} \over -s_{(a+1)\ldots (j-1)} } \biggr)
 \Biggr]\nn \\
&& \null \hskip 1.5 cm
+ \Biggl( {1\over 6} {c^m_{2,n}\over s_{12} }
      +  {f^m_{2,n} \over s_{12} }
      + {1\over 4} b^m_{2,n} \Biggr) \Kz(s_{12})
+ \Biggl( {1\over 6} {c^m_{n,1} \over s_{n1}}
      +   {f^m_{n,1} \over s_{n1} }
       + {1\over 4} b^m_{n,2} \Biggr) \Kz(s_{n1}) \nn \\
&& \null \hskip 1.5 cm
+ \Biggl( {1\over 6} {c^m_{m+1,m-1} \over s_{m(m+1)}}
       +  {f^m_{m+1,m-1} \over s_{m(m+1)} }
        + {1\over 4} b^m_{m+1,m-1}
            \Biggr) \Kz(s_{m(m+1)})\nn \\
&& \null \hskip 1.5 cm
+ \Biggl( {1\over 6} {c^m_{m-1,m} \over s_{(m-1)m}}
       +  {f^m_{m-1,m} \over s_{(m-1)m}}
         + {1\over 4} b^m_{m-1,m+1}
        \Biggr) \Kz(s_{(m-1)m} ) \Biggr\} \,.
\label{BBST}
\ea
In performing the cut completion, we
have rearranged the cuts somewhat before introducing the $\Ll_i$
functions, which automatically remove the spurious singularities.
The coefficients appearing in the cuts are,
%
\ba
b^m_{j_1, j_2} & = &
2 \, { \spa{1}.{j_1} \spa{1}.{j_2} \spa{m}.{j_1} \spa{m}.{j_2}
   \over {\spa{1}.{m}}^2 {\spa{j_1}.{j_2}}^2 }
\, , \\
%
c^m_{j,a} &=&
{\spa{m}.{j} \sandpp{j}.{\Ksl_{j\ldots a}}.1
 -\sandmm{m}.{\Ksl_{j\ldots a}}.j \spa{j}.1
    \over \spa1.{m}^2 }
             \spa{j}.1\spa{m}.j\,
      {{\spa{a}.{(a+1)}\over \spa{a}.{j}\spa{j}.{(a+1)}}}
\,, \\
%
%
d^m_{j,a} &=&
 -{1\over 3} c^m_{j,a}
\, { \spa{1}.{j} \spa{m}.{j} \spab{1}.{\Ksl_{(j+1)\ldots a}}.{j}
    \spab{m}.{\Ksl_{(j+1)\ldots a}}.{{j}} \over \spa{1}.{m}^2 }
  \,, \\
e^m_{j,a} &=&
   - {1\over 2}\, {\spa1.{j} \spa{m}.{j} \over  \spa1.{m}^2} \,
               \sand1.{\Ksl_{(j+1)\ldots a}}.{j}
               \sand{m}.{\Ksl_{(j+1)\ldots a}}.{j}
               ( b^m_{j,a} - b^m_{j,a+1})
 \,,  \hskip 1 cm \\
%
%
f^m_{j,a} &=&
 \Biggl({\spa1.{j} \spa{m}.{j} \over \spa1.{m}^2} \Biggr)^2
    \Biggl[ {\spa1.{a} \spa{m}.{a} \over \spa{a}.{j}^3}
      \Bigl( \spa1.{a} \sand{m}.{\Ksl_{(j+1)\ldots a}}.{j} +
                    \spa{m}.{a} \sand{1}.{\Ksl_{(j+1)\ldots a}}.{j} \Bigr)
       \nn \\
&& \null \hskip 1 cm
          - {\spa{1}.{(a+1)} \spa{m}.{(a+1)} \over \spa{(a+1)}.{j}^3 }
      \Bigl( \spa1.{(a+1)} \sand{m}.{\Ksl_{(j+1)\ldots a}}.{j}
      \nn \\
&& \null \hskip 5.4cm
           + \spa{m}.{(a+1)} \sand1.{\Ksl_{(j+1)\ldots a}}.j \Bigr) \Biggr]
 \,. \hskip 1 cm
\ea
The quantities $b^m_{j_1, j_2}$ and $c^m_{j,a}$ also
appear in the expression for the $\NeqOne$ supersymmetric
amplitude~\cite{Neq1Oneloop}.

\begin{figure}[t]
\centerline{\epsfxsize 3.5 truein\epsfbox{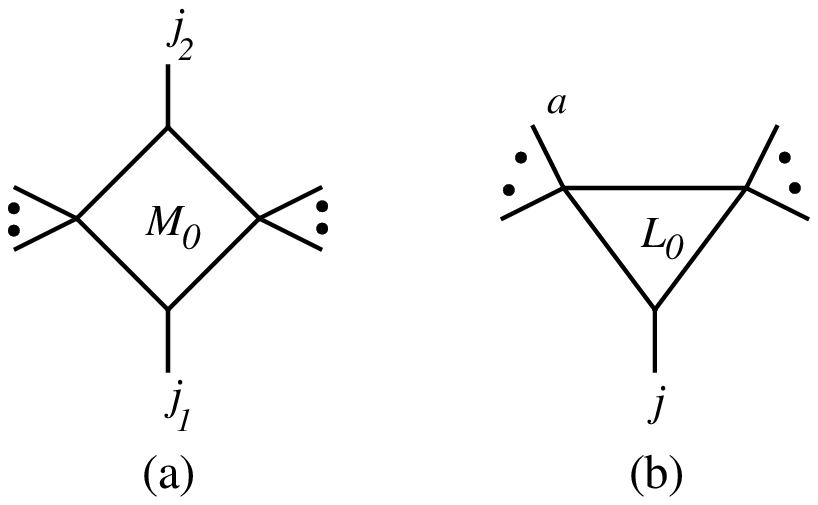}} \caption{
Kinematic configuration of the box and triangle functions
appearing in the completed-cut contribution given in \eqn{BBST}.  }
\label{triboxFigure}
\end{figure}

The sums over the $\Ll_i$ triangle functions run over the ranges
corresponding to all configurations in \fig{triboxFigure}(b) with
two massive corners, each containing a negative-helicity leg,
\ba
\SelectSeta &=& \left\{
\begin{array}{ll}
 \{1,2, \ldots ,m-2 \}, \hskip .5cm   j =m+1 \,,
\\
 \{1,2, \ldots ,m-1 \} , \hskip .5cm  m+1 <j < n \,,
\\
 \{2, \ldots ,m-1 \} , \hskip .8cm j =n \, ,
\end{array}  \right.
\label{CSum}  \\
[9pt]
\SelectSetb &=& \left\{
\begin{array}{ll}
 \{m, m+1, \ldots, n-1 \} , \hskip .4cm  j = 2 \,, \\
 \{m, m+1, \ldots, n \} , \hskip 1.0 cm 2 <j < m-1 \,, \\
 \{m+1, m+2, \ldots, n \} , \hskip .4cm j = m-1\, .
\end{array} \right.
\label{hatCSum}
\ea
For $\S_j$ when $m=n-1$ and $j=n$, and for $\hat\S_j$ when $m=3$ and
$j=2$, the kinematics degenerates and the intersection of the first
and third conditions in \eqns{CSum}{hatCSum} should be used.  The
$\Kz$ functions correspond to cases where the two-external-mass
triangles degenerate to bubbles.  For adjacent negative helicities,
there are no boxes, and the remaining double sum collapses to a
single sum, leaving the simpler expression given in
ref.~\cite{Neq1Oneloop}.

The box function is defined by
\def\Psq{P^2}
\def\Qsq{Q^2}
\ba
\Mz ( s_1, s_2 ; P^2, Q^2   ) & =&
\Li_2 \Bigl( 1 -{\Psq \Qsq \over s_1 s_2 } \Bigr)
-\Li_2 \Bigl( 1 -{\Psq\over s_1} \Bigr)
-\Li_2 \Bigl( 1 -{\Psq\over s_2} \Bigr)
-\Li_2 \Bigl( 1 -{\Qsq\over s_1 } \Bigr) \hskip .7 cm
\nn \\
&& \hskip 2 cm
 -\Li_2 \Bigl( 1 -{\Qsq \over s_2} \Bigr)
- {1\over2} \ln^2 \Bigl( {s_1\over s_2} \Bigr)\, .
\ea
The box function is equal to
minus the two-mass easy box function of ref.~\cite{ZFourPartons},
$\Mz(s,t,P^2,Q^2) = -\Ls_{-1}^{{\rm 2m}e}(s,t,P^2,Q^2)$,
corresponding to a $D=6$ box integral.
An alternative expression for this box integral may be found
in ref.~\cite{BST}.

The $\Ll_i$ functions~\cite{GGGGG} are given in
\eqn{Lsdef}. In the above we have replaced
$\Ll_1$ with a somewhat different function, modified so
as to better respect the symmetry properties of the amplitude,
\be
\Lnl_1(r) =
{\ln(r)-(r-1/r)/2\over (1-r)^2} \,.
\label{L1hat}
\ee
In addition, we have the bubble function,
\be
\Kz(s) =
{1 \over \epsilon\,{(1-2\,\epsilon)}}(-s/\mu^2)^{-\epsilon} \; =
\; \Big(\!\!-\ln(-s/\mu^2)\;+\;2\;+\;{1\over \epsilon}\Big)
\;+\; {\cal{O}}(\epsilon)\,.
\ee

By using the $\Lnl_1$ and $\Ll_2$ functions in the completed-cut 
expression~(\ref{BBST}) instead of logarithms,
we eliminate all of the spurious poles in $\Cuth^{\NeqZero}_n$
that develop $z$ dependence under the $\Shift1{m}$ shift.  
These poles arise from differences of four-momentum invariants,
for example $(s_{(j+1)\ldots a} - s_{j\ldots a})^3$
in the denominator of $\Ll_2((-s_{(j+1)\ldots a})/(- s_{j\ldots a}))$.
These invariants contain leg $m$ but not leg 1, so they are 
each shifted, leading to a $z$-dependent denominator.

Note that there are a host of other spurious poles in \eqn{BBST} that
do {\it not} develop dependence on $z$ under the $\Shift{1}{m}$ shift.
We do not need to worry about cancelling their spurious behavior;
it will happen automatically using the rational terms we construct.
We discussed this issue for the five-gluon amplitude in \sect{5ptSection}.
In the $n$-gluon case it is a bit more intricate.
The denominator factors of $\spa{1}.{m}$ are cancelled by a numerator factor
of ${\spa{1}.{m}}^4$ in $A_n^\tree(1,m)$, but in any case $\spa{1}.{m}$ is
left invariant by the $\Shift{1}{m}$ shift, so no $z$-dependent pole is
induced.  Now, the denominator factors of 
$\spa{j_1}.{j_2}$ in $b^m_{j_1,j_2}$ are not entirely
cancelled by numerator factors (the function $\Mz$ does cancel some 
of their singular behavior); but again, because 
neither $j_1$ nor $j_2$ can equal $1$ or $m$, these factors are 
left invariant by the $\Shift{1}{m}$ shift.
Finally, there are denominator factors of $\spa{a}.{j}$ and 
$\spa{(a+1)}.{j}$ in $c^m_{j,a}$, $e^m_{j,a}$ and $f^m_{j,a}$.
While $j$ cannot be equal to $1$ or $m$, $a$ or $(a+1)$ can be.
However, the relevant terms are protected from $z$-dependent spurious
poles by numerator factors of $\spa1.{j}$, $\spa{m}.{j}$, $\spa{m}.{a}$ or
$\spa{m}.{(a+1)}$.

Since the entire amplitude must be free of spurious poles, we can now
conclude that the rational remainder we wish to compute is free of 
$z$-dependent spurious poles.

\def\indentA{\hskip 7mm}
\def\indentB{\hskip 7mm}

The rational parts of the completed-cut terms are obtained
by setting all logarithms, polylogarithms and $\pi^2$ terms
to zero in \eqn{BBST},
\begin{equation}
\CuthRat_n(1,m) = \Cuth^{\NeqZero}_n(1,m)
  \Bigg|_{ \ln, \Li_2, \pi^2 \rightarrow  0}
\,. \label{CRAllndef}
\end{equation}
After some rearrangement, this expression can be written as,
\begin{eqnarray}
&&\CuthRat_n(1,m)
\label{eq:CR_hat_unshifted}
=  \At_n(1,m)  \Biggl[ {1 \over 3} \biggl( {1\over \e} +2 \biggr)  \\
&& \hskip 1 cm \null
+ \frac{1}{2}
\sum_{j_1=1}^{m-1}
\sum_{j_2=\max(m,j_1+2)}^{\min(n,j_1+n-2)}
\sum_{r=2}^3
\frac{1}{s_{(j_1+1) \ldots j_2}}
\nonumber\\
&&\hskip 4cm \times
\bigg(\Nccoeff_{+;j_1j_2}(r,j_1,j_2+1)+\Nccoeff_{-;j_1j_2}(r,j_1+1,j_2+1)
\nonumber\\
&&\hskip 4.25cm \null
-\Nccoeff_{+;j_1j_2}(r,j_1+1,j_2)-\Nccoeff_{-;j_1j_2}(r,j_1,j_2)\bigg)
\nonumber
\Biggr]
\, ,
\end{eqnarray}
where
\begin{eqnarray}
\mathcal{N}_{\pm;j_1j_2}(r,a,b)=
{N}_{r;j_1j_2}(a,b,m,1)\pm
{N}_{r;j_1j_2}(b,a,1,m) \, ,
\end{eqnarray}
with
\begin{eqnarray}
{N}_{2;j_1j_2}(a,b,m_1,m_2)&=&
\frac{S_{j_1 j_2}(a,b,m_1,m_2)}{\sand{a}.{\Ksl_{(j_1+1)\ldots j_2}}.{a}},\;\;
\nonumber\\
{N}_{3;j_1j_2}(a,b,m_1,m_2)&=&
\frac{C_{j_1 j_2}(a,b,m_1,m_2)}{\sand{a}.{\Ksl_{(j_1+1)\ldots j_2}}.{a}^2}.
\label{N2N3definition}
\end{eqnarray}
%
These quantities in turn depend upon,
\begin{eqnarray}
&&\hspace*{-0.8cm}C_{j_1j_2}(a,b,m_1,m_2)=
\nonumber\\
&&\indentA -\spa{m_1}.{a}\spa{m_2}.{b}
\sandmp{m_2}.{\ksl_{a}\Ksl_{(j_1+1)\ldots j_2}}.{m_1}
\nonumber\\
&&\indentA\indentB\times
\frac{\sandmp{m_2}.{\Ksl_{(j_1+1)\ldots j_2}\ksl_{a}}.{m_1}
\sandmp{m_2}.{[\ksl_{a}\Ksl_{(j_1+1)\ldots j_2}
              -\Ksl_{(j_1+1)\ldots j_2}\ksl_{a}]}.{m_1}}
{3\spa{m_1}.{m_2}^4\spa{a}.{b}},
\nonumber\\
&&\hspace*{-0.8cm}S_{j_1 j_2}(a,b,m_1,m_2) =
\nonumber\\ &&\indentA \spa{m_1}.{a}\spa{m_2}.{b}\spa{m_2}.{a}\spa{m_1}.{b}
\nonumber\\ &&\indentA\indentB\times
 \frac{\sandmp{m_2}.{\Ksl_{(j_1+1)\ldots j_2}\ksl_{a}}.{m_1}
\sandmp{m_1}.{\Ksl_{(j_1+1)\ldots j_2}\ksl_{a}}.{m_2}}
{\spa{m_1}.{m_2}^4\spa{a}.{b}^2}\, .
\label{SCdefinition}
\end{eqnarray}

Applying the $\Shift{1}{m}$ shift to $\Cuth_n^{\NeqZero}(1,m)$
in \eqn{BBST} and taking the $z\to\infty$ limit,
we can extract the value of $\Inf\Cuth_n$ required by
the basic formula~(\ref{BasicFormula}).  The result is,
\begin{eqnarray}
&&\hspace*{-0.8cm}
\label{eq:definition_of_InfC}
\Inf\Cuth_n(1,m) =
\At_n(1,m)
\\
&&
\times \sum_{j_1=2}^{m-1} \sum_{j_2=m+1}^{n}
\frac{\spa{(m-1)}.{m}\spa{m}.{(m+1)}}
     {\spa{(m-1)}.{1} {\spa{1}.{m}}^2 \spa{1}.{(m+1)}}
\Bigg(
\frac{\spa{1}.{j_1}^3\spa{1}.{j_2}^3\spb{j_2}.{j_1}^2}{\spa{j_1}.{j_2}^2
\sandpp{j_1}.{\ksl_m}.{1}\sandpp{j_2}.{\ksl_m}.{1}}
\nonumber\\
&& \null
-R^{mm}_{+;j_1j_2}(j_1,j_2)+R^{mm}_{+;j_1j_2}(j_1+1,j_2-1)
+R^{mm}_{-;j_1j_2}(j_1,j_2-1)-R^{mm}_{-;j_1j_2}(j_1+1,j_2)\Bigg) \,,
\nonumber
\end{eqnarray}
where we set $R^{mm}_{\pm;j_1j_2}(a,b)=0$ if $K^2_{a\ldots b}=0$; otherwise
$R^{mm}_{\pm;j_1j_2}(a,b)$ is,
\begin{eqnarray}
R^{mm}_{\pm;j_1,j_2}(a,b)&=&
\frac{\spa{1}.{j_1}^2\spa{1}.{j_2}^2}{2
\spa{j_1}.{j_2}^2\sandmp{1}.{\Ksl_{a\ldots b}\ksl_m}.{1}}
\Bigg(\frac{\sandmp{1}.{\Ksl_{a\ldots b} \ksl_{j_1}}.{1}^2}
           {\sandmp{1}.{\ksl_{j_1}\ksl_m}.{1}}
  \pm\frac{\sandmp{1}.{\Ksl_{a\ldots b} \ksl_{j_2}}.{1}^2}
          {\sandmp{1}.{\ksl_{j_2}\ksl_m}.{1}}
\nonumber\\
&&\null
\hskip 3 cm +\sandmp{1}.{\Ksl_{a\ldots b} \ksl_{j_1}}.{1}
   \pm\sandmp{1}.{\Ksl_{a\ldots b}\ksl_{j_2}}.{1}\Bigg) \,.
\label{Rcoeffammformula}
\end{eqnarray}
Up to a factor of $-A_n^\tree(1,m)$, this expression can
also be obtained from formula~(\ref{eq:Bndry_overall_def_a})
for $T_7^{i_1i_2}(n)$ by setting $i_1=i_2=m$.


\section{Rational Parts of Six-Gluon $\NeqZero$ MHV Amplitudes}
\label{SixpointAppendix}

There are three independent six-gluon MHV amplitudes. All others can
be obtained from these by cyclic permutations of the
external legs. The analytic form of the $\NeqZero$ amplitude with adjacent
negative helicities, $A_{6;1}^{\NeqZero}(1,2)$, may be found in
ref.~\cite{Bootstrap}. In this appendix we present
analytic forms for the rational remainders
of the other two independent MHV amplitudes.
Although the structure is rather intricate,
such forms can be useful for future phenomenological studies
of four-jet events.  Given their mild spurious singularities
(compared to more direct evaluations of Feynman diagrams),
as discussed in the conclusions,
we do not anticipate any significant complications arising from
round-off error when constructing an NLO program.

The amplitude $A_{6;1}^{\NeqZero}(1,3)$ is given by
\begin{equation}
A_{6;1}^{\NeqZero}(1,3)
= \cg \left[ \Cuth_6(1,3) + \Remaining_6(1,3)\, \right] \,,
\end{equation}
where $\Cuth_6(1,3)$ is given in \eqn{BBST},
\begin{equation}
\Remaining_6(1,3) =  - \Inf\Cuth_6(1,3) + \Remaining^a_6(1,3)\,,
\label{totR6hat13}
\end{equation}
and $\Inf\Cuth_6(1,3)$ is given in \eqn{eq:definition_of_InfC} with
$n=6$ and $m=3$.  After simplification, the result for
$\Remaining^a_6(1,3)$ is,
\begin{eqnarray}
 &&
\hskip - .5 cm
\Remaining^a_6(1,3) \nonumber\\
&&  =
i \Biggl[
-\frac{\spb{2}.{4}^{3}}{3 \spa{5}.{6}^{2}\spb{1}.{2}\spb{1}.{3}\spb{3}.{4}}
-\frac{2 \spa{1}.{3}^{4}}
    {9 \spa{1}.{2}\spa{1}.{6}\spa{2}.{3} \spa{3}.{4}\spa{4}.{5}\spa{5}.{6}}
\nonumber\\&& \null
+\frac{\spa{1}.{3}^{2}\spa{1}.{4}\spb{2}.{4}}
      {6 \spa{1}.{6}\spa{2}.{3}\spa{2}.{4}\spa{4}.{5}\spa{5}.{6}\spb{2}.{3}}
+\frac{\sand{1}.{(2+5)}.{4}\spa{1}.{3}^{2}\spa{1}.{5}}
{2 \spa{1}.{4}\spa{1}.{6}\spa{2}.{5}^{2}\spa{3}.{4}\spa{5}.{6}\spb{3}.{4}}
\nonumber\\&& \null
-\frac{\sand{1}.{(3+5)}.{4}^{3}\sand{1}.{(4+5)}.{3}\spb{3}.{5}}
 {3 \sand{2}.{(4+5)}.{3}\sand{6}.{(1+2)}.{3}\spa{1}.{2}\spa{1}.{6}
                                     \spa{4}.{5}\spb{3}.{4}^{2} s_{345}}
\nonumber\\&& \null
-\frac{\spb{2}.{4}^{3}\spb{3}.{6}^{3}}
      {3 \sand{5}.{(2+4)}.{3}^{2}\spb{1}.{3}\spb{1}.{6}\spb{2}.{3}\spb{3}.{4}}
-\frac{\spa{1}.{3}^{3}\spa{1}.{5}\spb{2}.{5}}
  {6 \sand{1}.{(3+4)}.{2}\spa{1}.{6}\spa{2}.{4}\spa{2}.{5}
                                           \spa{3}.{4}\spa{5}.{6}}
\nonumber\\&& \null
-\frac{\spa{1}.{3}^{2}\spa{1}.{5}(\sandmp{1}.{3(2+4)}.{5}+
4 s_{24}\spa{1}.{5})\spb{2}.{4}\spb{5}.{6}}
 {6 \sand{1}.{(2+3)}.{4}\sand{1}.{(3+4)}.{2}\spa{2}.{4}
 \spa{2}.{5}\spa{4}.{5}\spa{5}.{6}s_{234}}
\nonumber\\&& \null
-\frac{s_{26}\sand{1}.{(4+5)}.{3}^{3}\spb{4}.{5}^{3}}
 {3 \sand{1}.{(3+4)}.{5}\sand{2}.{(4+5)}.{3}^{2}
 \sand{6}.{(1+2)}.{3}^{2}\spb{3}.{4}s_{345}}
\nonumber\\&& \null
-\frac{\sand{5}.{(1+3)}.{6}\spb{2}.{6}s_{245}^{2}}
 {3 \sand{4}.{(1+3)}.{6}\sand{5}.{(2+4)}.{3}\spa{2}.{6}\spa{4}.{5}
                                      \spa{5}.{6}\spb{1}.{3}\spb{1}.{6}}
\nonumber\\&& \null
-\frac{\spa{1}.{3}^{3}\spa{4}.{6}\spb{2}.{6}}
   {6 \sand{4}.{(1+3)}.{2}\spa{1}.{2}\spa{2}.{6}\spa{3}.{4}
                                        \spa{4}.{5}\spa{5}.{6}}
+\frac{\sand{1}.{(2+3)}.{4}^{2}\spa{1}.{2}\spa{1}.{5}\spb{2}.{4}}
 {2 \sand{1}.{(2+4)}.{3}\spa{1}.{6}\spa{2}.{5}^{2}
                                      \spa{5}.{6}\spb{3}.{4}s_{234}}
\nonumber\\&& \null
+\frac{\spa{1}.{3}^{3}\spa{1}.{5}\spa{2}.{5}\spb{4}.{5}\spb{5}.{6}}
 {6 \sand{1}.{(2+3)}.{4}\sand{2}.{(1+3)}.{6}\spa{1}.{6}
                           \spa{2}.{3}\spa{2}.{4}\spa{4}.{5}\spa{5}.{6}}
\nonumber\\&& \null
+\frac{\sand{1}.{(3+5)}.{4}(\spa{1}.{5} \spb{5}.{4} + \sand{1}.{(3+5)}.{4})
        \spa{1}.{5}^{2}\spb{4}.{5}}
 {6 \spa{1}.{2}\spa{1}.{6}\spa{2}.{5}\spa{4}.{5}\spa{5}.{6}
                                            \spb{3}.{4}^{2}s_{345}}
\nonumber\\&& \null
-\frac{\sand{1}.{(3+5)}.{4}^{2}\spa{1}.{5}^{2}\spb{4}.{5}}
 {2 \sand{1}.{(4+5)}.{3}\spa{1}.{6}\spa{2}.{5}^{2}
            \spa{5}.{6}\spb{3}.{4}s_{345}}
-\frac{\sand{5}.{(1+3)}.{2}^{3}\spa{4}.{6}\spb{4}.{5}}
{3 \sand{6}.{(1+2)}.{3}\spa{4}.{5}^{2}\spa{5}.{6}^{2}
                                       \spb{1}.{2}\spb{1}.{3}s_{123}}
\nonumber\\&& \null
-\frac{(2 \spa{1}.{2} \spb{2}.{4} + \spa{1}.{3} \spb{3}.{4} )
     \sand{1}.{(2+3)}.{4}\spa{1}.{2}\spb{2}.{4}}
{6 \spa{1}.{6}\spa{2}.{4}\spa{2}.{5}\spa{5}.{6}\spb{3}.{4}^{2}s_{234}}
\nonumber\\&& \null
+\frac{\spa{1}.{3}^{2}(\spa{1}.{2}^{2}\spa{4}.{5}\spb{2}.{4}-
\spa{1}.{5}^{2}\spa{2}.{4}\spb{4}.{5})}
{6 \spa{1}.{2}\spa{1}.{6}\spa{2}.{4}\spa{2}.{5}\spa{3}.{4}\spa{4}.{5}\spa{5}.{6}\spb{3}.{4}}
\nonumber\\&& \null
-\frac{(\spa{1}.{3} \spb{3}.{2} + 2 \spa{1}.{4} \spb{4}.{2})
         \sand{1}.{(3+4)}.{2}\spa{1}.{4}\spb{2}.{4}}
{6 \spa{1}.{6}\spa{2}.{4}\spa{4}.{5}\spa{5}.{6}\spb{2}.{3}^{2}s_{234}}
\nonumber\\&& \null
+\frac{(s_{45}+s_{56})\spa{1}.{3}^{3}\spb{4}.{6}^{2}}
{3 \sand{1}.{(2+3)}.{4}\sand{2}.{(1+3)}.{6}\spa{2}.{3}\spa{4}.{5}
            \spa{5}.{6}s_{123}}
\nonumber\\&& \null
-\frac{\sand{1}.{(2+4)}.{3}^{3}\spb{2}.{4}^{3}}
{3 \sand{5}.{(2+4)}.{3}\spa{1}.{6}\spa{2}.{4}\spa{5}.{6}
\spb{2}.{3}^{2}\spb{3}.{4}^{2}s_{234}}
\nonumber\\&& \null
+\frac{\sand{6}.{(1+3)}.{2}^{2}\spb{2}.{6}s_{123}}
{3 \sand{4}.{(1+3)}.{2}\spa{2}.{6}^{2}\spa{4}.{5}\spa{5}.{6}
\spb{1}.{2}\spb{1}.{3}\spb{2}.{3}}
\nonumber\\&& \null
+\frac{\sand{5}.{(1+3)}.{2}\sand{6}.{(1+3)}.{2}^{2}\spb{2}.{6}}
{3 \sand{4}.{(1+3)}.{2}\sand{6}.{(1+2)}.{3}\spa{4}.{5}
\spa{5}.{6}^{2}\spb{1}.{2}\spb{1}.{3}}
\nonumber\\&& \null
-\frac{(s_{12}+s_{23})\sand{6}.{(1+3)}.{2}\spa{1}.{6}^{2}\spb{2}.{6}s_{123}}
{6 \sand{4}.{(1+3)}.{2}\sand{6}.{(1+2)}.{3}\spa{1}.{2}
\spa{2}.{6}^{2}\spa{4}.{5}\spa{5}.{6}\spb{1}.{2}\spb{2}.{3}}
\nonumber\\&& \null
-\frac{\sandmp{1}.{(5+6)(2+4)}.{5}^{2}\spa{1}.{5}^{2}\spb{2}.{4}\spb{5}.{6}}
{2 \sand{1}.{(2+4)}.{3}\sand{1}.{(3+4)}.{2}\sand{5}.{(2+4)}.{3}
\spa{2}.{5}^{2}\spa{4}.{5}\spa{5}.{6}s_{234}}
\nonumber\\&& \null
+\frac{\sand{1}.{(3+4)}.{2}\sand{5}.{(1+3)}.{6}\spa{1}.{5}
\spb{3}.{6}s_{245}^{2}}
{3 \sand{2}.{(4+5)}.{3}\sand{4}.{(1+3)}.{6}\sand{5}.{(2+4)}.{3}^{2}
\spa{1}.{6}\spa{4}.{5}\spa{5}.{6}\spb{1}.{6}}
\nonumber\\&& \null
+\frac{\sand{5}.{(1+3)}.{6}\spa{1}.{3}\spb{2}.{5}\spb{3}.{6}
        (\sand{5}.{(2+4)}.{3}\spa{1}.{3}+2 \spa{1}.{5}s_{245})}
{3 \sand{2}.{(4+5)}.{3}\sand{4}.{(1+3)}.{6}\sand{5}.{(2+4)}.{3}
             \spa{1}.{6}\spa{4}.{5}\spa{5}.{6}\spb{1}.{6}}
\nonumber\\&& \null
-\frac{\sandmp{1}.{(2+6)(4+5)}.{6}^{2}\spa{1}.{2}
\spa{1}.{6}\spb{2}.{6}\spb{4}.{5}}
{2 \sand{1}.{(3+4)}.{5}\sand{1}.{(4+5)}.{3}\sand{6}.{(1+2)}.{3}
            \spa{2}.{6}^{2}\spa{4}.{5}\spa{5}.{6}s_{345}}
\nonumber\\&& \null
+\frac{\sandmp{1}.{(2+6)(4+5)}.{6}\spa{1}.{6}
   (\sandmp{1}.{3(4+5)}.{6}+2 s_{45}\spa{1}.{6})\spb{2}.{6}\spb{4}.{5}}
{6 \sand{1}.{(3+4)}.{5}\sand{6}.{(1+2)}.{3}^{2}\spa{2}.{6}
          \spa{4}.{5}\spa{5}.{6} s_{345}}
\nonumber\\&& \null
-\frac{\sandmp{1}.{(2+6)(4+5)}.{2}\spa{1}.{2}
(\sandmp{1}.{3(4+5)}.{2}+2 s_{45}\spa{1}.{2})\spb{2}.{6}\spb{4}.{5}}
{6 \sand{1}.{(3+4)}.{5}\sand{2}.{(4+5)}.{3}^{2}\spa{2}.{5}
\spa{2}.{6}\spa{4}.{5}s_{345}}
\nonumber\\&& \null
+\frac{\sand{1}.{(2+3)}.{4}\sand{1}.{(3+4)}.{2}\spb{2}.{4}
(- \spa{1}.{2}^{2}\spa{4}.{5}\spb{2}.{3}+\spa{1}.{4}^{2}\spa{2}.{5}\spb{3}.{4})}
{2 \sand{1}.{(2+4)}.{3}\spa{1}.{6}\spa{2}.{4}\spa{2}.{5}\spa{4}.{5}
\spa{5}.{6}\spb{2}.{3}\spb{3}.{4}s_{234}}
\nonumber\\&& \null
+\frac{\spa{1}.{3}^{3}\spa{4}.{6}^{2}\spb{2}.{6}(2 \sandmp{1}.{2(5+6)}.{4}
     +\sandmp{1}.{65}.{4}+\spa{1}.{4}\spa{2}.{6}\spb{2}.{6})}
{2 \sand{1}.{(3+4)}.{5}\sand{4}.{(1+3)}.{2}\spa{1}.{4}\spa{2}.{6}^{2}
          \spa{3}.{4}\spa{4}.{5}^{2}\spa{5}.{6}}
\nonumber\\&& \null
-\frac{\sand{5}.{(1+3)}.{6}^{2}\spa{1}.{5}(3
   \sand{2}.{(4+5)}.{6}\spa{1}.{5}
     - \spa{1}.{3}\spa{2}.{5}\spb{3}.{6})\spb{5}.{6}s_{245}}
{6 \sand{2}.{(4+5)}.{6}\sand{4}.{(1+3)}.{6}\sand{5}.{(2+4)}.{3}
\spa{1}.{6}\spa{2}.{5}^{2}\spa{4}.{5}\spa{5}.{6}\spb{1}.{6}\spb{3}.{6}}
\nonumber\\&& \null
-\frac{\spa{1}.{3}^{3}\spa{1}.{5}(\spa{1}.{4}\spa{1}.{5}\spa{2}.{4}
   \spa{2}.{5}\spb{2}.{5}^{2}
   - \sandmp{1}.{52}.{4}\spa{1}.{6}\spa{4}.{5}\spb{5}.{6})}
{2 \sand{1}.{(3+4)}.{2}\sand{4}.{(1+3)}.{6}\spa{1}.{4}\spa{1}.{6}^{2}
          \spa{2}.{4}\spa{2}.{5}^{2}\spa{3}.{4}\spa{5}.{6}}
\nonumber\\&& \null
+\frac{\sandmp{1}.{(2+6)(4+5)}.{2}\spa{1}.{2}
\spb{2}.{6}\spb{4}.{5}}
{2 \sand{1}.{(3+4)}.{5}\sand{1}.{(4+5)}.{3}
    \sand{2}.{(4+5)}.{3}\spa{2}.{5}^{2}\spa{2}.{6}^{2}\spa{4}.{5}s_{345}}
\nonumber\\&& \null\hskip 6mm
\times\bigl(\sandmp{1}.{(2+6)(4+5)}.{2}\spa{1}.{6}\spa{2}.{5}
      +\sandmp{1}.{(2+6)4}.{5}\spa{1}.{2}\spa{2}.{6}\bigr)
\nonumber\\&& \null
-\frac{\sand{5}.{(1+3)}.{6}^{3}}
{3 \sand{2}.{(4+5)}.{3}\sand{2}.{(4+5)}.{6}\sand{5}.{(2+4)}.{3}
        \spa{4}.{5}\spa{5}.{6}\spb{1}.{6}}
\nonumber\\&& \null\hskip 6mm\times
\Biggl(\frac{\spa{4}.{6}\spb{3}.{6}\spb{4}.{5}}
{\spa{4}.{5}\spb{1}.{3}}
+\frac{\spa{1}.{2}\spb{5}.{6} s_{245}}
{\sand{4}.{(1+3)}.{6}\spa{2}.{5}}\Biggr)
\nonumber\\&& \null
+\frac{\spa{1}.{3}^{3}
\bigl(\spa{1}.{5}\spa{2}.{6}\spa{4}.{5}\spb{2}.{5} (
6 \spa{1}.{5}\spa{2}.{4}
 +\spa{1}.{4}\spa{2}.{5})
 +2 \spa{1}.{4}\spa{1}.{6}\spa{2}.{5}^{2}\spa{4}.{6}\spb{2}.{6})}
{6 \sand{4}.{(1+3)}.{6}\spa{1}.{4}\spa{1}.{6}^{2}\spa{2}.{5}^{2}
        \spa{2}.{6}\spa{3}.{4}\spa{4}.{5}\spa{5}.{6}}
\nonumber\\&& \null
-\frac{\spa{1}.{5}^{3}\spb{2}.{4}^{2}\spb{5}.{6}}
{6 \sand{1}.{(2+3)}.{4}\sand{1}.{(3+4)}.{2}
  \sand{5}.{(2+4)}.{3}\spa{2}.{5}\spa{4}.{5}\spa{5}.{6}^{2}
         \spb{2}.{3}\spb{3}.{4}s_{234}}
\nonumber\\&& \null\hskip 6mm\times
\Bigl(2 \spa6.4 \spb4.3\spa{1}.{4}\spa{2}.{5}\spb{2}.{4}^{2}+2
  \spa6.2 \spb2.3\spa{1}.{2}\spa{4}.{5}\spb{2}.{4}^{2}
 -5 s_{24}\spa{1}.{3}\spa{5}.{6}\spb{2}.{3}\spb{3}.{4}\Bigr)
\nonumber\\&& \null
+\frac{\sand{2}.{(1+3)}.{6}\spa{1}.{2}\spb{2}.{6}s_{245}}
{6 \sand{2}.{(4+5)}.{3}\sand{4}.{(1+3)}.{6}\spa{1}.{6}\spa{2}.{5}^{2}
        \spa{2}.{6}^{2}\spa{4}.{5}\spa{5}.{6}\spb{1}.{6}\spb{3}.{6}}
\nonumber\\&& \null\hskip 6mm\times
\Bigl(3 \sand{6}.{(1+3)}.{6}\spa{1}.{6}\spa{2}.{5}^{2}
      - 3 \sand{5}.{(1+3)}.{6}\spa{1}.{5}\spa{2}.{6}^{2}
\nonumber\\&& \null\hskip 2 cm
      - \spa{1}.{3}\spa{2}.{5}\spa{2}.{6}\spa{5}.{6}\spb{3}.{6}\Bigr)
\nonumber\\&& \null
-\frac{\spa{1}.{3}^{3}}
{3 \sand{1}.{(3+4)}.{5}\sand{4}.{(1+3)}.{6}\spa{3}.{4}\spa{5}.{6}}
\Biggl(
\frac{\spa{1}.{2}\spa{1}.{5}^{2}\spb{2}.{5}^{3}}
{\sand{1}.{(3+4)}.{2}\spa{1}.{6}^{2}\spa{2}.{5}}
\nonumber\\&& \null\hskip 3 cm
-\frac{\spa{1}.{4}\spb{2}.{5}^{2}\spb{2}.{6}^{2}}
{\sand{1}.{(3+4)}.{2}\sand{4}.{(1+3)}.{2}}
+\frac{\spa{2}.{4}\spa{4}.{6}^{2}\spb{2}.{6}^{3}}
{\sand{4}.{(1+3)}.{2}\spa{2}.{6}\spa{4}.{5}^{2}}\Biggr)
\nonumber\\&& \null
-\frac{\spb{2}.{6}s_{245}^{2}}
{3 \sand{2}.{(4+5)}.{3}\sand{4}.{(1+3)}.{6}\sand{5}.{(2+4)}.{3}
 \spa{1}.{6}\spa{2}.{5}\spa{2}.{6}\spa{4}.{5}\spa{5}.{6}\spb{1}.{3}\spb{1}.{6}}
\nonumber\\&& \null\hskip 6mm\times\Bigl(
\sand{5}.{(1+3)}.{6}\spa{1}.{2}\spb{1}.{3}(\spa{1}.{2}\spa{5}.{6}
                                        - \spa{1}.{6}\spa{2}.{5})
\nonumber\\&& \null\hskip 6mm\hphantom{\times}
- \sandmp{1}.{(3+6)(1+3)}.{2}\spa{2}.{5}\spa{5}.{6}\spb{3}.{6}\Bigr)
\nonumber\\&& \null
+\frac{\spa{1}.{3}^{3}}
{6 \sand{1}.{(3+4)}.{5}\spa{1}.{4}
\spa{1}.{6}\spa{2}.{5}^{2}\spa{2}.{6}^{2}\spa{3}.{4}
\spa{4}.{5}^{2}\spa{5}.{6}}
\nonumber\\&& \null\hskip 6mm\times
\Bigl(6 \spa{1}.{5}^{2}\spa{2}.{4}\spa{2}.{6}^{2}\spa{4}.{5}\spb{2}.{5}-
 2 \spa{1}.{4}\spa{1}.{5}\spa{2}.{5}\spa{2}.{6}^{2}\spa{4}.{5}\spb{2}.{5}
\nonumber\\&& \null\hskip 6mm\hphantom{\times}
- 6 \spa{1}.{6}^{2}\spa{2}.{4}\spa{2}.{5}^{2}\spa{4}.{6}\spb{2}.{6}
 + 5 \spa{1}.{4}\spa{1}.{6}\spa{2}.{5}^{2}\spa{2}.{6}\spa{4}.{6}
                              \spb{2}.{6} \Bigr)
\nonumber\\&& \null
+  { \spa1.2 \spa1.3^3 \spa1.5^2 \spa2.4 \spb2.5^2 \over
   \sand1.{(3+4)}.5 \sand4.{(1+3)}.6 \spa1.4 \spa1.6^2
   \spa2.5^2 \spa3.4 \spa5.6 }
\nonumber\\&& \null
 - {\spa1.2 \spa1.3^3 \spa2.4^2 \spb2.5 \spb2.6 \over
    2 \sand1.{(3+4)}.5 \sand4.{(1+3)}.6 \spa1.4 \spa2.5
       \spa2.6^2 \spa3.4 \spa4.5}
\nonumber\\&& \null
  + {\spa1.2 \spa1.3^3 \spa1.5 \spa2.4 \spb2.5 \spb2.6 \over
       2 \sand1.{(3+4)}.5 \sand4.{(1+3)}.6 \spa1.4 \spa1.6
          \spa2.5^2 \spa3.4 \spa5.6}
\nonumber\\&& \null
-  {\spa1.3^3 \spa1.4 \spb2.5 \spb2.6 \over 6 \sand1.{(3+4)}.5
    \sand4.{(1+3)}.6 \spa1.6 \spa3.4 \spa4.5 \spa5.6 }
\nonumber\\&& \null
- { \spa1.3^3 \spa2.4 \spb2.5 \spb2.6 \over 2 \sand1.{(3+4)}.5
    \sand4.{(1+3)}.6 \spa2.6 \spa3.4 \spa4.5 \spa5.6 }
\nonumber\\&& \null
 + {\spa1.2^2 \spa1.3^3 \spa4.6 \spb2.5 \spb2.6 \over
  2 \sand1.{(3+4)}.5 \sand4.{(1+3)}.6 \spa1.4 \spa1.6
   \spa2.5  \spa2.6 \spa3.4 \spa5.6} \Biggr]
   \label{Rhmpmppp}
 \,.
\end{eqnarray}

The amplitude $A_{6;1}^{\NeqZero}(1,4)$
can be assembled analogously from eqs.~(\ref{BBST}),
(\ref{eq:definition_of_InfC}) and (\ref{Rhmppmpp}),
\begin{equation}
A_{6;1}^{\NeqZero}(1,4)
= \cg \left[ \Cuth_6(1,4) + \Remaining_6(1,4) \right]\, ,
\end{equation}
where
\begin{equation}
 \Remaining_6(1,4) =  - \Inf\Cuth_6(1,4)
+ \Remaining_6^a(1,4) + \Remaining_6^a(1,4)
\Bigl|_{\rm flip}\,,
\end{equation}
and
where we have used the flip symmetry,
\be
X(1,2,3,4,5,6)\Bigr|_{\rm flip} \equiv X(1,6,5,4,3,2) \, .
\ee
After simplification, we find,
\begin{eqnarray}
&& \hskip -.5 cm \Remaining_6^a(1,4)
\nonumber\\
&& \null =
i \Biggl[
   -\frac{8 \spa{1}.{3}\spa{1}.{4}^{3}}
      {9 \spa{1}.{2}\spa{1}.{6}\spa{2}.{3}\spa{3}.{4}\spa{3}.{5}\spa{5}.{6}}
-\frac{( \spa{1}.{4}\spb{4}.{3} + 2\spa{1}.{5}\spb{5}.{3})\sand{1}.{(4+5)}.{3}
                 \spa{1}.{5}^{2}\spb{3}.{5}}
{6 \spa{1}.{2}\spa{1}.{6}\spa{2}.{5}\spa{3}.{5}\spa{5}.{6}
                                             \spb{3}.{4}^{2}\, s_{345}}
\nonumber\\&& \null
+\frac{\sand{2}.{(1+4)}.{6}\sand{5}.{(1+4)}.{6}\spa{1}.{5}^{2}
                    \spb{5}.{6}s_{146}}
   {2 s_{16}\sand{3}.{(1+4)}.{6}\sand{5}.{(2+3)}.{4}\spa{2}.{3}
                            \spa{2}.{5}^{2}\spa{5}.{6}\spb{4}.{6}}
\nonumber\\&& \null
-\frac{\spb{3}.{5}^{3}\spb{4}.{6}^{3}}
{3 \sand{2}.{(3+5)}.{4}^{2}\spb{1}.{4}\spb{1}.{6}\spb{3}.{4}\spb{4}.{5}}
+\frac{\sand{1}.{(2+4)}.{3}^{3}\spb{2}.{4}}
{3 \sand{5}.{(2+3)}.{4}\spa{1}.{6}\spa{2}.{3}\spa{5}.{6}\spb{3}.{4}^{2}s_{234}}
\nonumber\\&& \null
+\frac{\sand{1}.{(2+4)}.{3}\spa{1}.{2}\spb{2}.{3}
       (\sand{1}.{(2+4)}.{3}+\spa{1}.{2}\spb{2}.{3})}
{6 \spa{1}.{6}\spa{2}.{3}\spa{2}.{5}\spa{5}.{6}\spb{3}.{4}^{2}s_{234}}
\nonumber\\&& \null
+\frac{\sand{1}.{(2+4)}.{3}^{2}\spa{1}.{2}\spa{1}.{5}\spb{2}.{3}}
{2 \sand{1}.{(2+3)}.{4}\spa{1}.{6}\spa{2}.{5}^{2}\spa{5}.{6}\spb{3}.{4}s_{234}}
- \frac{\spa1.4^4} {3 \spa1.2 \spa2.3 \spa3.4 \spa4.5 \spa5.6 \spa6.1}
\nonumber\\&& \null
-\frac{\sand{5}.{(1+4)}.{6}\spa{1}.{4}\spa{1}.{5}\spb{5}.{6}s_{146}}
{6 s_{16}\sand{3}.{(1+4)}.{6}\sand{5}.{(2+3)}.{4}\spa{2}.{3}
                                          \spa{2}.{5}\spa{5}.{6}}
\nonumber\\&& \null
-\frac{\spa{1}.{2}\spa{1}.{4}^{2}\spb{2}.{3}}
{6 \spa{1}.{6}\spa{2}.{3}\spa{2}.{5}\spa{3}.{4}\spa{5}.{6}\spb{3}.{4}}
\Biggl(1+\frac{3 \spa{1}.{5}\spa{2}.{3}}
{\spa{1}.{3}\spa{2}.{5}}\Biggr)
\nonumber\\&& \null
+\frac{\sand{2}.{(1+4)}.{6}^{3}\sand{3}.{(2+5)}.{4}\spb{2}.{3}\spb{4}.{6}}
{3 \sand{2}.{(3+5)}.{4}^{2}\sand{5}.{(1+4)}.{6}
        \sand{5}.{(2+3)}.{4}\spa{2}.{3}^{2}\spb{1}.{4}\spb{1}.{6}}
\nonumber\\&& \null
+\frac{\spa{1}.{4}^{2}\spa{1}.{5}^{2}\spb{3}.{5}}
      {6 \spa{1}.{2}\spa{1}.{6}\spa{2}.{5}\spa{3}.{4}\spa{3}.{5}
                                                \spa{5}.{6}\spb{3}.{4}}
\Biggl(1+\frac{3 \spa{1}.{2}\spa{3}.{5}}
              {\spa{1}.{3}\spa{2}.{5}}\Biggr)
\nonumber\\&& \null
+\frac{\spa{1}.{4}\spa{1}.{5}\spb{2}.{3}\spb{5}.{6}}
     {6 \sand{1}.{(3+4)}.{2}\spa{2}.{5}\spa{5}.{6}s_{234}}
\Biggl(\frac{\spa{1}.{4}}
{\spa{2}.{3}}+\frac{3 \sand{1}.{(2+4)}.{3}\spa{1}.{5}}
{\sand{1}.{(2+3)}.{4}\spa{2}.{5}}\Biggr)
\nonumber\\&& \null
-\frac{\sand{1}.{(3+5)}.{4}^{4}\spb{3}.{5}^{3}}
      {6 \sand{2}.{(3+5)}.{4}\sand{6}.{(3+5)}.{4}\spa{1}.{2}
         \spa{1}.{6}\spa{3}.{5}\spb{3}.{4}^{2}\spb{4}.{5}^{2}s_{345}}
\nonumber\\&& \null
+\frac{\sand{1}.{(3+5)}.{4}^{4}\spa{2}.{6}\spb{2}.{6}\spb{3}.{5}^{4}}
{6 \sand{1}.{(3+4)}.{5}\sand{1}.{(4+5)}.{3}
  \sand{2}.{(3+5)}.{4}^{2}\sand{6}.{(3+5)}.{4}^{2}
                  \spb{3}.{4}\spb{4}.{5} s_{345}}
\nonumber\\&& \null
+\frac{\sandmp{1}.{(2+6)(3+5)}.{6}^{2}\spa{1}.{6}\spb{2}.{6}\spb{3}.{5}}
{3 \sand{1}.{(3+4)}.{5}\sand{1}.{(4+5)}.{3}\sand{6}.{(3+5)}.{4}
    \spa{2}.{6}\spa{3}.{6}\spa{5}.{6}s_{345}}
\nonumber\\&&\null \hskip 8mm\times
\Biggl(\frac{\spa{1}.{4}}
           {2 \spa{3}.{5}}
   -\frac{\spa{1}.{6}\spb{3}.{5}}
         {\sand{6}.{(3+5)}.{4}}\Biggr)
\nonumber\\&& \null
-\frac{\sand{5}.{(2+3)}.{6}\spb{5}.{6}s_{146}^{2}
   (\spb{1}.{4}\spa{1}.{2}\spa{1}.{5}
        + \spa{2}.{5}\spa{1}.{6}\spb{4}.{6} ) }
   {3 \sand{2}.{(3+5)}.{4}\sand{3}.{(1+4)}.{6}\sand{5}.{(2+3)}.{4}
    \spa{1}.{6}\spa{2}.{3}\spa{5}.{6}\spb{1}.{6}\spa{2}.{5} \spb{1}.{4} }
\nonumber\\&& \null
-\frac{\spa{1}.{5}^{2}\spb{2}.{3}^{2}\spb{5}.{6}}
   {6 \sand{1}.{(3+4)}.{2}\sand{5}.{(2+3)}.{4}\spa{2}.{5}\spa{5}.{6}s_{234}}
\nonumber\\&& \null\hskip8mm\times
\Biggl(3 \spa{1}.{4}
+\frac{3 \sand{1}.{(2+4)}.{3}\spa{1}.{5}\spa{2}.{3}}
                          {\sand{1}.{(2+3)}.{4}\spa{2}.{5}}
-\frac{2 \spa{1}.{5}\spa{2}.{6}\spb{2}.{3}}
                          {\spa{5}.{6}\spb{3}.{4}}\Biggr)
\nonumber\\&& \null
+\frac{\spa{1}.{4}^{3}\spa{1}.{5}\spb{2}.{5}}
    {6 \sand{1}.{(3+4)}.{2}\sand{3}.{(1+4)}.{6}\spa{1}.{3}\spa{1}.{6}^{2}
                          \spa{2}.{3}\spa{2}.{5}^{2}\spa{3}.{4}\spa{5}.{6}}
\nonumber\\&& \null\hskip 8mm\times
       \Bigl((\sandmp{1}.{6(2+5)}.{3}+\sandmp{1}.{(5+6)2}.{3})
      (\spa{1}.{3}\spa{2}.{5}-3 \spa{1}.{5}\spa{2}.{3})
\nonumber\\&& \null\hskip8mm\hphantom{\times}
-3 \spa{1}.{2}\spa{1}.{5}\spa{2}.{3}\spa{3}.{5}\spb{2}.{5}\Bigr)
\nonumber\\&& \null
+\frac{\spa{1}.{4}^{3}\spa{3}.{6}\spb{2}.{6}}
    {6 \sand{1}.{(3+4)}.{5}\sand{3}.{(1+4)}.{2}\spa{1}.{2}\spa{1}.{3}
        \spa{2}.{6}^{2}\spa{3}.{4}\spa{3}.{5}^{2}\spa{5}.{6}}
\nonumber\\&& \null\hskip 8mm\times
\Bigl( 2 (\sandmp{1}.{2(5+6)}.{3}+\sandmp{1}.{(2+6)5}.{3})
   \spa{1}.{3}\spa{2}.{6}
\nonumber\\&& \null\hskip 2cm\hphantom{\times}
-3 (2 \sandmp{1}.{2(5+6)}.{3}+\sandmp{1}.{65}.{3})\spa{1}.{6}\spa{2}.{3}
\Bigr)
\nonumber\\&& \null
-\frac{\sand{1}.{(4+5)}.{3}\spa{1}.{5}^{2}\spb{3}.{5}}
{2 \sand{1}.{(3+5)}.{4}\spa{1}.{2}\spa{1}.{6}\spa{2}.{3}\spa{2}.{5}
\spa{3}.{5}\spa{5}.{6}\spb{3}.{4}s_{345}}
\nonumber\\&& \null\hskip 8mm\times
\Biggl(\frac{\sandmp{1}.{(2+6)(3+5)}.{2}\spa{1}.{2}\spa{3}.{5}}
{\spa{2}.{5}}
- \sand{1}.{(3+4)}.{5}\spa{1}.{3}\spa{2}.{5}
\Biggr)
\nonumber\\&& \null
+\frac{\sand{2}.{(1+4)}.{6} s_{146}^{2}}
 {3 s_{16}\sand{2}.{(3+5)}.{4}^{2}\sand{3}.{(1+4)}.{6}\spa{2}.{3}
                                    \spa{2}.{5}\spa{2}.{6}\spb{1}.{4}}
\nonumber\\&& \null\hskip 8mm\times
\Biggl(\frac{\spa{1}.{6}\spa{2}.{5}\spa{2}.{6}\spb{4}.{6}^{2}\spb{5}.{6}}
    {\sand{5}.{(2+3)}.{4}}
-\frac{\sand{2}.{(1+4)}.{6}\spa{1}.{2}^{2}\spb{1}.{4}\spb{2}.{6}}
                {\sand{5}.{(2+3)}.{6}}
\Biggr)
\nonumber\\&& \null
+\frac{\sand{2}.{(1+4)}.{6}^{2}\spa{1}.{2}\spb{2}.{6} s_{146}}
      {2 s_{16}\sand{2}.{(3+5)}.{4}\sand{3}.{(1+4)}.{6}
  \sand{5}.{(1+4)}.{6}\spa{2}.{3}\spa{2}.{5}
  \spa{2}.{6}\spa{5}.{6}\spb{4}.{6}}
\nonumber\\&& \null\hskip 8mm\times
\Biggl(\frac{\sand{5}.{(1+4)}.{6}\spa{1}.{5}\spa{2}.{6}}
       {\spa{2}.{5}}
-\frac{\sand{6}.{(1+4)}.{6}\spa{1}.{6}\spa{2}.{5}}
        {\spa{2}.{6}}
+\frac{\spa{1}.{4}\spa{5}.{6}\spb{4}.{6}}
          {3}\Biggr)
\nonumber\\&& \null
+\frac{\sandmp{1}.{(2+6)(3+5)}.{6}^{2}\spa{1}.{6}\spb{2}.{6}\spb{3}.{5}}
     {\sand{1}.{\!(3+4)\!}.{5}\! \sand{1}.{\!(3+5)\!}.{4}\!
        \sand{1}.{\!(4+5)\!}.{3} \!
 \sand{6}.{\!(3+5)\!}.{4} \!\spa{2}.{3}\spa{2}.{6}\spa{3}.{5} \spa{3}.{6}
      \spa{5}.{6}}
\nonumber\\&& \null\hskip 8mm\times
\frac{1}{2\, s_{345}}
\Biggl(
\frac{\sand{1}.{(3+4)}.{5}\spa{1}.{3}\spa{2}.{6}\spa{3}.{5}}
                  {\spa{3}.{6}}
-\frac{\sandmp{1}.{(2+6)(3+5)}.{2}\spa{1}.{2}\spa{3}.{6}}
                                                {\spa{2}.{6}}\Biggr)
\nonumber\\&& \null
%
%
+ {\spa1.4^3 \spa1.2 \spa2.3 \over \spa3.4 \spa5.6 \spa6.1^2
    \spa1.3 \spa3.5^2 \sand1.{(3+4)}.5  \sand3.{(1+4)}.6 }
\nonumber\\&& \null\hskip 8mm\times
    \Biggl(
  \frac{1}{3} {\spa1.3^2 \spa3.5^2 \spa1.6^2 \spb2.5^2 \spb2.6^2 \over
      \spa1.2 \spa2.3 \sand1.{(3+4)}.2 \sand3.{(1+4)}.2}
  - \frac{1}{3} {\spa1.3 \spa1.5^2 \spa3.5^2 \spb2.5^3 \over
        \spa2.3 \spa2.5 \sand1.{(3+4)}.2 }
\nonumber\\&& \null\hskip 12mm
  - \frac{1}{3} { \spa1.3 \spa3.6^2 \spa1.6^2 \spb2.6^3 \over
              \spa1.2 \spa2.6 \sand3.{(1+4)}.2 }
\nonumber\\&& \null \hskip 12mm
  + \frac{1}{3}  {\spa1.3 \over \spa1.2 \spa2.3 \spa2.5 \spa2.6}
         \Bigl(\sandmp1.{(2+6)5}.3 + \sandmp1.{6(2+5)}.3 \Bigr)
\nonumber\\&& \null \hskip 22mm \times
         \Bigl(\spa1.6 \spa3.6 \spa2.5 \spb2.6
          + \spa1.5 \spa3.5 \spa2.6 \spb2.5 \Bigr)
 \nonumber\\&& \null\hskip 12mm
 + \frac{1}{2} \spb2.5 \spb2.6 \spa6.1 \spa3.5
        \biggl( {1\over 3} { \spa1.3^2 \over \spa1.2 \spa2.3 }
     + {\spa1.5\spa3.5 \over \spa2.5^2}
     + {\spa1.3\spa3.5 \over \spa2.3 \spa2.5}
    + {\spa1.6 \spa3.6 \over \spa2.6^2}
 \nonumber\\&& \null\hskip 51mm
    + {\spa1.3\spa1.6 \over \spa1.2 \spa2.6} \biggr)
 \nonumber\\&& \null\hskip 12mm
 + {\sandmp1.{(2+6)5}.3  \spa3.5^2 \spb2.5 \over \spa2.5^2 }
      { (\spa1.5 \spa2.3 + \spa1.3 \spa2.5 ) \over \spa2.3^2}
 \nonumber\\&& \null\hskip 12mm
  + {\sandmp1.{6(2+5)}.3  \spa1.6^2  \spb2.6 \over \spa2.6^2}
      {(\spa1.2 \spa3.6 + \spa1.3 \spa2.6 ) \over \spa1.2^2}
\nonumber\\&& \null\hskip 12mm
  +  {\spa1.3^2 \over \spa1.2^2 \spa2.3^2}
      \Bigl(\sandmp1.{(2+6)5}.3  \sandmp1.{25}.3
       + \sandmp1.{62}.3  \sandmp1.{25}.3
\nonumber\\&& \null \hskip 37mm
       + \sandmp1.{6(2+5)}.3  \sandmp1.{62}.3 \Bigr) \Biggr)\Biggr]
\,. \label{Rhmppmpp}
\end{eqnarray}


\end{document}